\documentclass[twocolumn]{emulateapj}
\usepackage[bookmarks,bookmarksopen,colorlinks,linkcolor={blue},
  citecolor={blue},urlcolor={blue}]{hyperref}
\usepackage{amsmath}
\usepackage{natbib}
\usepackage{color}
\usepackage[x11names]{xcolor}
\usepackage{graphics}
\usepackage{enumerate}
\usepackage{setspace}

\newcommand{\nobs}[1]{$n_{\text{obs}}$#1}
\newcommand{\mps}[1]{m s$^{-1}$#1}
\newcommand{\prot}[1]{$P_{\text{rot}}$#1}
\newcommand{\vsini}[1]{$v\sin{i_s}$#1}
\newcommand{\msini}[1]{$m_p\sin{i}$#1}

\newcommand{\cdbox}[1]{%
  \colorlet{currentcolor}{.}%
  {\color{Blue1}%
    \dbox{\color{currentcolor}#1}}%
}
\usepackage{dashbox,framed,color,ocg-p}
\newcommand{\ToggleLayer}[2]{%
  \leavevmode
  \pdfstartlink user {
    /Subtype /Link
    /Border [0 0 0]%
    /A <<
      /S/JavaScript
      /JS (
         var aOCGs = this.getOCGs(), Layer;
         var Layers = "#1".split(","), Active = -1, i, l;
         for (l=0; l<Layers.length; l++) {
           Layer = Layers[l];
           for (i=0; aOCGs && i<aOCGs.length; i++) {
             if (aOCGs[i].state && aOCGs[i].name == Layer) {
               Active = l;
               aOCGs[i].state = false;
             }
           }
           if (Active >= 0) break;
         }
         if (Active == -1) {
           for (l=0; l<Layers.length; l++) {
             if (Layers[l] == "") Active = l;
           }
         }
         Active = Active + 1;
         if (Active == Layers.length) Active = 0;
         Layer = Layers[Active];
         for (i=0; aOCGs && i<aOCGs.length; i++) {
           if (aOCGs[i].name == Layer) aOCGs[i].state = true;
         }
      )
    >>
  }#2%
  \pdfendlink
}

\shortauthors{Cloutier et al.}
\shorttitle{SPIRou Legacy Survey-Planet Search Simulations}

\begin{document}
\title{Predictions of planet detections with near infrared radial velocities in the
  up-coming SPIRou legacy survey-planet search}
\author{Ryan Cloutier\altaffilmark{1,2,3}}
\author{\'{E}tienne Artigau\altaffilmark{3}}
\author{Xavier Delfosse\altaffilmark{4}}
\author{Lison Malo\altaffilmark{3,5}}
\author{Claire Moutou\altaffilmark{5,6}}
\author{Ren\'{e} Doyon\altaffilmark{3}}
\author{Jean-Francois Donati\altaffilmark{7,8}}
\author{Andrew Cumming\altaffilmark{9}}
\author{Xavier Dumusque\altaffilmark{10}}
\author{\'Elodie H\'ebrard\altaffilmark{11}}
\author{Kristen Menou\altaffilmark{2,1}}

\altaffiltext{1}{Dept. of Astronomy \& Astrophysics, University
of Toronto. 50 St. George Street, Toronto, Ontario, M5S 3H4, Canada}
\altaffiltext{2}{Centre for Planetary Sciences,
Dept. of Physical \& Environmental Sciences, University of
Toronto Scarborough. 1265 Military Trail, Toronto, Ontario, M1C 1A4, Canada}
\altaffiltext{3}{Institut de recherche sur les exoplan\`{e}tes,
D\'{e}partement de physique, Universit\'{e} de Montr\'{e}al.
2900 boul. Édouard-Montpetit, Montr\'{e}al, Quebec, H3T 1J4, Canada}
\altaffiltext{4}{Universit\'{e} Grenoble Alpes, CNRS, IPAG, F-38000 Grenoble, France}
\altaffiltext{5}{CFHT Corporation, 65-1238 Mamalahoa Hwy, Kamuela, HI 96743, USA}
\altaffiltext{6}{Aix Marseille Universit\'e, CNRS, LAM, Laboratoire d'Astrophysique de Marseille, Marseille, France} 
\altaffiltext{7}{Universit\'e de Toulouse, UPS-OMP, IRAP, 14 avenue E. Belin, Toulouse, F-31400 France}
\altaffiltext{8}{CNRS, IRAP / UMR 5277, Toulouse, 14 avenue E. Belin, F-31400 France}
\altaffiltext{9}{Department of Physics and McGill Space Institute, McGill University, 3600 rue University, Montr\'eal, Quebec H3A 2T8, Canada}
\altaffiltext{10}{Observatoire Astronomique de l'Universit\'{e} de Gen\`{e}ve, 51 Chemin des Maillettes, 1290 Versoix, Switzerland}
\altaffiltext{11}{Department of Physics and Astronomy, York University, Toronto, Ontario L3T 3R1, Canada}

\begin{abstract}
  The SPIRou near infrared spectro-polarimeter is
  destined to begin science operations at the Canada-France-Hawaii Telescope
  in mid-2018. One of the instrument's primary science goals is to discover the closest
  exoplanets to the Solar System by conducting a 3-5 year long radial velocity
  survey of nearby M dwarfs at an expected precision of $\sim 1$ \mps{;} the SPIRou
  Legacy Survey-Planet Search (SLS-PS). In this study we conduct a detailed Monte-Carlo
  simulation of the SLS-PS using our current understanding of the occurrence rate
  of M dwarf planetary systems and physical models of stellar activity.
  From simultaneous modelling of planetary signals and activity, we predict the population
  of planets detected in the SLS-PS. With our fiducial survey strategy and expected
  instrument performance over a nominal survey length of $\sim 3$ years,
  we expect SPIRou to detect $85.3^{+29.3}_{-12.4}$ planets including
  $20.0^{+16.8}_{-7.2}$ habitable zone planets and $8.1^{+7.6}_{-3.2}$ Earth-like planets
  from a sample of 100 M1-M8.5 dwarfs out to 11 pc. By studying
  mid-to-late M dwarfs previously inaccessible to existing optical velocimeters, SPIRou will
  put meaningful constraints on the occurrence rate of planets around those stars
  including the value of $\eta_{\oplus}$ at an expected level of precision of $\lesssim 45$\%.
  We also predict a subset of $46.7^{+16.0}_{-6.0}$ planets may be accessible with
  dedicated high-contrast imagers on the next generation of ELTs including $4.9^{+4.7}_{-2.0}$ potentially imagable
  Earth-like planets. Lastly, we compare the results of our fiducial survey strategy to other foreseeable survey
  versions to quantify which strategy is optimized to reach the SLS-PS science goals.
  The results of our simulations are made available to the community on
  \href{https://github.com/r-cloutier/SLSPS\_Simulations}{github}. 
\end{abstract}

\section{Introduction} 
The radial velocity method of detecting exoplanets is one of the most
successful methods of exoplanet detection
and has been widely used since the first discovery of an
exoplanet around a main-sequence star over two decades ago \citep{mayor95}. 
Since then numerous international teams have successfully built and used
precision velocimeters to grow the population of radial velocity (RV) planets
(e.g. HARPS; \citealt{mayor03}, HARPS-N; \citealt{cosentino12}, HIRES; \citealt{vogt94}).
Up until recently the majority of 
these precision velocimeters have operated in the visible wavelength
regime where their sensitivity is maximized for the discovery of planets
around Sun-like stars. 

In recent years much interest has been generated regarding the population of
exoplanets around M dwarfs. M dwarfs, with effective temperatures
$\lesssim 3800$ K and masses $\lesssim 0.6$ M$_{\odot}$, outnumber Sun-like
stars in the Solar neighbourhood (within $\sim 10$ pc) nearly 4:1
\citep{henry09}. Furthermore, M dwarfs are known to frequently host multiple
small ($r_p \leq 4$ R$_{\oplus}$) planets \citep[e.g.][]{dressing15a, gaidos16}
including a large fraction of planets within the star's habitable zone (HZ)
which itself spans shorter orbital periods than around the more luminous
Sun-like stars \citep{kasting93, kopparapu13}.
Lastly the amplitude of the radial velocity signal induced by a given planet
is larger around M dwarfs than around Sun-like stars owing to their smaller
masses. These favorable qualities have many astronomers 
committed to uncovering the M dwarf exoplanet population with purpose-built
transit (e.g. MEarth; \citealt{irwin15}, ExTrA \citealt{bonfils15},
TRAPPIST; \citealt{gillon11}, SPECULOOS; \citealt{gillon13})
and radial velocity instrumentation (e.g.
SPIRou; \citealt{delfosse13, artigau14}, NIRPS; \citealt{bouchy17}, CARMENES;
\citealt{quirrenbach14}, HPF; \citealt{mahadevan12}, IRD; \citealt{tamura12}).

One particular precision velocimeter optimized for the detection of
exoplanets around M dwarfs in radial velocity is \emph{SPIRou} \citep[Un
  Spectro-Polarim\`{e}tre Infra-Rouge;][]{delfosse13, artigau14}. SPIRou
is a high-resolution near-infrared velocimeter whose first-light is scheduled on
the Canada-France-Hawaii Telescope (CFHT) in 2018. A significant fraction of SPIRou's
allocated time will be spent surveying
nearby M dwarfs searching for new exoplanets in a campaign known as the
\emph{SPIRou Legacy Survey-Planet Search} (SLS-PS). With the increased
sensitivity to cool M dwarfs enabled by nIR detectors, the planet detections
resulting from the
SLS-PS will be able to constrain the occurrence rate of planets around stars
later than $\sim$ M4.5 and find the closest exoplanetary
systems, beyond Proxima Centauri \citep[1.3 pc;][]{angladaescude16}, which
may be amenable to direct imaging with the next generation of imagers on-board
an Extremely Large Telescope (ELT).

In this study we present a comprehensive simulation of the SLS-PS to estimate its
planet yield as well as the bulk properties of the detected SPIRou planet population.
These simulations were performed for a variety
of survey strategies which enabled the SPIRou science team to establish an
experimental setup which optimizes both the detection sensitivity and survey
yield given the nominal time allocation of the SLS-PS. The main results of this
study are based on the survey version deemed to be optimal for meeting the 
science goals of SPIRou. However, the results of the
various surveys are summarized in the final Sect.~\ref{sect:surveys}. Comparison
of the various survey versions may be useful to inform other up-coming radial velocity
planet searches similar to the SLS-PS.

The paper is organized as follows:

\begin{itemize}
\item Sect.~\ref{sect:spirou} gives an overview of the important aspects of the
  SPIRou spectro-polarimeter.
\item Sect.~\ref{sect:starsample} describes the stellar input catalog.
\item Sect.~\ref{sect:survey} describes the simulated SLS-PS.
\item Sect.~\ref{sect:planetsample} describes the population of simulated
  planetary systems. 
\item Sects.~\ref{sect:GP}-\ref{sect:detection} describe how we mitigate the
  effects of stellar activity and detect planets.
\item Sects.~\ref{sect:sensitivity}-\ref{sect:yield} describe the results of
  the survey.
\item Sect.~\ref{sect:2occ} considers the effect of an increased planet frequency
  on the survey results.
\item Sect.~\ref{sect:measurements} describes how well we can measure
  planet occurrence rates based on the results of the SLS-PS.
\item Sect.~\ref{sect:imaging} discusses the potential for targeting SPIRou
  planets in direct imaging campaigns with ELTs
\item and Sect.~\ref{sect:surveys} compares the merits of various potential
  versions of SLS-PS with the fiducial version presented throughout this paper.
\end{itemize}

\section{Un Spectro-Polarim\`{e}tre Infra-Rouge} \label{sect:spirou}
SPIRou is an up-coming nIR \'echelle spectro-polarimeter and high-precision velocimeter whose first
light is scheduled for 2018 on the Canada-France-Hawaii Telescope on Maunakea.
The instrument is optimized to observe exoplanets via the radial
velocity technique around low mass stars and to study the magnetic fields of young embedded protostars
\citep{delfosse13}. 
The design of SPIRou is intended to address its main science goals of detecting and characterizing
M dwarf exoplanetary systems and to investigate the role that magnetic fields have on the processes of
star and planet formation. SPIRou can be considered a heritage instrument which is built upon the success
of the previous generation of optical spectro-polarimeters and high-precision velocimeters such as the ESPaDOnS
spectro-polarimeter \citep{donati06} as well as the RV spectrographs SOPHIE \citep{bouchy06}
and HARPS \citep{mayor03}.

Here we provide a brief overview of the main instrument specifications as they pertain to the detection of
new exoplanetary systems around nearby M dwarfs.  Details of the optical and mechanical design
of the instrument can be found in \cite{artigau14}. 
The instrument itself is a fiber-fed, bench-mounted, double-pass, cross-dispersed, spectro-polarimeter
that is cryogenically cooled to an operation temperature of 80 K and provides simultaneous spectroscopic and
polarimetric observations. The optical fiber-link connecting the
Cassegrain unit---used for polarimetric analysis and guiding---to the calibration module and spectrograph is made
from purified fluoride; a special optical material featuring improved transmission at wavelengths $>2.0$ $\mu$m thus
enabling the inclusion of the $K$ band. The inclusion of the $K$ band in the SPIRou spectral coverage
is unique among most nIR velocimeters and is highly desirable for the velocimetry of late M dwarfs as
a large fraction of the RV information content is contained in the $K$ band (Artigau et al. 2017 in prep).
With the broad continuous spectral coverage of SPIRou spanning the nIR $YJHK$ bands
($0.98-2.35 \mu$m), SPIRou will pioneer infrared planet searches by targeting low mass stars whose flux
peaks in the nIR wavelength domain. In order to detect Earth-size planets, SPIRou is required to achieve a
long-term RV precision of 1 \mps{} while simultaneously monitoring the star's intrinsic activity which is enabled
by its high spectral resolution ($\lambda/ \Delta \lambda = 70,000$).

Thanks to its spectro-polarimetric capabilities, SPIRou is also optimized for characterizing stellar activity---an
obvious asset for studying M dwarfs---and particularly late-type M dwarfs which are known for their significant levels
of magnetic activity \citep{west15}. This will allow users to i) minimize the impact of activity on RV curves and
ease planet detections and ii) to characterize the impact of stellar activity on the close-in habitable zone
planets that SPIRou will detect.

\section{Stellar Input Catalog} \label{sect:starsample}
\subsection{Stellar Sample} \label{sect:starsamplesub}
The SPIRou input catalog used in our simulated SLS-PS (see Sect.~\ref{sect:survey})
contains 100 stars visible from CFHT on Maunakea
($\delta \gtrsim -30^{\circ}$). We note that the stellar sample used in these simulations is intended to be an
approximation to the true SPIRou input catalog which has yet to be formalized exactly. The stars chosen
were selected based on their high scores in the
SPIRou merit function. The merit function is based on being able to detect the RV semi-amplitude
from the gravitational pull of an 3 M$_{\oplus}$ planet at an equilibrium temperature of 250 K (Malo et al. in prep).
Of the optimum selection of 120 stars, approximately 20 were
deemed to result in poor detection sensitivities based on 
the measured fractions of simulated planets detected around those stars in preliminary simulations
of the SLS-PS. The primary
culprit for the rejection of these stars was their large projected rotation
velocities \vsini{,} which have a large detrimental effect on the RV measurement
precision and hence on our ability to detect planets. 

\begin{figure*}
  \centering
  \includegraphics[width=\hsize]{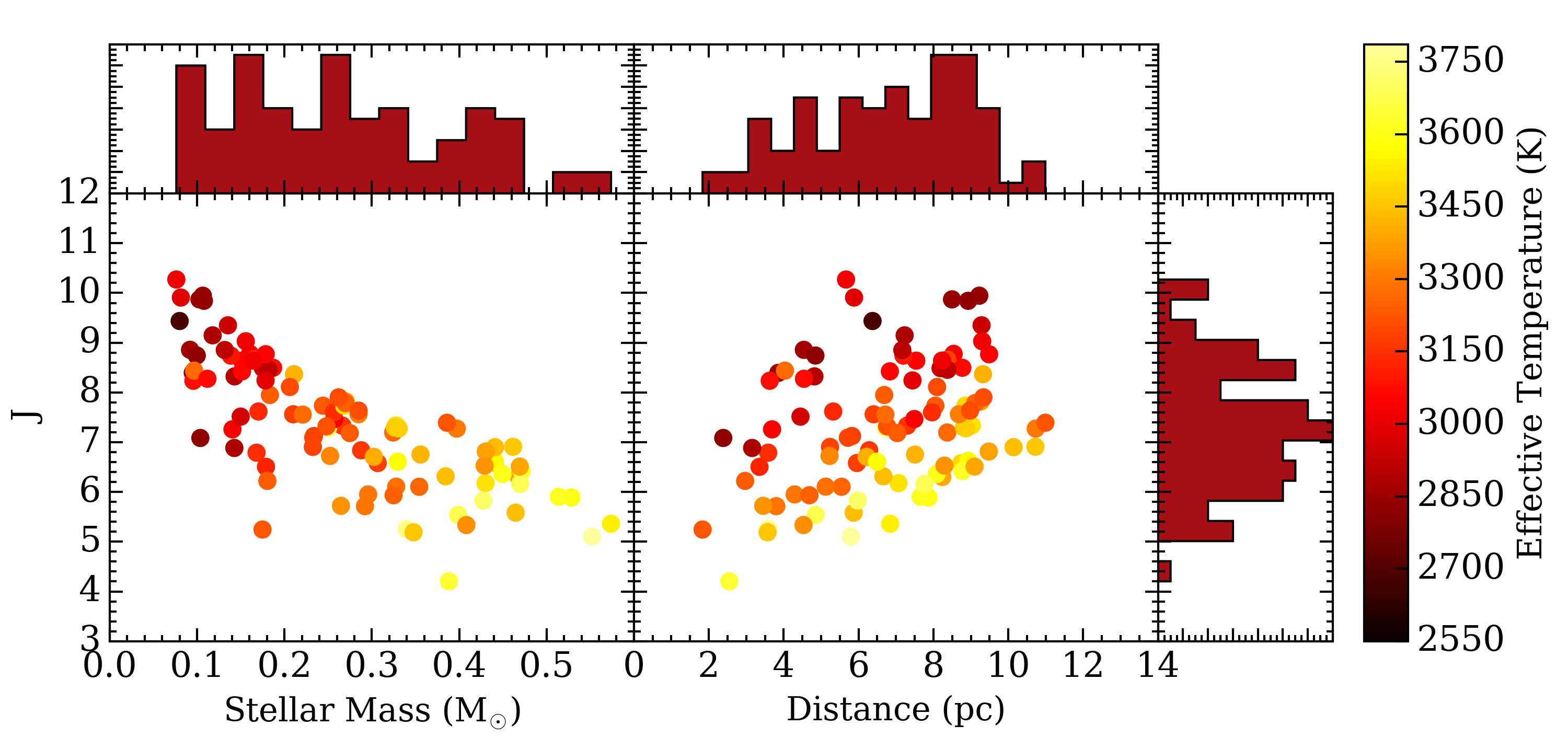}%
  \hspace{-\hsize}%
  \begin{ocg}{fig:staroff}{fig:staroff}{0}%
  \end{ocg}%
  \begin{ocg}{fig:staron}{fig:staron}{1}%
   \includegraphics[width=\hsize]{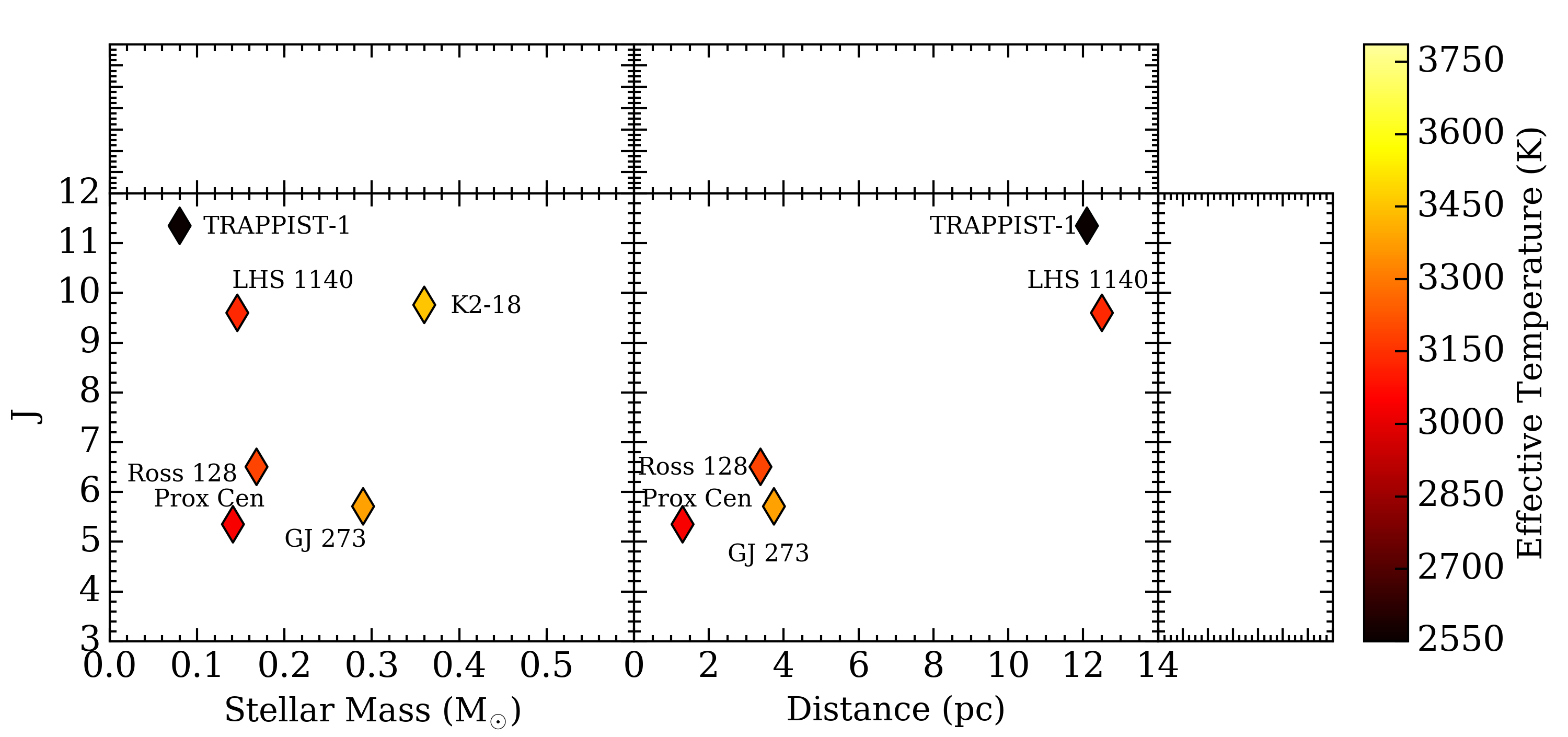}%
  \end{ocg}
  \hspace{-\hsize}%
  \caption{Scatter plots and histograms depicting the distribution of SPIRou input catalog
    $J$ band magnitudes, stellar masses, distances, and effective temperatures for our fiducial 
    version of the \emph{SPIRou Legacy Survey-Planet Search}. Histograms are
    in linear units. Nearby M dwarf planetary systems with at least one known HZ planet
    are depicted with
    \ToggleLayer{fig:staron,fig:staroff}{\protect\cdbox{\emph{diamonds}}}. K2-18 at 34 pc does not
    appear in the scatter plot in right panel.}
  \label{fig:stellardist}
\end{figure*}

Properties of stars in the SPIRou input catalog are presented in Fig.~\ref{fig:stellardist}.
For comparison purposes, in Fig.~\ref{fig:stellardist} we include six nearby M dwarf
planetary systems with at least one known planet in or near the HZ: 
Proxima Centauri \citep{angladaescude16}, Ross 128 \citep{bonfils17a},
GJ 273 \citep{astudillodefru17}, LHS 1140 \citep{dittmann17}, TRAPPIST-1 \citep{gillon17},
and K2-18 \citep{montet15, cloutier17b}.
The global properties of the 100 stars were defined from the \cite{boyajian12} relation (effective
temperature and radii, when [Fe/H] is fixed to the solar value) and the \cite{delfosse00} relation
(estimated stellar mass based on absolute $J$ magnitude).
Projected rotational velocities and rotation periods were taken
from an exhaustive literature search or derived from the CFHT-CoolSnap program
\citep{moutou17}. In our sample
we consider stars with masses between 0.08-0.57 M$_{\odot}$ with $J$ band magnitudes $4.2-10.3$ in
a range of distances spanning $\sim 1.8-11$ pc.

Rotational information for stars in our sample is an important characteristic for RV modelling
as stellar rotation strongly affects the activity arising from rotationally modulated active
regions observed in radial velocity \citep{saar97, meunier10, aigrain12, dumusque14}.
Rotation timescales also restrict the periodicities at which we can detect planets
due to difficulties in detecting RV planets with 
orbital periods close to the stellar rotation period or its harmonics \citep{vanderburg16}. 
However, the rotational information for our stellar sample is incomplete.
For stars with no known available rotation measurements (8/100 stars),
or with a \vsini{} upper limit only (2/100 stars),
we sample \prot{} from a \emph{modified} empirical distribution of M dwarf
rotation periods from ground-based photometry as a function of stellar mass
\citep{newton16a}. The nature of this `modification' is discussed in the subsequent
paragraph. The corresponding \vsini{} is then computed
from the sampled value of \prot{,} the known stellar radius, and the
inclination of the stellar spin--axis to the line-of-sight $i_s$ which we
draw from a geometrical distribution (i.e. uniform in $\cos{i_s}$).
For stars which only have a measured upper limit on \vsini{,} 
the \emph{modified} empirical distribution from which \prot{} is sampled is
truncated at the minimum \prot{} corresponding to the upper limit on \vsini{.}

The necessary modification to the empirical \prot{} distribution arises from
an observational bias in the \cite{newton16a} sample which favors rapidly rotating stars.
The detection of short photometric rotation periods (i.e. rapid rotators) is attained more
easily than long rotation periods because full phase coverage is more readily obtained over many
rotation cycles. Furthermore, in the case of early M dwarfs ($M_s > 0.25$ M$_{\odot}$), 
there is evidence for a positive correlation between the star's rotation rate and the
amplitude of its photometric variability \citep{newton16a}. Rapid rotators therefore
tend to exhibit larger amplitudes of variability thus making the signal more easily detectable.
The raw empirical distribution therefore does not represent the true underlying distribution of M
dwarf rotation periods in the Solar neighborhood. We attempt to account for this bias in a simplified
way by modifying the empirical \prot{} distribution by insisting that only
$\sim 25$\% of sampled rotation periods can be $<10$ days, as estimated from the volume-limited sample
of field M dwarfs with measured rotation velocities from \cite{delfosse98}. This constraint
reduces the fraction of fast rotators with \prot{} $< 10$ days by a factor of $\sim 2$.

\begin{figure}
  \centering
  \includegraphics[width=\hsize]{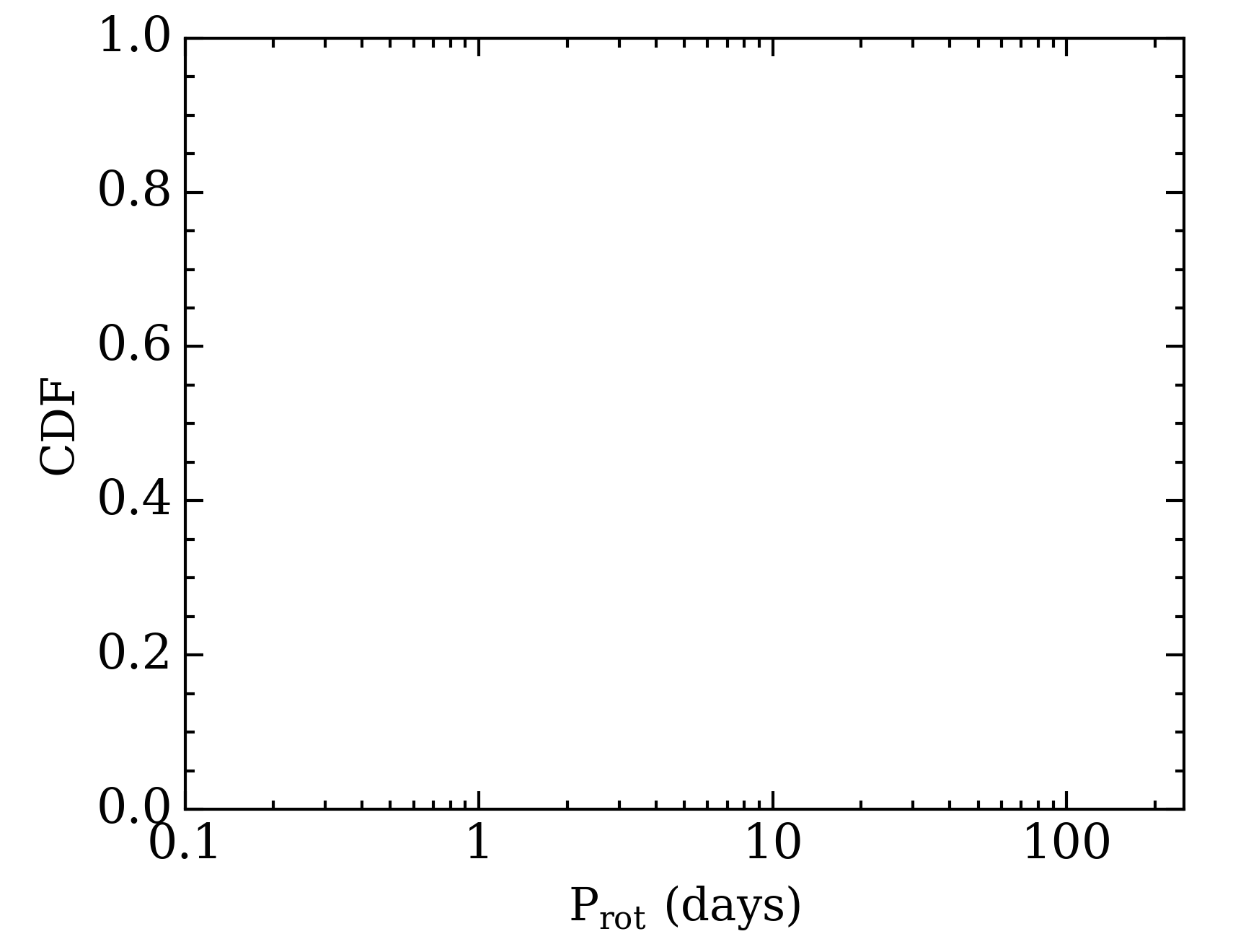}%
  \hspace{-\hsize}%
  \begin{ocg}{fig:1off}{fig:1off}{0}%
  \end{ocg}%
  \begin{ocg}{fig:1on}{fig:1on}{1}%
    \includegraphics[width=\hsize]{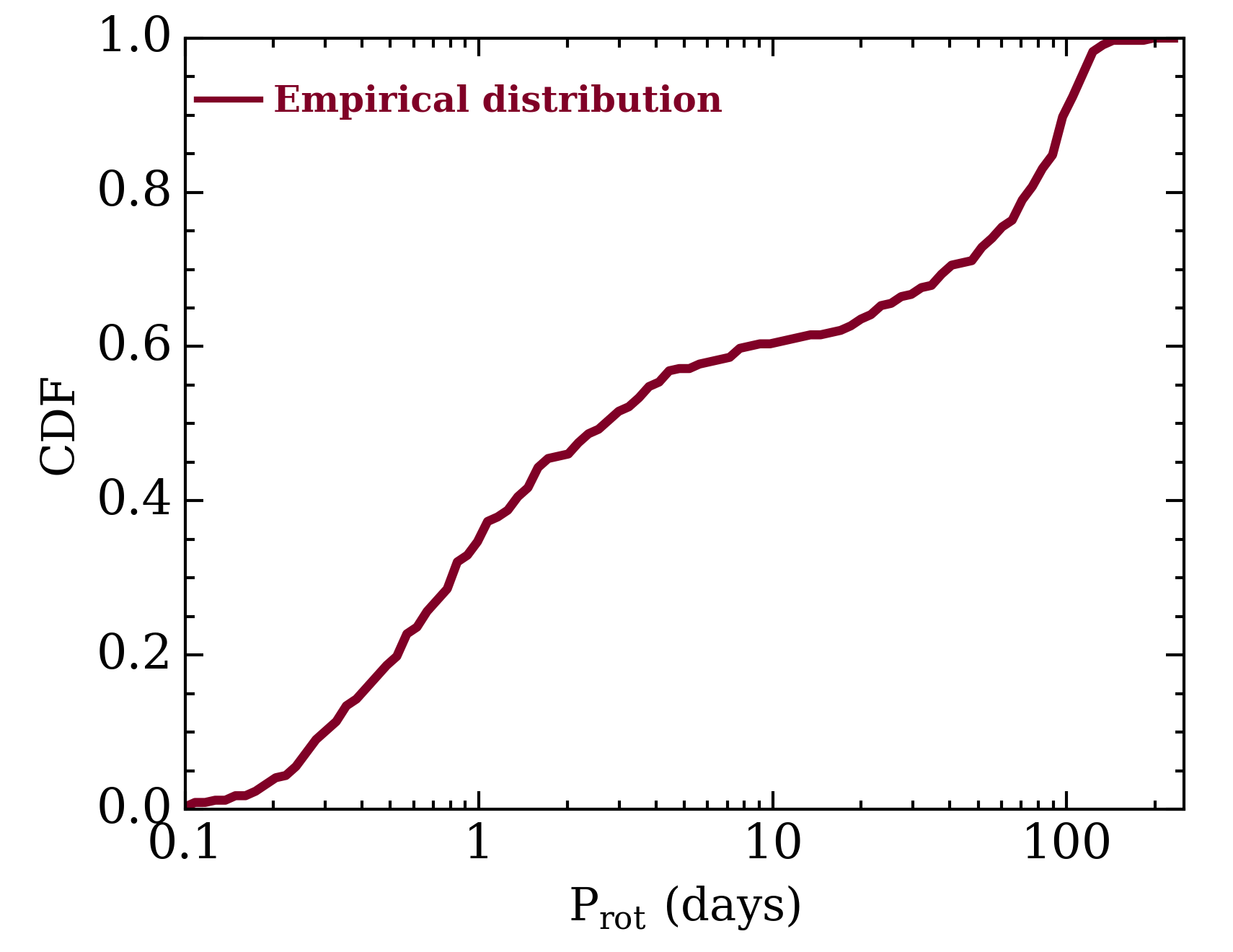}%
  \end{ocg}
  \hspace{-\hsize}%
  \begin{ocg}{fig:2off}{fig:2off}{0}%
  \end{ocg}%
  \begin{ocg}{fig:2on}{fig:2on}{1}%
    \includegraphics[width=\hsize]{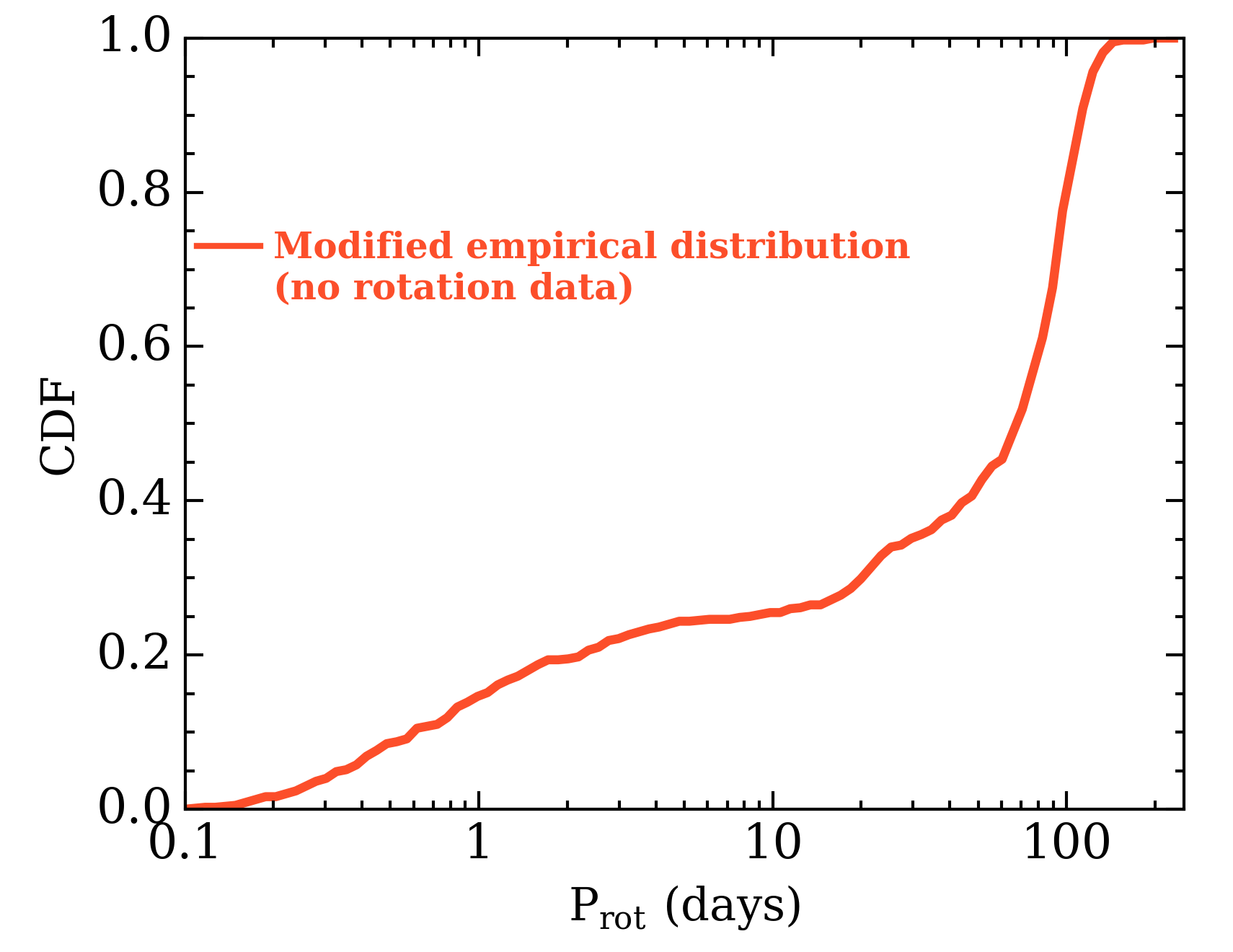}%
  \end{ocg}
  \hspace{-\hsize}%
  \begin{ocg}{fig:3off}{fig:3off}{0}%
  \end{ocg}%
  \begin{ocg}{fig:3on}{fig:3on}{1}%
    \includegraphics[width=\hsize]{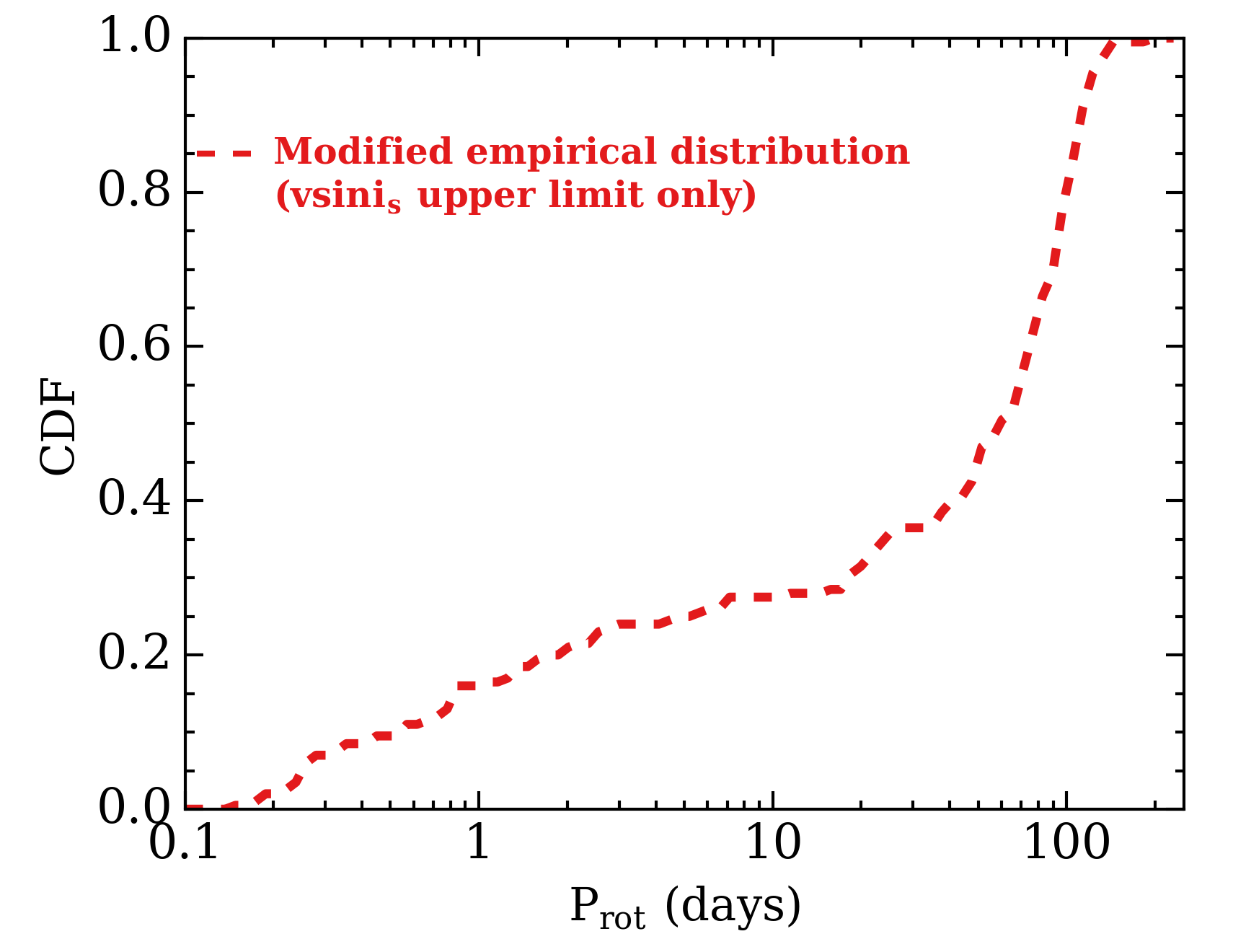}%
  \end{ocg}
  \hspace{-\hsize}%
  \begin{ocg}{fig:4off}{fig:4off}{0}%
  \end{ocg}%
  \begin{ocg}{fig:4on}{fig:4on}{1}%
    \includegraphics[width=\hsize]{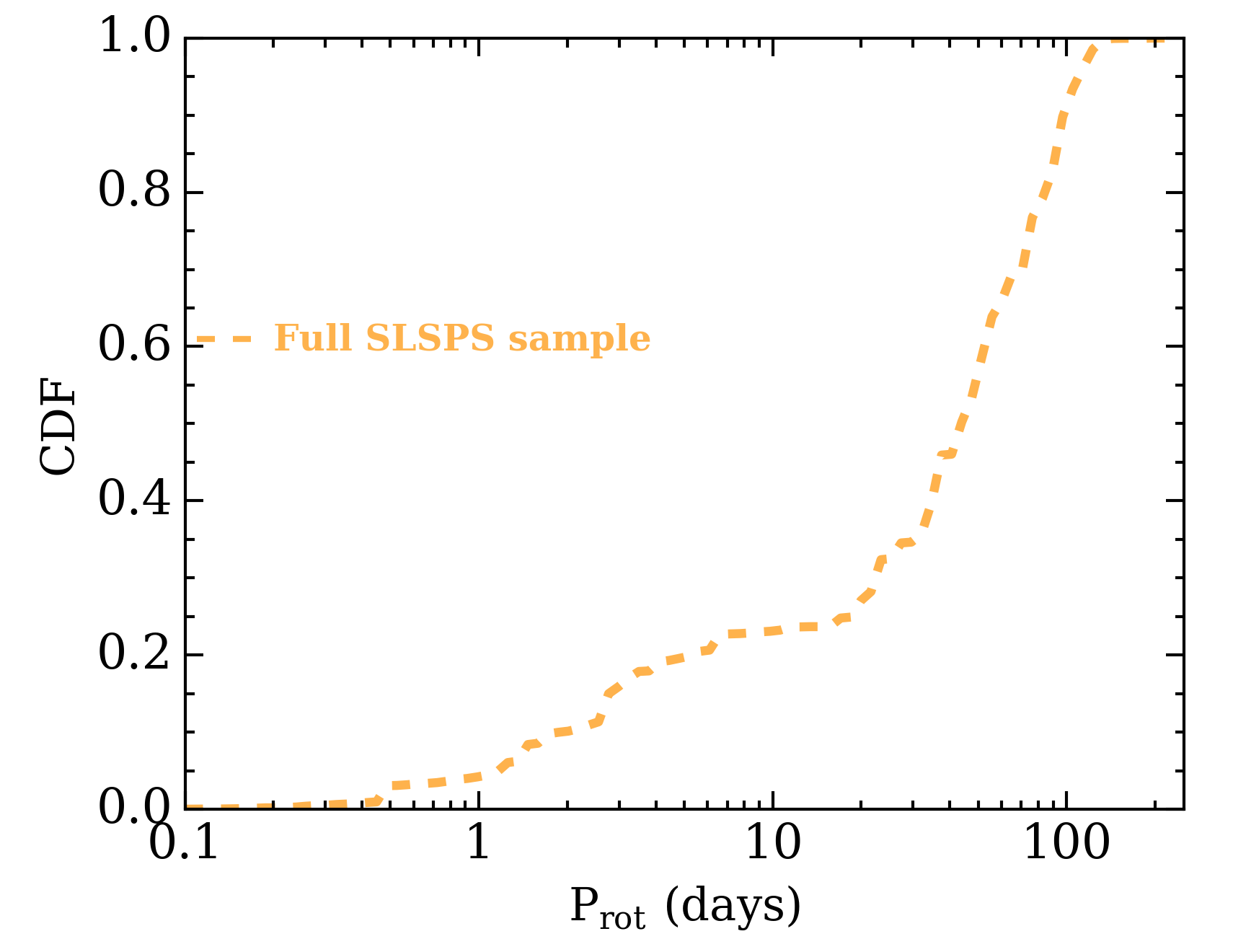}%
  \end{ocg}
  \hspace{-\hsize}%
  \caption{Cumulative distribution functions of the 
    \ToggleLayer{fig:1on,fig:1off}{\protect\cdbox{empirical distribution}} of M dwarf rotation periods
    from \cite{newton16a}, the modified empirical distribution of SPIRou stars with a 
    \ToggleLayer{fig:3on,fig:3off}{\protect\cdbox{\vsini{} upper limit}} measured, the
    modified empirical distribution of SPIRou stars with
    \ToggleLayer{fig:2on,fig:2off}{\protect\cdbox{no rotation data}} available, and the
    \ToggleLayer{fig:4on,fig:4off}{\protect\cdbox{full SLS-PS sample}}.}
  \label{fig:protcdf}
\end{figure}

The modification to the empirical \prot{} distribution is visualized in Fig.~\ref{fig:protcdf}.
Here we compare the empirical \prot{} distribution, based on the full stellar sample with detected
\prot{} from \cite{newton16a}, with the modified distributions of stars in the SPIRou input catalog
with either a \vsini{}
upper limit only or no available rotation data. The latter two distributions are nearly equivalent as
the majority of \vsini{} upper limits do not provide substantial new information regarding the star's \prot{}
and in both cases we insist that only $\sim 25$\% of sampled \prot{} can be $< 10$ days. We impose this
condition by noting that $\sim 60$\% of stars in the empirical distribution have \prot{} $<10$ days
and resample a particular fraction of those stars from the subset of the empirical distribution restricted
to \prot{} $\geq 10$ days. The fraction of stars with \prot{} $<10$ days that get resampled is
$1-0.25/0.6 \approx 0.58$.

\subsection{Physical Models of Stellar RV Activity} \label{sect:activity}
\subsubsection{\texttt{SOAP 2.0}: activity simulations} \label{sect:soap}
Active regions (ARs) in the stellar photosphere (e.g. star spots and faculae) and in
the hot chromosphere (e.g. plages) are expected to be present in M dwarfs. These surface inhomogeneities
have characteristic temperatures that differ from the star's effective temperature and therefore 
disrupt the symmetry of the visible stellar disk as they rotate in and out of view at the stellar rotation
period. One resulting source of RV activity from ARs, known as the \emph{flux effect} \citep{dumusque14},
results in an anomalous RV signal as the ARs block a fraction of Doppler-shifted
photons from the rotating stellar limbs. The strong local magnetic fields associated with ARs
at the stellar photospheric boundary also act to inhibit the upward flow of hot convective material
in an effect known as the suppression of \emph{convective blueshift} \citep{dravins81}.

In the limit of simple distributions of ARs, the observed RV structure from the two aforementioned effects 
is dependent on the fractional coverage of the visible stellar disk by the ARs and its first time
derivative \citep{aigrain12}. For each simulated RV time-series, we sample the relevant physical
parameters of the ARs (i.e. AR sizes and spatial distribution) and 
simulate the corresponding RV activity, full width at half maximum (FWHM), bi-sector inverse slope (BIS),
and photometric time-series arising from both the flux effect and from the suppression of convective blueshift 
using the \texttt{SOAP 2.0} code \citep{dumusque14}. The FWHM and BIS time-series are shape parameters of
the cross-correlation function between the observed stellar spectra and the template spectrum used to measure
the stellar RVs in an observing campaign. These ancillary time-series are sensitive to the presence of ARs but not to planets
making them useful diagnostics for distinguishing activity-induced RV signals from planetary signals.
In particular, the FWHM time-series will be used in
Sect.~\ref{sect:GP} to train our RV activity model and disentangle RV activity signals from planetary signals.

The \texttt{SOAP 2.0} code outputs time-series that are phase-folded to the input 
stellar rotation period. These time-series are initially treated as strictly periodic
and interpolated to the epochs of observation. In
this way we ignore any contribution from differential rotation whose amplitude has been shown to
decrease with decreasing stellar mass \citep{donati08, morin08, kitchatinov11}
before evoking rigid-body rotation in fully convective M dwarfs
($M_s \lesssim 0.2$ M$_{\odot}$). In Sect.~\ref{sect:lifetimes},
the strictly periodic condition is relaxed to account for the finite lifetimes of ARs and the existence of
long-term magnetic activity cycles.

Unfortunately, very little is presently known about the physical nature of ARs on M dwarfs but we do observe
quasi-periodic photometric variability which arises from evolving ARs \citep{oneal05}. We use the
empirical distribution of photometric amplitudes as a function of stellar mass from \cite{newton16a} to sample
the \emph{average} photometric variability amplitudes\footnote{For photometric variations measured in the near-IR
  with MEarth over a custom passband spanning $\sim 0.7-1 \mu$m known as the $i+z$ band \citep{nutzman08}.}
$A$ for each star in the simulated SLS-PS. The sampled value of $A$ is interpreted as an
average value because the phase in the star's magnetic activity cycle at the time of the \cite{newton16a}
observations is unknown. We then use $A$ to constrain the size of up to four ARs with each AR being treated as either
a cool spot or bright plage. The spatial distribution of
ARs is determined from random draws in latitude and longitude as  
Doppler imaging provides evidence for more uniformly distributed ARs on M dwarfs than on
Sun-like (FGK) stars \citep{barnes01, barnes04}, whose ARs tend to be more localized along the
stellar equator. We note however that such Doppler imaging observations are limited to rapid rotators
which are largely avoided in the selection of the SPIRou input catalog. 
The temperature contrast between the ARs and the stellar effective
temperature is fixed to 200 K in all realizations \citep{berdyugina05}. The \texttt{SOAP 2.0} code is
designed to model Sun-like stars at an optical wavelength of $\lambda \sim 529$ nm.
The resulting activity is then scaled from the default \texttt{SOAP 2.0} wavelength to the approximate
central $H$ band wavelength of $\lambda' \sim 1.6$ $\mu$m via the ratio of blackbody emission at $\lambda$ to $\lambda'$
with a characteristic temperature of $T_{\text{eff}}$. 
This scaling decreases the amplitude of the flux and convective blueshift effects by a typical
factor of a few in the nIR compared to at optical wavelengths
\citep{martin06, huelamo08, prato08, reiners10, mahmud11}.

\subsubsection{Zeeman broadening}
Unlike the flux and convective blueshift effects, the RV activity due to Zeeman broadening tends to
\emph{increase} towards the nIR. Zeeman broadening of spectral features in unpolarized light occurs in
the presence of strong magnetic
fields that cause Zeeman splitting; an effect that grows with wavelength. \cite{reiners13} and
\cite{hebrard14} used polarized
radiative transfer at nIR wavelengths to compute the effect of Zeeman splitting from both atomic and
molecular sources on stellar line profiles. \cite{reiners13} report the following simplified model for the
RV signal resulting from Zeeman broadening in M dwarfs ($T_{\text{eff}} \in [2800, 3700]$ K)

\begin{equation}
  \text{RV}_{\text{Z}}(t) = 300 \text{ m s}^{-1} f(t) \left( \frac{B}{\text{1 kG}} \right)^2
  \left( \frac{\lambda}{1 \mu\text{m}} \right)^a, \label{eq:zeeman}
\end{equation}

\noindent where $f$ is the filling factor or fraction of the visible stellar disk that is
spanned by ARs, $B$
is the local magnetic field strength within the AR, and $\lambda$ is the wavelength of
observation. The powerlaw index $a \in [0,2]$, describes the increase of $\text{RV}_{\text{Z}}$ with
$\lambda$ and is variable as a result of the apparent distribution of molecular Land\'{e} g-values
in cool stars. Albeit only the FeH and CO bands are considered in the stellar atmospheric model from
which Eq.~\ref{eq:zeeman} is derived \citep{reiners13}.

Computing $\text{RV}_{\text{Z}}$ to add to our complete physical RV activity model requires knowledge of the
local $B$ field strength within
ARs. The empirical distribution of this quantity in M dwarfs is incomplete despite contributions from
various observing campaigns \citep[e.g.][]{reiners07, shulyak14, hebrard16, moutou17, shulyak17}. 
Instead of sampling $Bf$ from an empirical distribution, we use an ad hoc method of sampling $Bf$ which
exploits what is known about small-scale $Bf$
fields in M dwarfs as a function of spectral type and rotation. Namely, the fraction of M dwarfs
that are magnetically active as a function of rotation period differs between early-type
and late-type M dwarfs (M5-M8) as later M dwarfs are able to remain magnetically
active late into their lives even after considerable spin-down \citep{west15}. For each star we assign an
activity flag indicative of being an \emph{active} or \emph{inactive} star where the probability of being
flagged as an active star is equal to the measured activity fraction of M dwarfs from \cite{west15} and 
is dependent on the star's spectral type and \prot{.} If magnetically inactive, we sample the localized
magnetic field strength from $B \sim \mathcal{U}(.1,1)$ kG \citep{moutou17}. If magnetically active, instead
we draw from $B \sim \mathcal{U}(1,3.1)$ kG \citep{moutou17}. We then calculate the value of $a$ based on the
star's sampled $B$ and spectral type before evaluating the Zeeman broadening model (Eq.~\ref{eq:zeeman}) at the
central $H$ band wavelength of $1.6$ $\mu$m as a function of the time-evolving filling fraction which is known
from our \texttt{SOAP 2.0} simulations.

\subsubsection{Active region lifetimes} \label{sect:lifetimes}
The ARs giving rise to stellar activity in our simulations are short-lived compared to the baseline
of our observations. RV observations of M dwarfs have suggested that the lifetimes of individual ARs
may persist from one to a few stellar rotations and up to $\gtrsim 10$
\citep[e.g.][]{bonfils07,forveille09,hebrard16}. RV observations have also elucidated that M dwarfs undergo
long-term magnetic
activity cycles similarly to the Sun \citep[e.g.][]{gomesdasilva12, route16}. Following the prescription
of \cite{dumusque16a} for Sun-like stars, we proceed in deriving the temporal variation of AR sizes by scaling the
total RV activity signal according to each AR's appearance rate $\lambda(t)$. \cite{dumusque16a} also included
the time-dependent latitude of ARs which we neglect here due to the more uniform distribution of ARs observed
on M dwarfs \citep{barnes01, barnes04}. Furthermore, we assume that the appearance rate for both star spots and
bright plages are consistent.

The probability that an AR appears at a time $t$ is governed by the Poisson distribution

\begin{equation}
  P(t) = \frac{e^{-\lambda(t) \tau} (\lambda(t) \tau)^k}{k!} \label{eq:poisson}
\end{equation}

\noindent where $\tau$ is the time step in days and $k=0,1,2,3$ as we only consider a maximum of
four ARs. Next, as a function of time we draw from the probability distribution in Eq.~\ref{eq:poisson} 
which dictates at which epochs an AR is formed. For each newly formed AR 
we insist that it spends the first third of its lifetime evolving linearly to its maximum size before
shrinking towards zero over the remaining two thirds \citep{dumusque16a}.
Each AR's lifetime is sampled from a truncated Gaussian distribution with mean 3\prot{} and standard
deviation \prot{.} The Gaussian distribution is truncated at \prot{} such that all sampled ARs persist
for a minimum of one stellar rotation \citep[e.g.][]{bonfils07,forveille09,hebrard16}.
Also recall that the maximum size of the AR is determined by the star's sampled amplitude of photometric
variability.

The AR appearance rate per unit time is

\begin{equation}
  \lambda(t) = (\lambda_{max,act} - 0.5) \left[ -0.5 \cos{\left( \frac{2\pi t}{P_{\text{cycle}}} + \phi \right)} + 0.5 \right] + 0.5
\label{eq:lambda}
\end{equation}

\noindent where $\lambda_{max,act}$ describes the maximum appearance rate during the maximum of the star's
magnetic activity cycle whose period is $P_{\text{cycle}}$. We set $\lambda_{max,act}=10$ ARs per day and sample
$P_{\text{cycle}}$ from $\mathcal{U}(6,10)$ years \citep{mascareno16, wargelin17} which is a factor of two or more
greater than the baseline of the observations. The added term of 0.5 ARs per day
to Eq.~\ref{eq:lambda} ensures that we maintain a low but non-zero probability of forming an AR close to the
minimum of the stellar activity cycle.

Recall that the sampled amplitude of photometric variability sets the size of ARs in our simulations.
Furthermore, because the phase within a star's magnetic activity cycle at the time of photometric observations
is unknown, we treat the observed photometric variability amplitude as an average value. To account for this
approximately, we rescale our
derived AR lifetime scaling to the interval $0.1-2$---instead of $0-1$---to account for the varying levels of
stellar activity up to a factor of two greater than the maximum value and down to a minimum value slightly greater
than zero. We then use this scaling to rescale the injected activity in both the RVs and in the ancillary time-series.

\begin{figure}
  \centering
  \includegraphics[width=\hsize]{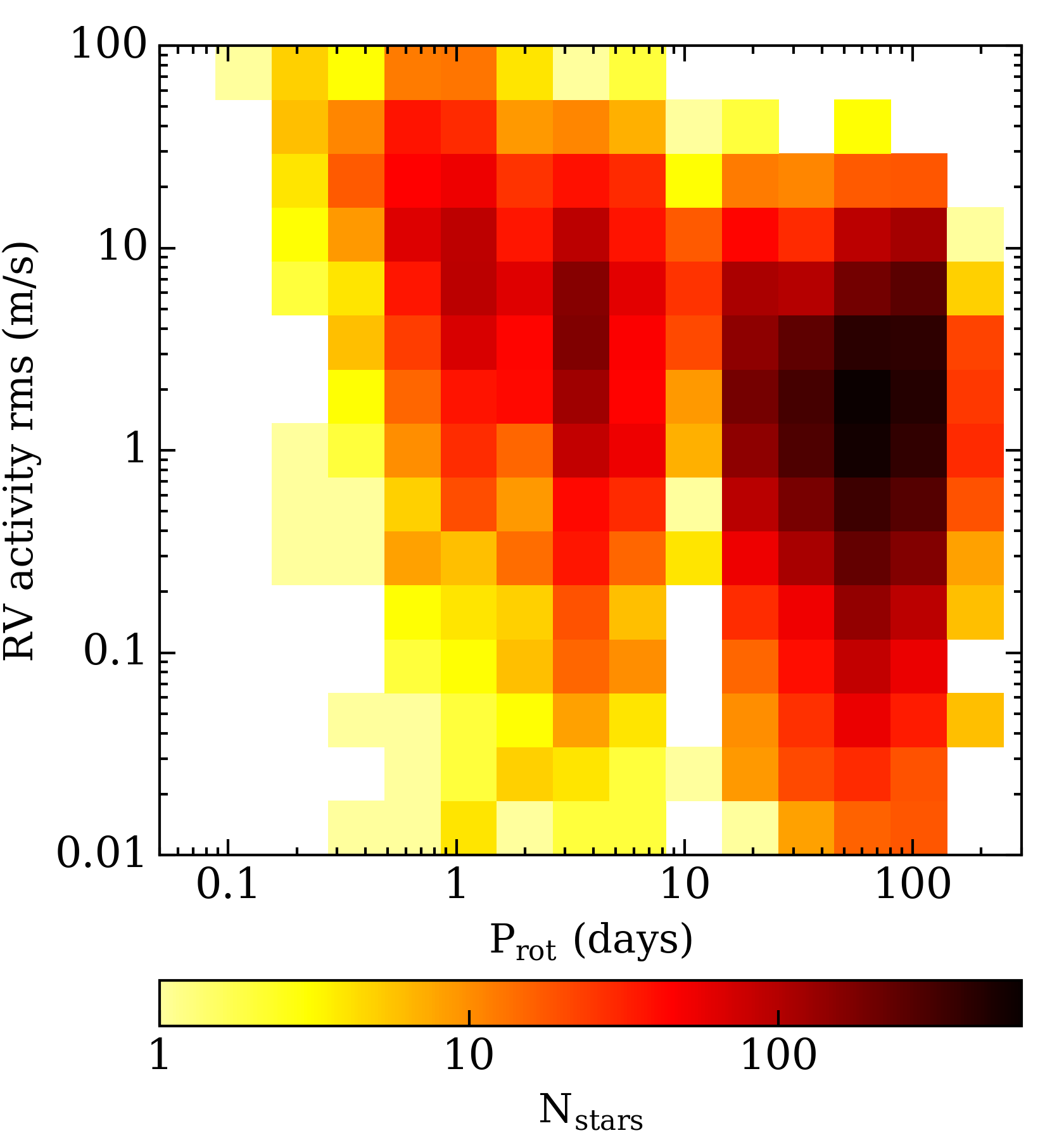}%
  \hspace{-\hsize}%
  \begin{ocg}{fig:curveoff}{fig:curveoff}{0}%
  \end{ocg}%
  \begin{ocg}{fig:curveon}{fig:curveon}{1}%
    \includegraphics[width=\hsize]{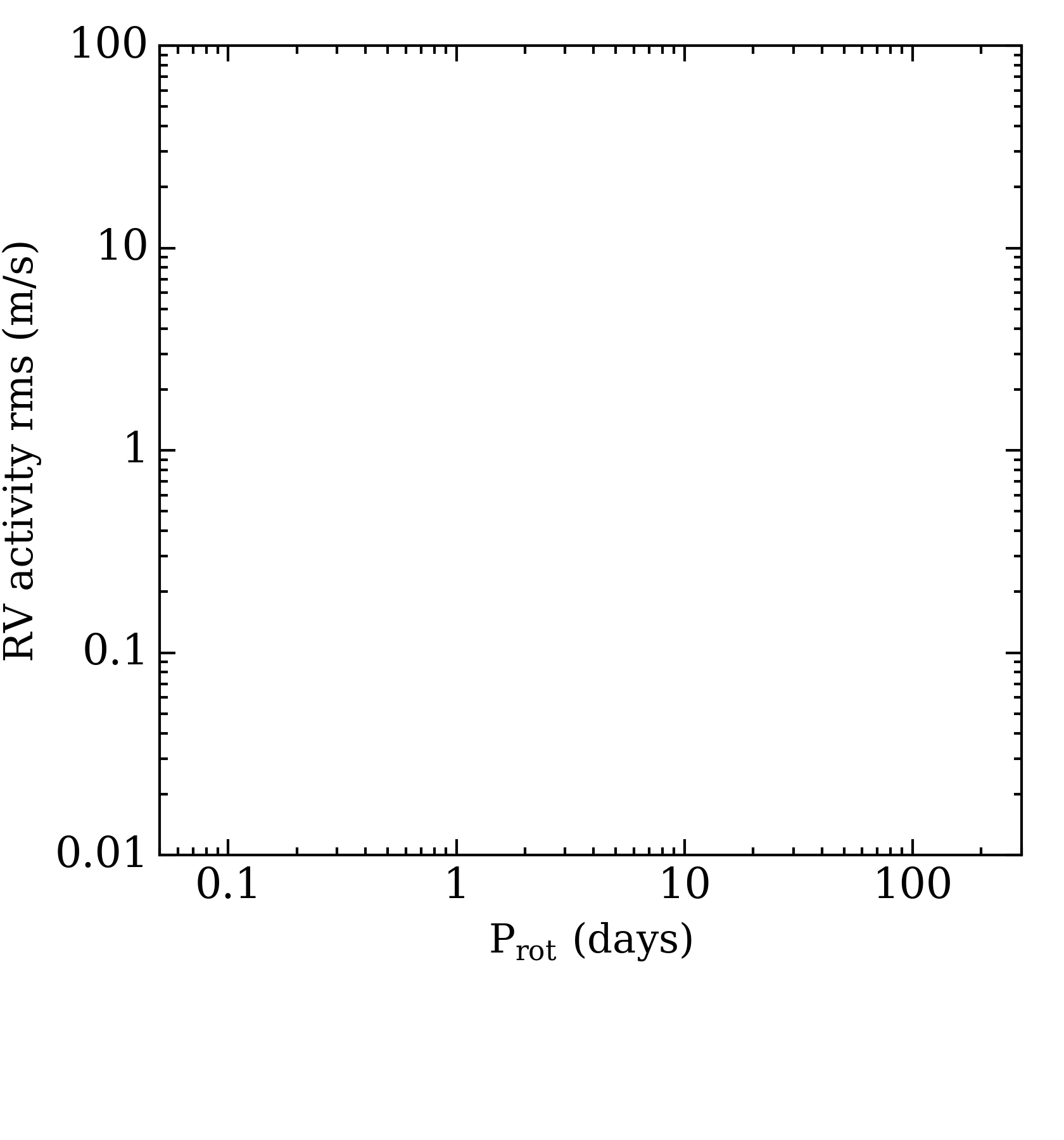}%
  \end{ocg}
  \hspace{-\hsize}%
  \begin{ocg}{fig:Hoff}{fig:Hoff}{0}%
  \end{ocg}%
  \begin{ocg}{fig:Hon}{fig:Hon}{1}%
    \includegraphics[width=\hsize]{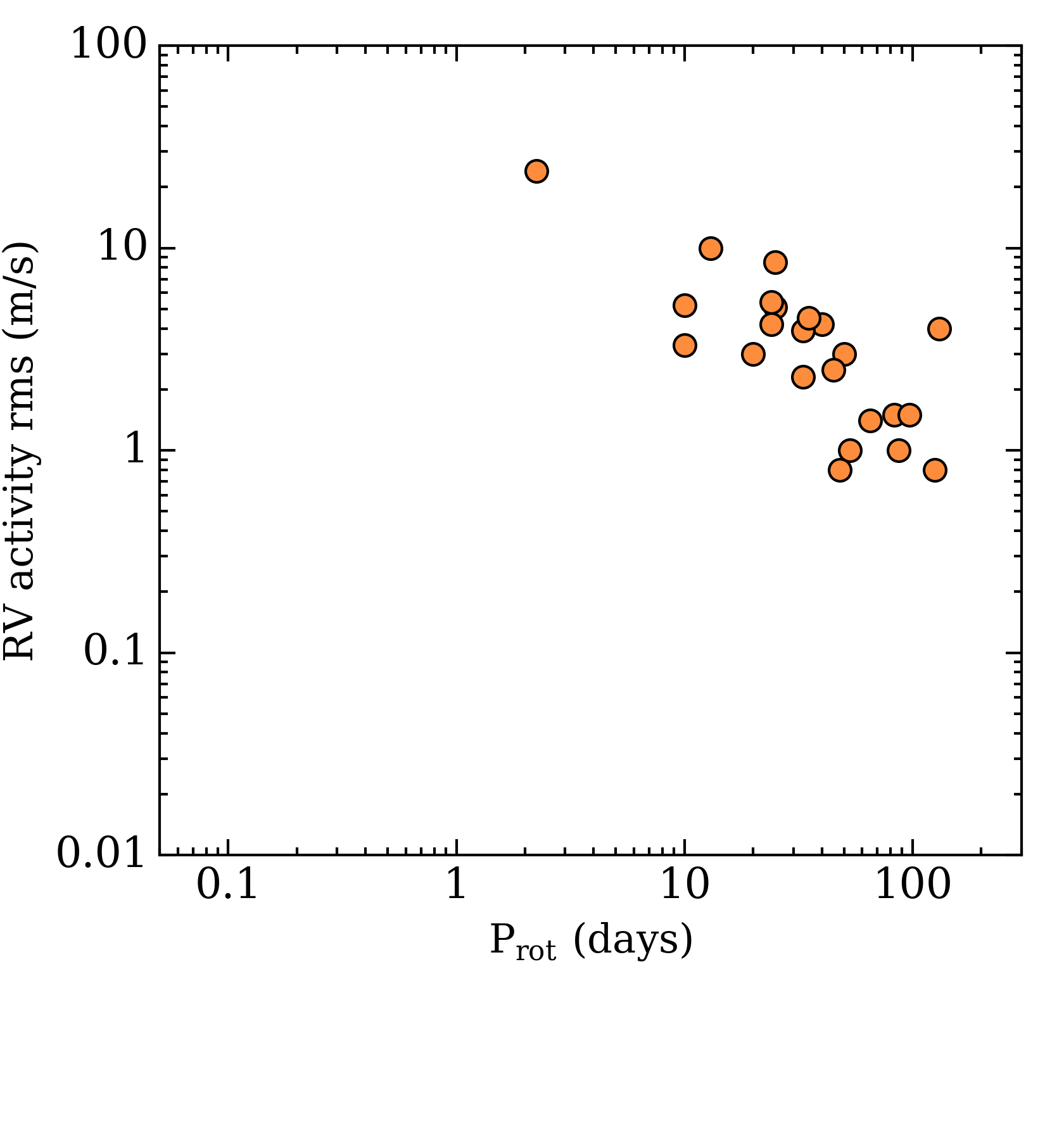}%
  \end{ocg}
  \hspace{-\hsize}%
  \caption{The RV activity rms as a function of stellar rotation period for stars in the simulated SLS-PS compared
    to a population of \ToggleLayer{fig:Hon,fig:Hoff}{\protect\cdbox{M dwarfs}} observed with the HARPS
    spectrograph. The \ToggleLayer{fig:curveon,fig:curveoff}{\protect\cdbox{\emph{dashed white curve}}} depicts the mean
    RV activity rms in each \prot{} bin.}
  \label{fig:activityrms}
\end{figure}

In general, our rescalings were optimized such that the root-mean-square (rms) of the injected RV activity is
roughly consistent with the upper envelope of the RV activity rms observed with HARPS.
Fig.~\ref{fig:activityrms} depicts the distribution of RV activity rms in our simulated stellar
sample as a function of \prot{} and directly compares it to the values of M dwarfs observed with HARPS.
The HARPS M dwarfs with measured RV activity rms in Fig.~\ref{fig:activityrms} include
Proxima Centauri \citep{angladaescude16},
GJ 3293, GJ 3341, GJ 3542 \citep{astudillodefru15},
GJ 1132 \citep{berta15}, 
GJ 876 \citep{correia10},
GJ 674 \citep{bonfils07},
Gl 205, Gl 358, Gl 388, Gl 479, Gl 526, Gl 846 \citep{bonfils13},
GJ 163 \citep{bonfils13b},
Gl 433, Gl 667C \citep{delfosse13b},
Gl 176 \citep{forveille09},
GJ 205, GJ 358, GJ 410, GJ 479, GJ 846 \citep{hebrard16},
and GJ 436 \citep{lanotte14}. 

We note that the comparison depicted in Fig.~\ref{fig:activityrms}
is not one-to-one as RV activity is an intrinsically chromatic effect and our
simulated time-series are computed at nIR wavelengths whereas the HARPS measurements are taken in the
visible. The effect of temperature contrast between ARs and the stellar photosphere on observed RV activity
is known to decrease towards longer wavelengths. Conversely, activity from Zeeman broadening increases towards
longer wavelengths and is known to be an important source of activity in the nIR \citep{hebrard14,moutou17}.
Our adopted activity scaling is chosen to best match the HARPS observations at the most frequently sampled rotation periods
in our stellar sample of $\sim 50-120$ days. In Fig.~\ref{fig:activityrms} it is clear that the mean RV activity rms
of our sample closely matches the HARPS stars, albeit with a large dispersion.
However at smaller \prot{} ($\lesssim 5$ days),
the mean RV activity rms in our simulated sample becomes slightly under-estimated relative to
the small sample of HARPS stars at those rotation periods. A larger sample of M dwarfs with measured RV activity rms at 
small \prot{} will be required to precisely characterize the correlation between the activity rms and
\prot{.}

\subsubsection{Ignored sources of RV activity}
Various other minor sources of stellar activity in M dwarfs are ignored. These include radial pulsations which 
do not appear to persist with any significant amplitude within the interiors of M dwarfs
\citep{rodriguezlopez15}. Secondly, we disregard the effect of granulation whose amplitude scales
with the velocity of convective cells which itself decreases towards later spectral types
\citep{dumusque11, meunier17}.
Lastly, we ignore flares in our activity model because their distinctive
spectral signature allows them to be flagged and removed from the subsequent analysis
\citep{schmidt12, angladaescude16}.
However, the high occurrence rate of flaring events on many M dwarfs may prove costly when attempting to
construct large time-series of uncontaminated RVs.

\section{Simulated Survey} \label{sect:survey}
In this study we conduct a detailed Monte-Carlo (MC) simulation of the SLS-PS for the purpose of
predicting the SPIRou planet detection yield. To do so, we must construct a statistically significant
number of unique RV time-series for each of the 100 stars that we target throughout the simulated SLS-PS.
In practice we simulate 100 unique RV time-series for each star totalling $10^4$ realizations in the full
simulated SLS-PS. Each RV time-series contains unique signals from planets, which are sampled from their
known occurrence rates around early M dwarfs (see Sect.~\ref{sect:planetsample}), from physical models
of stellar activity (as described in Sect.~\ref{sect:activity}), and from instrumental noise.
Each contribution is sampled
in time using a unique window function spanning $\sim 300$ nights over $\sim 3$ years,
which represents the typical time baseline of observations for a single star
in accordance with the expected subset of SPIRou's time allocation that will be dedicated to the
discovery of new exoplanetary systems in the SLS-PS. We note however that at this time the exact number of available
nights dedicated to the SLS-PS, nor the duration of the full SLS-PS, have been established absolutely. In addition
to the RVs, we also derive various spectroscopic activity indicators arising from stellar activity
which are contemporaneous with the RVs. One such ancillary time-series is the full width at half maximum
(FWHM) of the cross-correlation function which will be used to
train non-parametric Gaussian process models of the RV activity based on its common covariance structure with
the FWHM (see Sect.~\ref{sect:GP}). SPIRou is also unique in that it simultaneously operates its spectroscopic and
polarimetry modes thus providing a contemporaneous diagnostic of the star's magnetic topology. Such time-series
may also
be used to model RV signals from magnetically ARs \citep{hebrard16} although we do not consider
such time-series in this study. 

In each
simulated RV time-series we attempt to recover the injected planets to form an estimate of the expected
planet population that will be discovered with SPIRou. This is facilitated by the joint modelling of stellar
activity and planetary signals. This allows for the self-consistent characterization of each RV signal and the
detection of a subset of planets which are nominally hidden by their host star's intrinsic RV activity.

\subsection{Window Functions} \label{sect:wf}
In each MC realization, the unique window function $\mathbf{t}$ is the vector of length \nobs{} containing
the epochs of observation in barycentric julian dates (BJD). The window function describes the time sampling of
our time-series. To derive a set of window functions for each star in the SPIRou input catalog
we run a separate MC simulation of stellar observing sequences for all targets taking into account
when each
star is visible from CFHT on Maunakea with an airmass of $<2.5$ based on its celestial coordinates.
During every available night, all visible stars are observed up to two times,
each with an integration time required to achieve a
S/N per resolution element of 150---at the central $J$ band wavelength of 1.25 $\mu$m---with a minimum integration
time of 15 minutes. The imposed lower limit on the
integration time may also be necessary to mitigate the effects of granulation \citep{lovis05} which is
expected to be low on M dwarfs. 
Integration times required to achieve at least the target S/N are computed for
each star based using their $YJHK$ magnitudes. An overhead of 5 minutes is added to each
integration for guiding and setup purposes.
The output from these simulations is a set of window functions each pertaining to a star in the
SPIRou input catalog. Multiple MC simulations are run for various observing
sequences and thus provide unique window functions for the simulated SLS-PS. We then sample from
these window functions for the purpose of investigating
the sensitivity of our planet detection results to the exact form of the window function.

\begin{figure}
  \centering
  \includegraphics[width=\hsize]{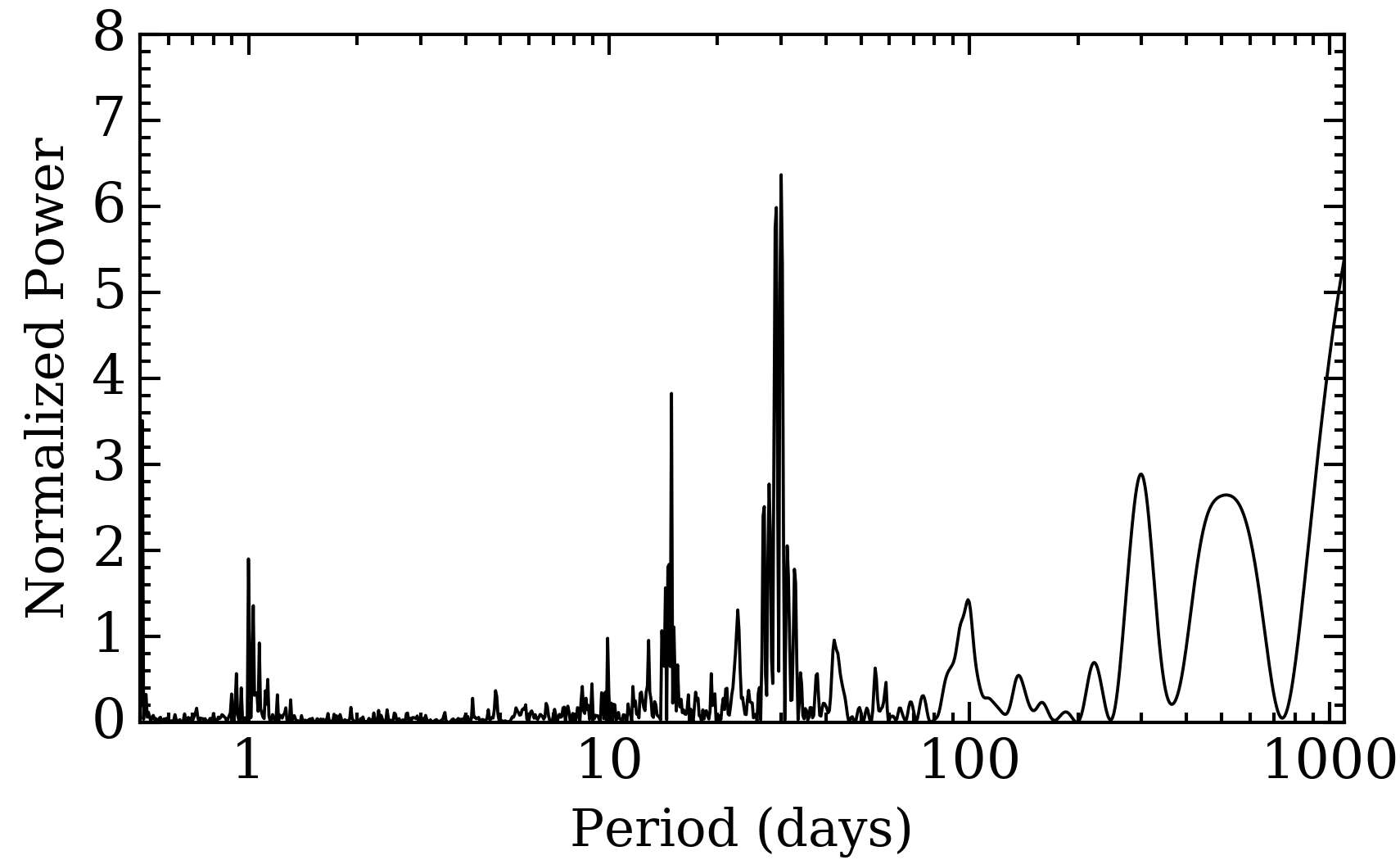}%
  \hspace{-\hsize}%
  \begin{ocg}{fig:Poff}{fig:Poff}{0}%
  \end{ocg}%
  \begin{ocg}{fig:Pon}{fig:Pon}{1}%
    \includegraphics[width=\hsize]{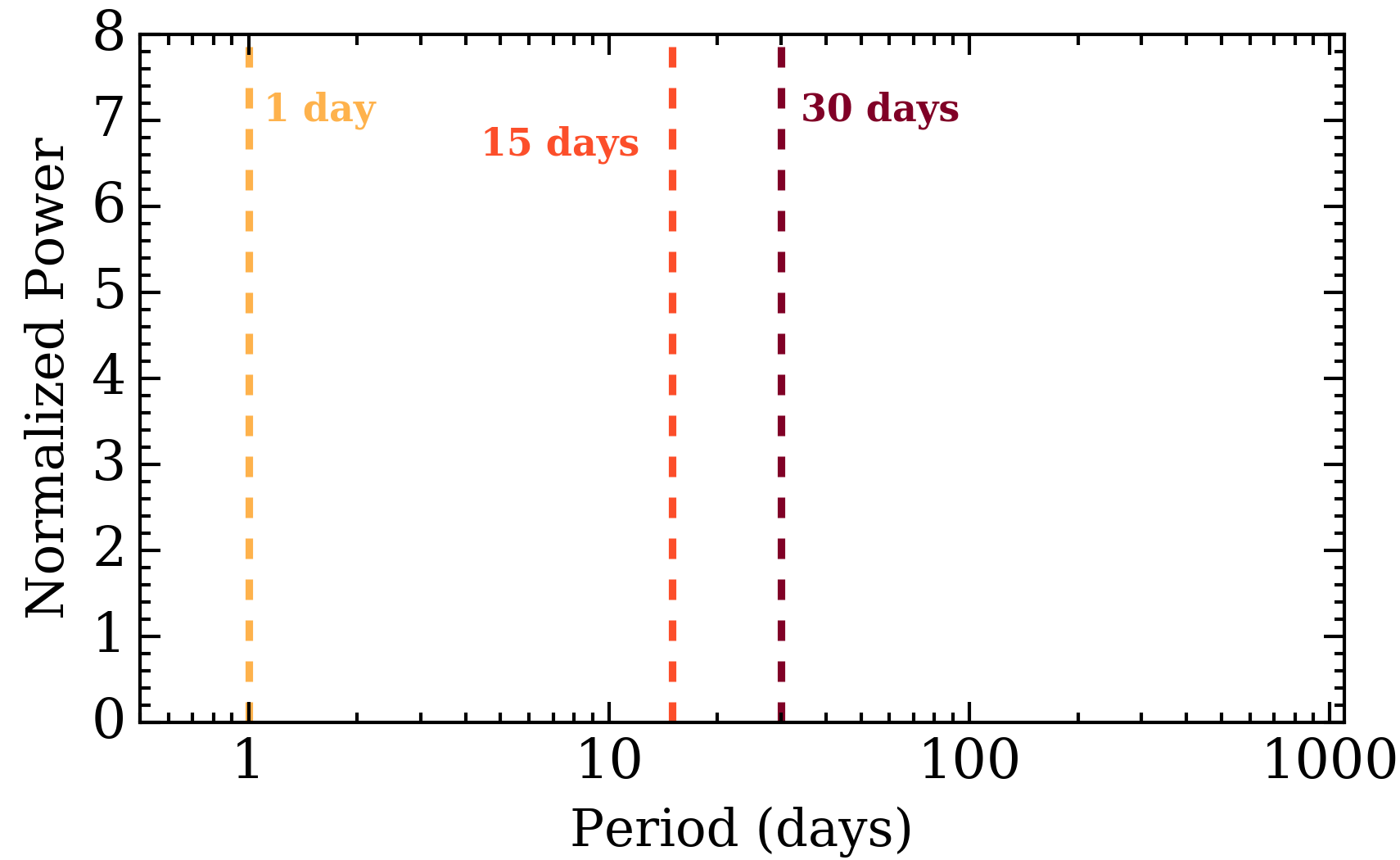}%
  \end{ocg}
  \hspace{-\hsize}%
  \begin{ocg}{fig:FAPoff}{fig:FAPoff}{0}%
  \end{ocg}%
  \begin{ocg}{fig:FAPon}{fig:FAPon}{1}%
    \includegraphics[width=\hsize]{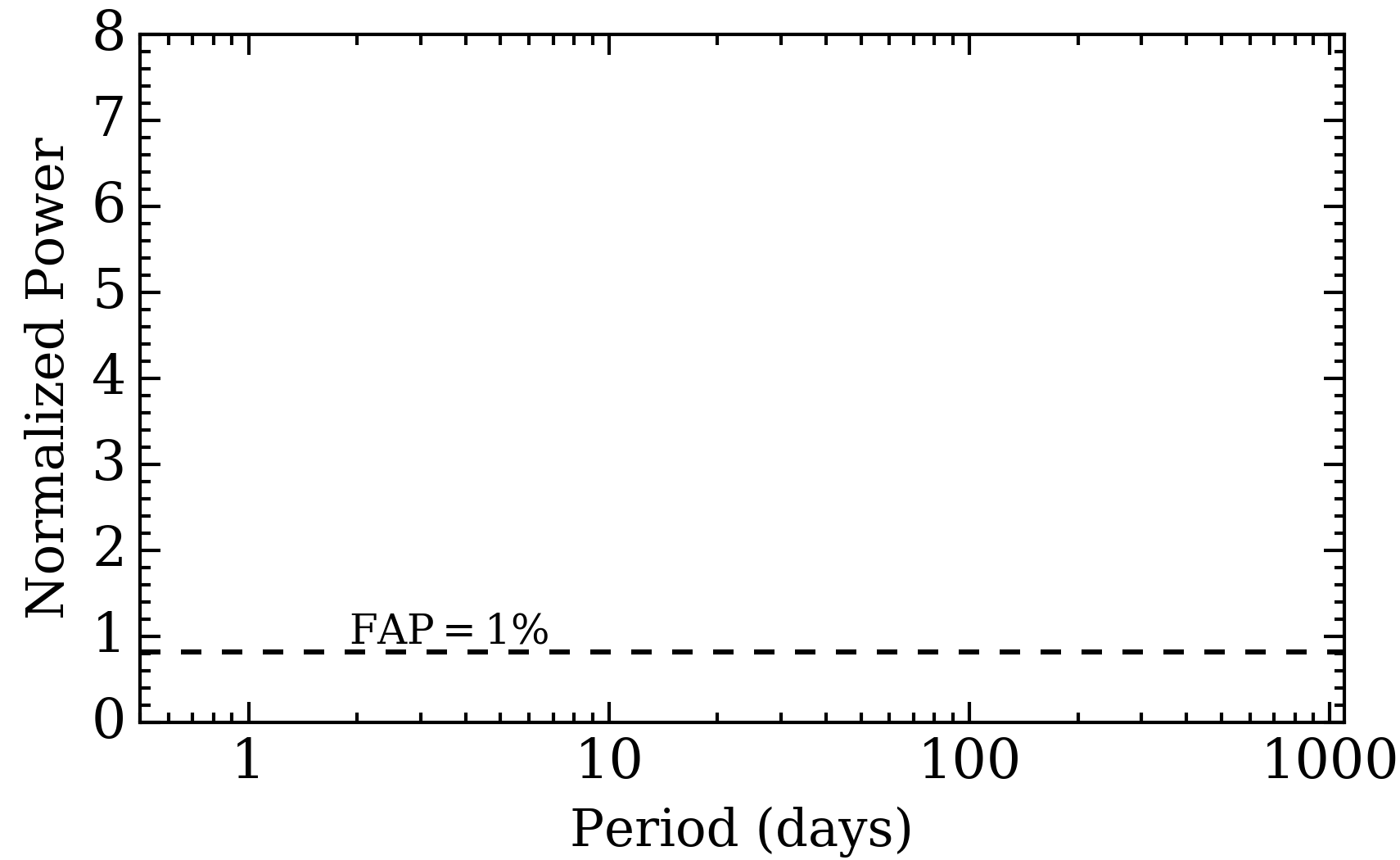}%
  \end{ocg}
  \hspace{-\hsize}%
  \caption{Lomb-Scargle periodogram of an example window function from the simulated SLS-PS.
    The periodogram exhibits strong aliases from our time sampling at the periodicities
    highlighted by the 
    \ToggleLayer{fig:Pon,fig:Poff}{\protect\cdbox{\emph{vertical lines}}}. The power
    corresponding to a 1\% false alarm probability in the periodogram is highlighted by the  
    \ToggleLayer{fig:FAPon,fig:FAPoff}{\protect\cdbox{\emph{horizontal dashed line}}}.}
  \label{fig:wfs}
\end{figure}

The available nights for observation with SPIRou are limited by two important considerations.
The first being the effect
of stochastic weather which limits the number of epochs in our derived window functions according to the
CFHT observatory's weather statistics. The second effect is somewhat unique to CFHT as the telescope hosts
a suite of instruments that do not operate simultaneously. In particular, the telescope's wide-field
optical imager requires dark-time to conduct its observations whereas the SPIRou spectro-polarimeter
does not. As such, we proceed with constructing SPIRou window functions that only include
non-dark-time observation and thus correspond to higher levels of lunar contamination. This represents
a worst case scenario for SPIRou as aliases from the window function will undoubtedly arise at periodicities
close to the cadence of the non-dark-time observing sequences. This cadence 
evolves with a period close to the period of the lunar cycle at $\sim 30$ days.
Because the SPIRou time sampling occurs in windows separated by $\sim 30$ days,
at that period and its first harmonic at 15 days, significant aliases in the Lomb-Scargle (LS)
periodogram \citep{scargle82}
of window function can arise as is shown in Fig.~\ref{fig:wfs}. The effect of these aliases are
detrimental to the detection of periodic planetary signals at these periods because
in the LS periodogram of the RVs, one cannot
distinguish a-priori between these periodic signals as a planet or as an alias of the time sampling
\citep{dawson10}.
This effect has already been shown to mimic planetary signals \citep[e.g.][]{rajpaul16} and is particularly
detrimental to finding HZ planets around $\sim$ M2-M4 dwarfs whose HZ span $\sim 30$ day orbital periods.
We quantify the magnitude of this aliasing effect on the SPIRou planet detection sensitivity in
Sect.~\ref{sect:sensitivity}.

For the purpose of performing accurate statistical inference of the SPIRou planet population
following the full SLS-PS (see Sect.~\ref{sect:occurrence}),
we attempt to split the total observing time evenly among the stars in the
SPIRou input catalog. Doing so, while taking into account stochastic weather and non-dark-time restrictions,
limits the size of our window functions over an a-priori 3 year long survey to an average of
198.1 RV measurements per star for 100 stars over $\sim 300$ nights. The variance in the number of RV
measurements per star is relatively small and ranges from 181-212. 
In the other versions of the SLS-PS containing either more or less stars than in our fiducial survey version
(see Sect.~\ref{sect:surveys}),
our MC calculations of the window functions consequently contain less and more RV measurements per star, respectively.

From preliminary simulations of the SLS-PS using 10 unique window functions per sampled planetary system per star,
we found that the net planet detection results are largely independent of the exact window function used. Note that
all sampled window functions contained the same restrictions discussed previously but do not include the exact
same epochs of observation for each star. Explicitly,
the SPIRou planet yield was found to vary by only $\sim 1$\% across the various window functions used.
This dispersion is much
less than the uncertainties on the resulting estimates of the SPIRou planet yield due solely to uncertainties in
the input planet occurrence rates. Therefore we conclude that using multiple window functions for each planetary
system in our simulated
SLS-PS is an unnecessary computational expense that can be mitigated by considering a single window function
per planetary system and not significantly affect the results of our study. However each unique window function considered
was still assumed to be restricted to non-dark-time observations thus preserving the aliasing affect on planets with
orbital periods of $\sim 30$ days.

\subsection{Radial Velocity Time-Series Construction} \label{sect:timeseries}
In each MC realization, 
the RV contribution from $N_p$ injected planets in the simulated planetary system (see Sect.~\ref{sect:planetsample})
is calculated via the superposition of $N_p$ keplerian orbital solutions. Each keplerian RV$_{kep}$ is parameterized by
the planet's orbital period $P$, time of inferior conjunction $T_0$, orbital eccentricity $e$,
argument of periastron $\omega$, and RV semi-amplitude $K$ according to

\begin{equation}
 \text{RV}_{kep}(t) = K [\cos{(\nu(t) +\omega)} + e \cos{\omega}],
\end{equation}

\noindent where $\nu(t)$ is the true anomaly and is computed by solving Kepler's equation
and the eccentric anomaly as a function of time $t$ contained in the window function
$\mathbf{t}$. The RV semi-amplitude $K$ is computed from the planet's minimum mass
\msini{,} the stellar mass $M_s$, $P$, and $e$ using the standard formula

\begin{multline}
  K = 1.05 \text{ m s}^{-1} \left(\frac{m_p\sin{i}}{2\text{ M}_{\oplus}} \right)
  \left( \frac{P}{20 \text{ days}} \right)^{-1/3} \\
  \left( \frac{M_s}{0.3 \text{ M}_{\odot}} \right)^{-2/3}
  \frac{1}{\sqrt{1-e^2}}.
\end{multline}

The keplerian RV approximation
is valid in all single-planet systems and the majority of multi-planet systems considered
and is best motivated by its ability to negate the
need to perform costly numerical integrations of each sampled planetary system. However, we note that
this approximation naturally excludes certain dynamical effects in multi-planet systems such as planet-planet
interactions and mean-motion resonances which can affect the planet-induced periodicities within
the RV time-series and therefore also affect our ability to detect those planets.
Although the former effect is not accounted for in our simulated time-series, its amplitude in real
systems is typically small compared to the RV measurement uncertainty as a result of dynamical restrictions
on multi-planetary systems making tightly-packed systems less stable over long time-scales and therefore
rarely seen in nature.
The significance of the latter effect is also expected to be small given the dearth of 
multi-planet systems at low-order period ratio commensurabilities \citep{lissauer11, fabrycky14}.

All RV components are evaluated at $\mathbf{t}$ and contain the additive i) RV activity RV$_{act}$ derived from 
physical models (see Sect.~\ref{sect:activity}) ii) $N_p$ keplerian models, and iii) a white noise term with an rms equal
to the median RV measurement uncertainty $\sigma_{\text{RV}}$ expected for the host star. Each star's value of
$\sigma_{\text{RV}}$ is calculated using the nIR RV
information content calculations from \cite{figueira16} corrected using the empirical spectra from Artigau et al. in prep.
We use the results from the condition 3 in \cite{figueira16} to estimate $\sigma_{\text{RV}}$:
the photon-noise contribution to the spectrum
being amplified by the limited spectral window due to atmospheric telluric transmission.
The information content calculations for each star are computed using SPIRou's spectral resolution,
spectral coverage, and the expected S/N per resolution element obtained during an integration of the star.

The RV activity model contains additive contributions from the flux effect, the suppression of convective blueshift,
and Zeeman broadening from ARs. The complete RV model is therefore

\begin{equation}
  \text{RV}_{model}(t) = \text{RV}_{act}(t) + \sum_{i=1}^{N_p} \text{RV}_{kep,i}(t) + \mathcal{N}(0,\sigma_{\text{RV}}).
\end{equation}

\begin{figure}
  \centering
  \includegraphics[width=\hsize]{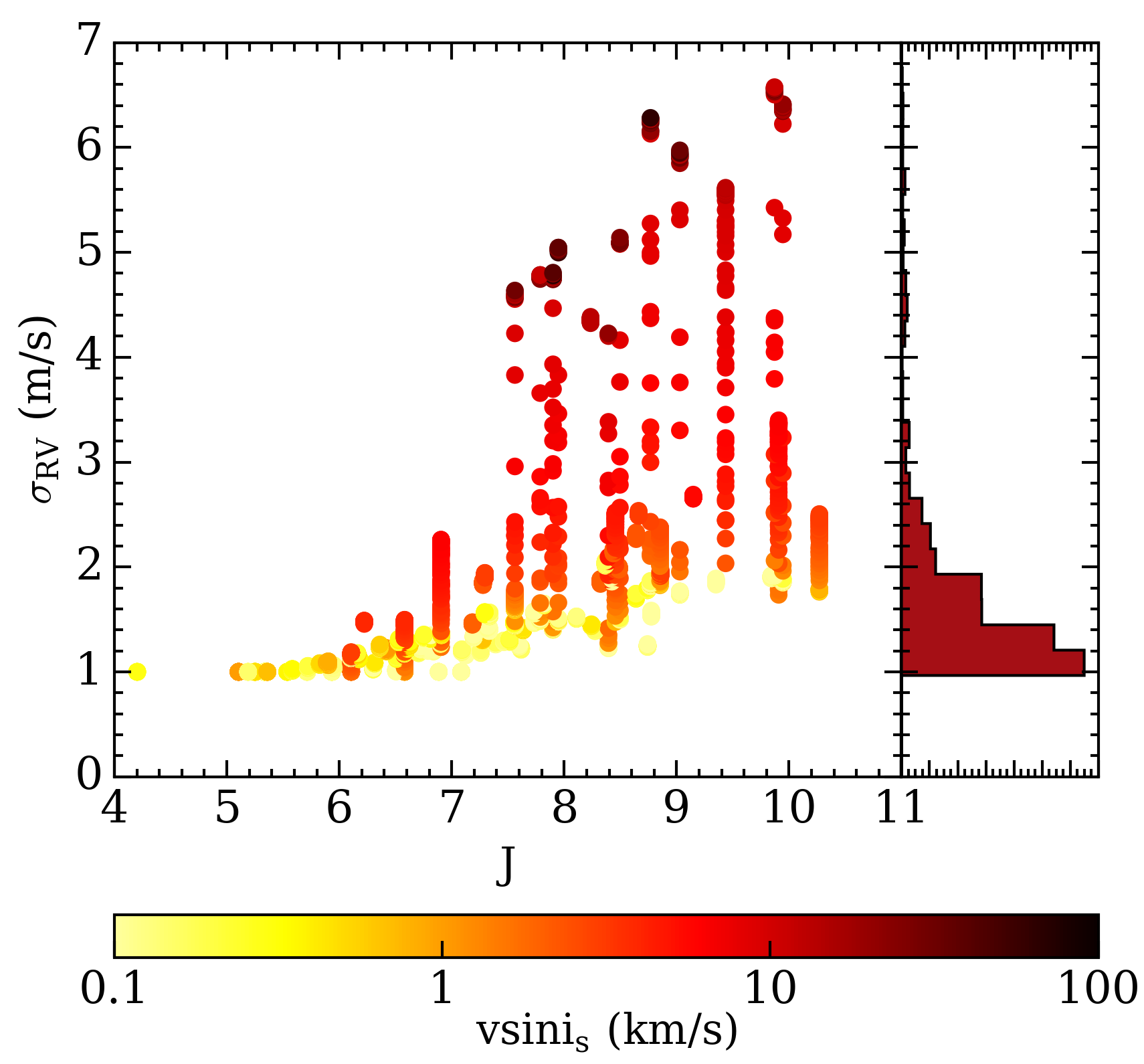}%
  \caption{The median RV measurement uncertainty in each simulated RV time-series $\sigma_{\text{RV}}$
    as a function of the star's $J$ band magnitude and projected stellar rotation velocity \vsini{.} A noise floor is
    imposed at 1 \mps{.} The histogram shown in the right panel is in linear units.}
  \label{fig:sigmaRV}
\end{figure}

\noindent The RV measurement uncertainties pertaining to each measured RV is contained in the vector
$\boldsymbol{\sigma_{\text{RV}}}$ and is modified from the scalar value of $\sigma_{\text{RV}}$
based on the variable absorption by terrestrial
water vapor. The water vapor correction at each epoch in $\mathbf{t}$ is the product of the airmass and the zenith
water column from the well-documented CFHT observing condition statistics throughout the calendar year. 
The distribution of median $\sigma_{\text{RV}}$ in our MC simulation for our full
stellar sample is shown in Fig.~\ref{fig:sigmaRV} as a function
of the stellar $J$ band magnitude and \vsini{.} A noise floor is imposed at the expected long-term RV precision limit
of SPIRou at 1 \mps{.} This results in a median
$\sigma_{\text{RV}}=1.52$ \mps{.} The maximum $\sigma_{\text{RV}}$ is 6.58 \mps{.} 
Not depicted in Fig.~\ref{fig:sigmaRV} is the dependence of
$\sigma_{\text{RV}}$ on spectral type as the RV measurement uncertainty tends to decrease towards later spectral types
due to the increased number of available spectral features and corresponding increase in RV information content.

Similarly to the construction of the RV activity time-series at the epochs in $\mathbf{t}$, the FWHM and BIS
ancillary times-series are constructed from our \texttt{SOAP 2.0} simulations (see Sect.~\ref{sect:soap}) 
at the same epochs as the RVs. These contemporaneous time-series will be used in Sect.~\ref{sect:GP} to
model the RV activity and help to detect underlying planetary signals.

\section{Planet Sample} \label{sect:planetsample}
\subsection{Planet Sample from known Occurrence rates}
In each MC realization we populate the simulated M dwarf planetary system with planets according to
their known occurrence rates from the \emph{Kepler} transit survey \citep{dressing15a}. In practice
we sample from a grid in orbital period $P$ and planetary radius $r_p$
according to the measured planetary occurrence rate at each point in the parameter space $f(P,r_p)$.
We assume that each point in the parameter space is
uncorrelated aside the dynamical stability constraints that we will apply to closely packed systems 
(see Sect~\ref{sect:dynam}).

The occurrence rates $f(P,r_p)$ (or equivalently $f$) from which we sample planets have
associated asymmetric uncertainties which are derived from data-driven posterior distributions
rather than from analytically-defined distributions which are more easily sampled from. \cite{dressing15a}
only present the modes and $1\sigma$ dispersions of their $f$ posteriors which is insufficient information
to reconstruct their $f$ posteriors in each $P$ and $r_p$ bin. To account for their reported uncertainties on
$f$ we must assume a functional form of the $f$ posteriors.
For each bin in $P$ and $r_p$ we sample planets with a probability drawn
from $\mathcal{N}(\mu, \bar{\sigma})$ where $\mu$ is the most likely value
of $f$ and $\bar{\sigma} = \text{mean}(\sigma_{\text{upper}},\sigma_{\text{lower}})$ where
$\sigma_{\text{upper}}$ and $\sigma_{\text{lower}}$ represent the asymmetric uncertainties on $f$.
The values of $\mu$, $\sigma_{\text{upper}}$, and $\sigma_{\text{lower}}$ are provided in \cite{dressing15a}
(see their Table 4) for all $P$ and $r_p$ bins for which $f>0$. For bins in which $f$ is consistent with zero,
we set $\mu=0$ and $\bar{\sigma}$ equal to the $1\sigma$ upper limit on $f$. For bins in
which $f$ is completely unconstrained as a result of a low detection sensitivity, we assume that $f(P,r_p)$ 
evolves smoothly such that we can extrapolate the values of $f$ from surrounding
bins with constraints in order to estimate the values of $f$ where it is unconstrained by the data
(e.g. $r_p \leq 1$ R$_{\oplus}$ and $P \geq 18.2$ days).
The extrapolated values are treated as $1\sigma$ upper limits and $\mu$ is set to zero such that when
integrating the most likely values of
$f(P,r_p)$ over the range of periods and planet radii considered by \citealt{dressing15a} ($P \in [0.5,200]$ days,
$r_p \in [.5,4]$ R$_{\oplus}$) we recover their cumulative
planet occurrence rate of $2.5 \pm 0.2$ planets per M dwarf. However our dynamical stability constraints will
reject a subset of sampled planetary systems thus effectively reducing the cumulative planet occurrence rate to
$<2.5$.

The dispersion in mutual inclinations among planets in multi-planet 
systems is related to the dispersion in eccentricities by $\langle i^2 \rangle \sim \langle e^2 \rangle /4$ 
\citep{stewart00, quillen07} with the mean orbital inclination 
to the plane of the sky $i$ being set to the value of $i_s$ obtained for the host star (i.e. from a geometrical
distribution). Although the distribution of spin--orbit angles for small planets around
M dwarfs has yet to be established, the observed low dispersion in orbital eccentricities \citep{vaneylen15}
and mutual inclinations \citep{figueira12, fabrycky14} in these types of planetary systems suggests that
they are dynamically cold. If this is indeed the case then the normal vector
to the mean planetary orbital plane is expected to be close to parallel to the stellar spin axis; $i \approx i_s$.
In each planetary system we sample each planet's orbital eccentricity $e$ from the $\beta$ probability
distribution describing the high detection significance sample of RV planets reported in \cite{cloutier15}
\citep[see also ][]{kipping13}. 

\subsection{Modifications to the Planet Sample} \label{sect:dynam}
Our adopted approach for sampling planetary systems is accompanied by four important caveats.

\emph{Converting planet radii to masses}. Firstly, the Kepler-derived $f$ directly samples
planetary radii whereas RV surveys are only sensitive to the planetary minimum
masses. Therefore an assumption must be made regarding the 
planetary mass-radius relation required to convert the sampled planetary radii into masses.
We opt for the following empirical mass-radius relation derived in \cite{weiss14} from an unbiased
sample of known transiting planets:

\begin{equation}
  \frac{m_p}{\text{M}_{\oplus}} =
  \begin{cases}
    0.440 \left( \frac{r_p}{\text{R}_{\oplus}} \right)^3 + 0.614 \left( \frac{r_p}{\text{ R}_{\oplus}} \right)^4, &
    r_p < 1.5 \text{ R}_{\oplus} \\
    2.69 \left( \frac{r_p}{\text{R}_{\oplus}} \right)^{0.93}, & r_p \ge 1.5 \text{ R}_{\oplus}. \\
  \end{cases}
 \label{eq:mr}
\end{equation}

\noindent This piece-wise mass-radius relation distinguishes between small rocky planets with
$r_p < 1.5$ R$_{\oplus}$ and larger gaseous planets \citep[e.g.][]{rogers15, dressing15b, fulton17}. The
intrinsic dispersion about the \emph{mean} mass-radius relation in Eq.~\ref{eq:mr} has characteristic
rms values of

\begin{equation}
  \sigma_{\text{rms}} =
  \begin{cases}
    2.7 \text{ M}_{\oplus}, & r_p < 1.5 \text{ R}_{\oplus} \\
    4.7 \text{ M}_{\oplus}, & r_p \ge 1.5 \text{ R}_{\oplus}. \\
  \end{cases}
  \label{eq:mrscat}
\end{equation}

\begin{figure}
  \centering
  \includegraphics[width=\hsize]{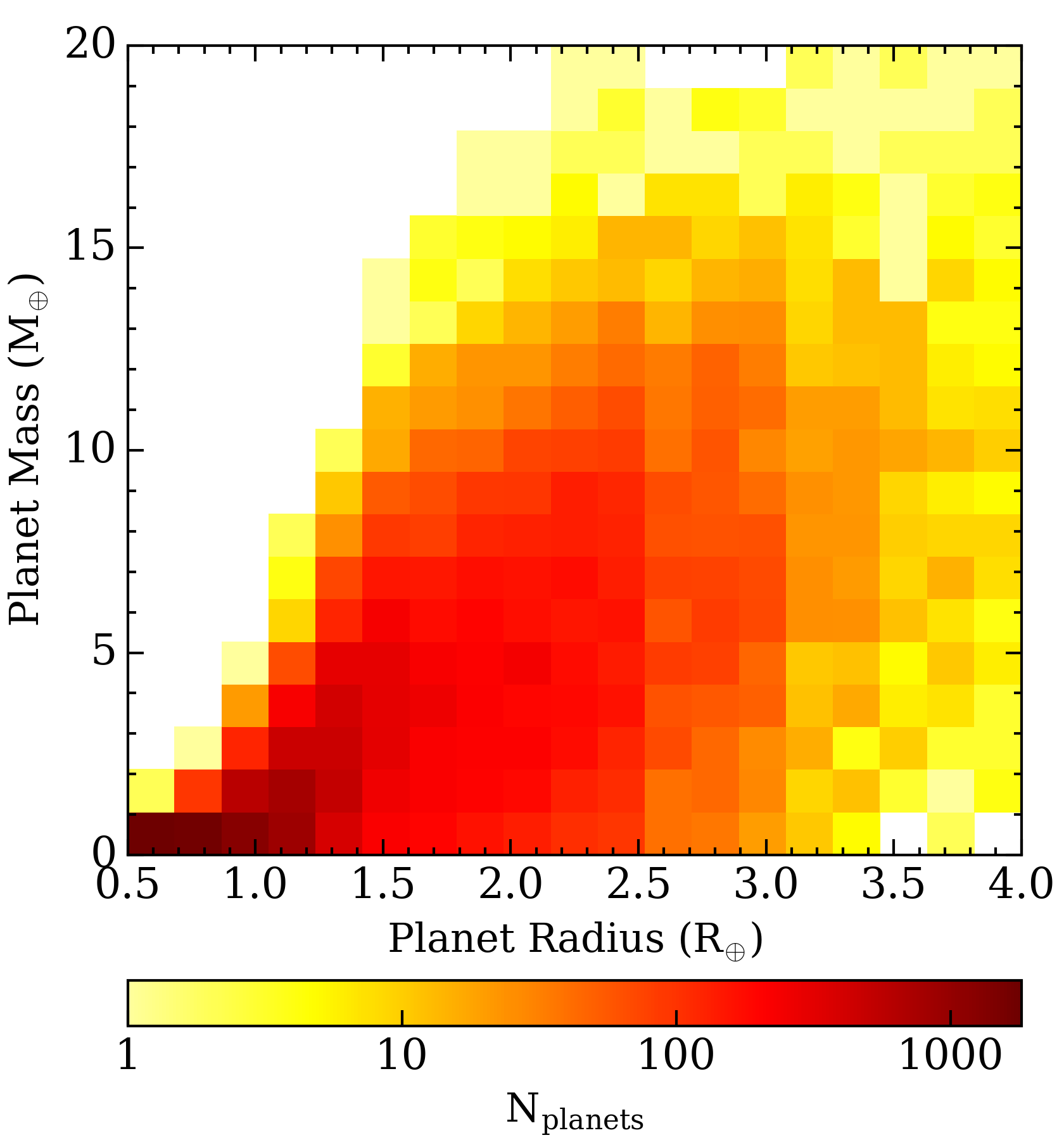}%
  \hspace{-\hsize}%
  \begin{ocg}{fig:curveoff}{fig:curveoff}{0}%
  \end{ocg}%
  \begin{ocg}{fig:curveon}{fig:curveon}{1}%
    \includegraphics[width=\hsize]{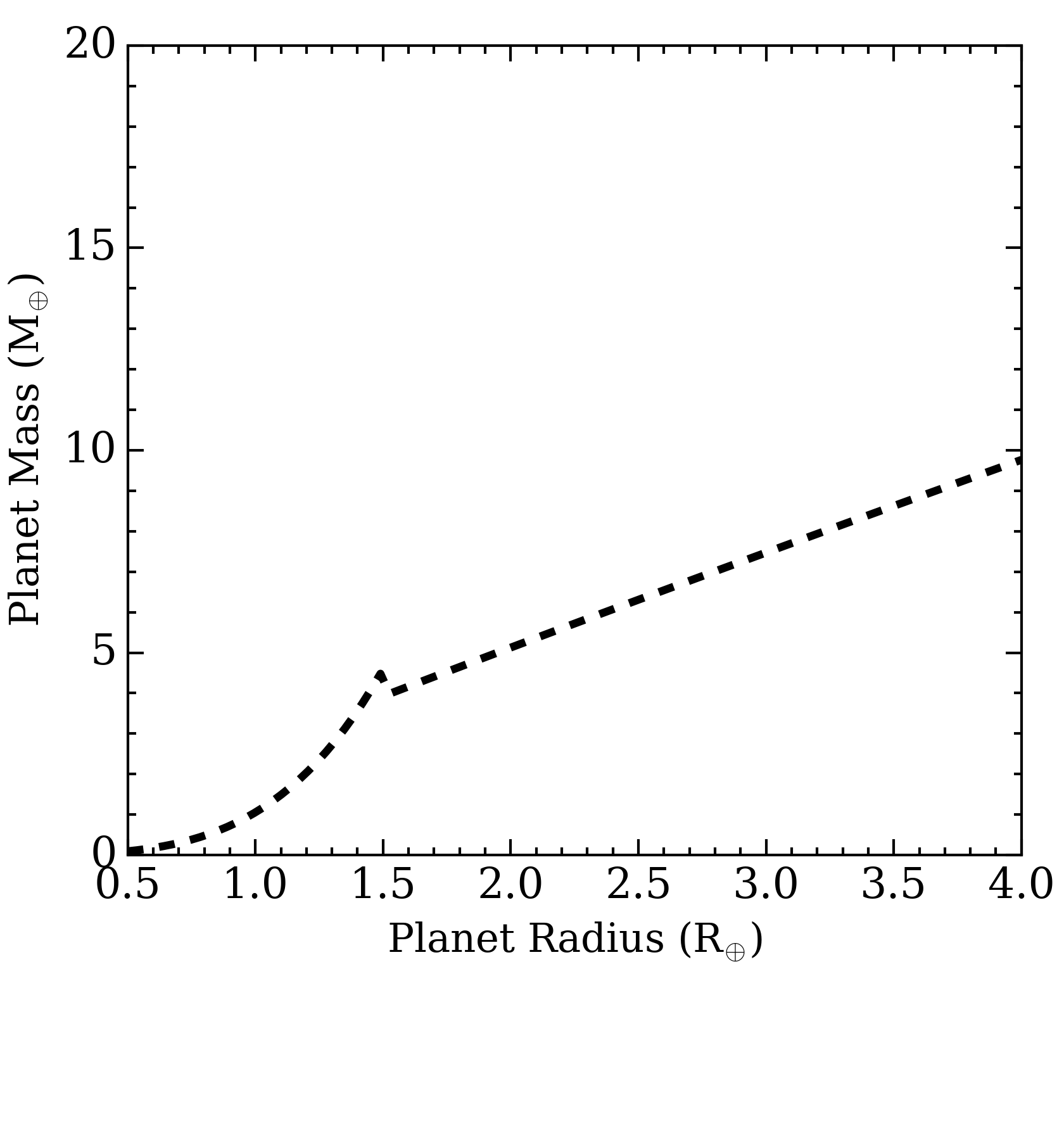}%
  \end{ocg}
  \hspace{-\hsize}%
  \begin{ocg}{fig:eboff}{fig:eboff}{0}%
  \end{ocg}%
  \begin{ocg}{fig:ebon}{fig:ebon}{1}%
    \includegraphics[width=\hsize]{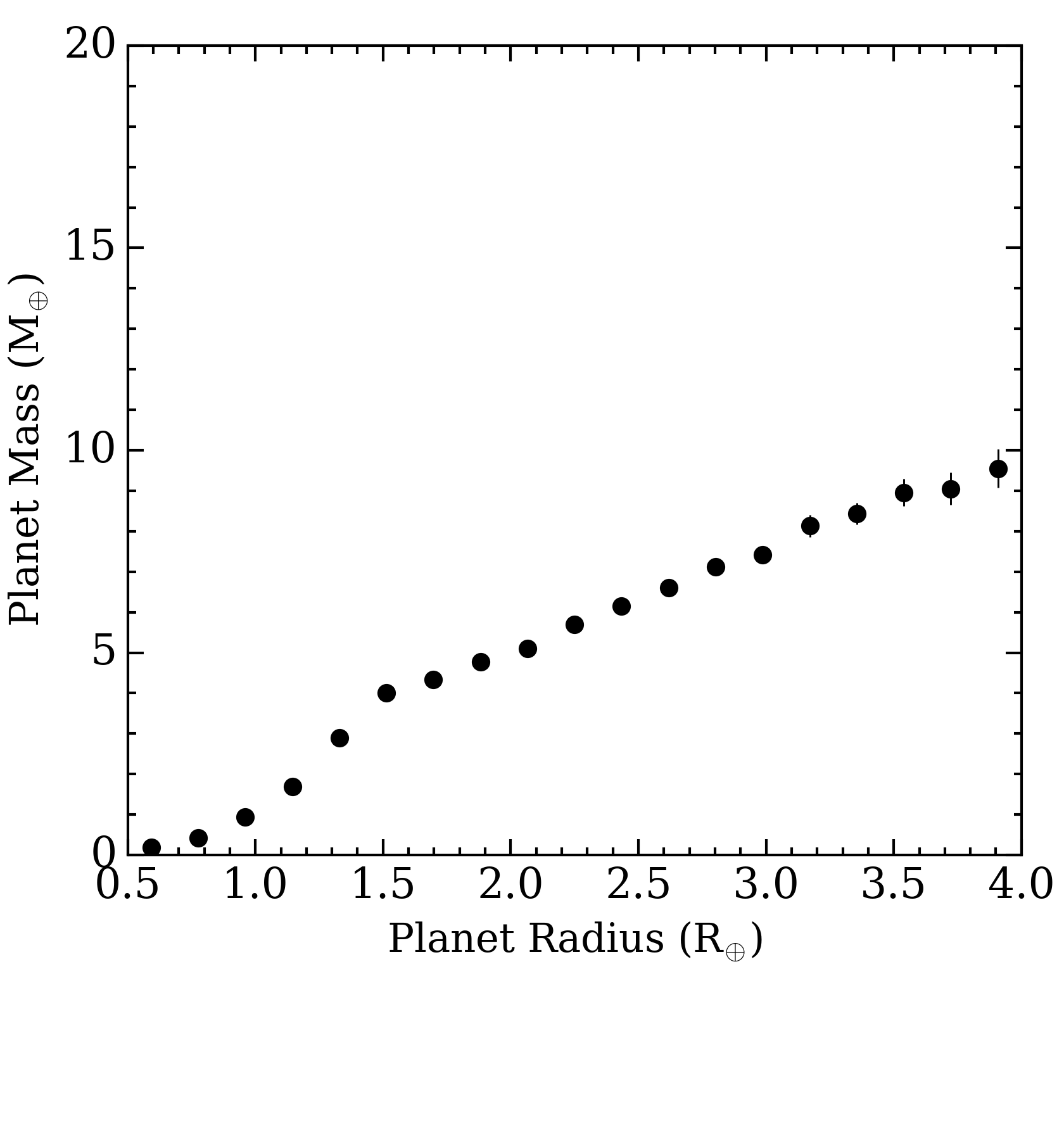}%
  \end{ocg}
  \hspace{-\hsize}%
  \caption{2D histogram depicting the masses and radii of the sampled
    small planet population with $r_p \leq 4$ R$_{\oplus}$. Planetary radii are drawn from the
    planet occurrence rates derived from Kepler and are mapped to 
    planetary masses using the mass-radius relation and Gaussian
    dispersion given in Eqs.~\ref{eq:mr} and \ref{eq:mrscat}. The mean
    mass-radius relation is over-plotted as the 
    \ToggleLayer{fig:curveon,fig:curveoff}{\protect\cdbox{\emph{dashed curves}}}. The mean
    planet mass and standard deviation of the mean in each planetary radius bin are plotted as
    \ToggleLayer{fig:ebon,fig:eboff}{\protect\cdbox{\emph{black circles}}} to
    show their close correspondence with the mean mass-radius relation.}
  \label{fig:mr}
\end{figure}

\noindent To include this intrinsic dispersion in our sampled planet population, for each sampled $r_p$
we compute the mean $m_p$ using Eq.~\ref{eq:mr} and add an additional offset drawn from
$\mathcal{N}(0,\sigma_{\text{rms}})$
where the value of $\sigma_{\text{rms}}$ is given by Eq.~\ref{eq:mrscat}. We reject planets with unphysical
negative masses
which naturally biases our sample to larger planet masses than are predicted by the \emph{mean} mass-radius
relation. We then apply a unique correction factor to each radius bin $r_p$ with a width of 0.2 R$_{\oplus}$
such that we recover the \emph{mean} mass-radius relation given in Eq.~\ref{eq:mr} in our final planet sample.
Our Kepler-derived planet sample in the mass-radius plane is shown in Fig.~\ref{fig:mr} along with the 
\emph{mean} mass-radius relation. Over-plotted are the mean planet masses in each $r_p$ bin demonstrating the
accuracy of our unique correction factors used to recover the mean mass-radius relation. Note the large
uncertainties in the mean planet masses resulting from the large dispersion observed in the empirical mass-radius
relation \citep{weiss14}.

\emph{Restricted range of $P$ and $r_p$}. The second caveat with the Kepler-derived $f$ is that
we are restricted in
our range of sampled orbital periods and planetary radii as imposed by the size of the parameter space
for which $f$ is robustly measured with Kepler. Specifically, $f(P,r_p)$ from \cite{dressing15a}
is restricted to the orbital period domain of $P \in [0.5,200]$ days and the planetary radius domain of 
$r_p \in [0.5,4]$ R$_{\oplus}$ which includes small Mars-sized planets up to about the size of Neptune. 
However, given that the duration of the SLS-PS is longer than 200 days it is important to also consider planets
with orbital periods $> 200$ days. To inform the inclusion of such long period planets we use the RV M dwarf
survey detection of the non-zero frequency of giant planets 
(\msini{} $\ge 100$ M$_{\oplus}$) with orbital 
periods of $>200$ days albeit with a low $f$ \citep[$\lesssim 5$\%;][]{bonfils13}.
We therefore supplement the Kepler occurrence rates with draws 
from the HARPS occurrence rates of giant planets \citep{bonfils13}.
Long period planet sampling is carried out similarly to how the Kepler occurrence rates are sampled.
Here we draw the planetary minimum masses rather than planetary radii; $f \to f(P,m_p\sin{i})$.
We also note that in \cite{bonfils13} $f$ for giant planets is computed over a coarse grid
in minimum mass spanning an order-of-magnitude 
($10^2 \le m_p\sin{i}/\text{M}_{\oplus} \le 10^3$) despite the most massive HARPS detection having only
\msini{} $\sim 112$ M$_{\oplus}$. Therefore it is possible that giant planets with minimum masses in excess
of $\sim 112$ M$_{\oplus}$ ($0.3$ M$_{Jup}$) do not exist in nature around M dwarfs.
We note that throughout this study we will
primarily focus on the population of small planets from the Kepler-derived $f(P,r_p)$ 
because of their much larger frequency compared to giant planets around M dwarfs.

\emph{Applicability of occurrence rates to the full input catalog}.
Robust statistics regarding $f$ were derived from the 
sample of small Kepler stars which almost exclusively contained early-to-mid M dwarfs and late K-dwarfs.
The assumption that the resulting $f$ extend to later spectral
types is still largely uncertain but the early discovery of seven transiting Earth-sized planets around the
ultracool dwarf TRAPPIST-1 \citep{gillon17}, from a small sample of observed stars, hints that small planets
around late M dwarfs might be as common, and potentially more common than around the early M dwarfs observed
with Kepler. Preliminary estimates of $f$ around
late M dwarfs with \emph{K2} suggests a potential lack of super-Earth-sized planets on close-in orbits but have
been insufficient to probe the population of Earth-sized planets \citep{demory16}. Theoretically,
planet formation scenarios have also predicted the existence of many such small planets on close-in orbits
around late M dwarfs \citep[e.g.][]{alibert13,alibert17}. In Sect.~\ref{sect:2occ} we will investigate the
effect of increasing $f$ on our planet detections in the SLS-PS.

\emph{Dynamical considerations of multi-planet systems}.
The final caveat arises from $f(P,r_p)$ being derived in uncorrelated bins whereas dynamical constraints 
will prevent certain types of planetary systems from existing in nature. For example, close pairs
of massive planets. To ensure that sampled multi-planetary systems in our simulations are dynamically stable
we impose two priors on each system with multiplicity $> 1$.
The first constraint is the analytic assessment of Lagrange
stability \citep{barnes06} which only depends on the masses of the central star and planets, the planets'
semi-major axes, and eccentricities. For adjacent planet pairs that are Lagrange stable, by definition
their ordering remains fixed, both planets remain bound to the central star, and the criterion limits 
permissible changes in planets’ semi-major axes. Lagrange stability can be thought of as a more stringent
extension of Hill stability \citep{gladman93}. However we
note that the analytic treatment of Lagrange stability is only applicable to the three-body system.
In planetary systems from our simulations with $>2$ planets, we apply the Lagrange stability criterion
to every adjacent planet pair. We then supplement the Lagrange stability
criterion with the heuristic criterion from \cite{fabrycky12} which is applicable to systems with multiplicity
$>2$. The resulting criterion for stability\footnote{Long-term stability requires that
  $a_{\text{out}} - a_{\text{in}} \gtrsim 3.5$ $R_{\text{Hill}}$ where $a_{\text{in}}$ and $a_{\text{out}}$ are the
  semi-major axes of the inner and outer planet respectively and $R_{\text{Hill}}$ is their mutual Hill radius.}
is derived from the stability analysis of a set of numerical integrations. In our MC simulations
we only include multi-planetary systems which satisfy both aforementioned stability criteria.

\begin{figure}
  \centering
  \includegraphics[width=\hsize]{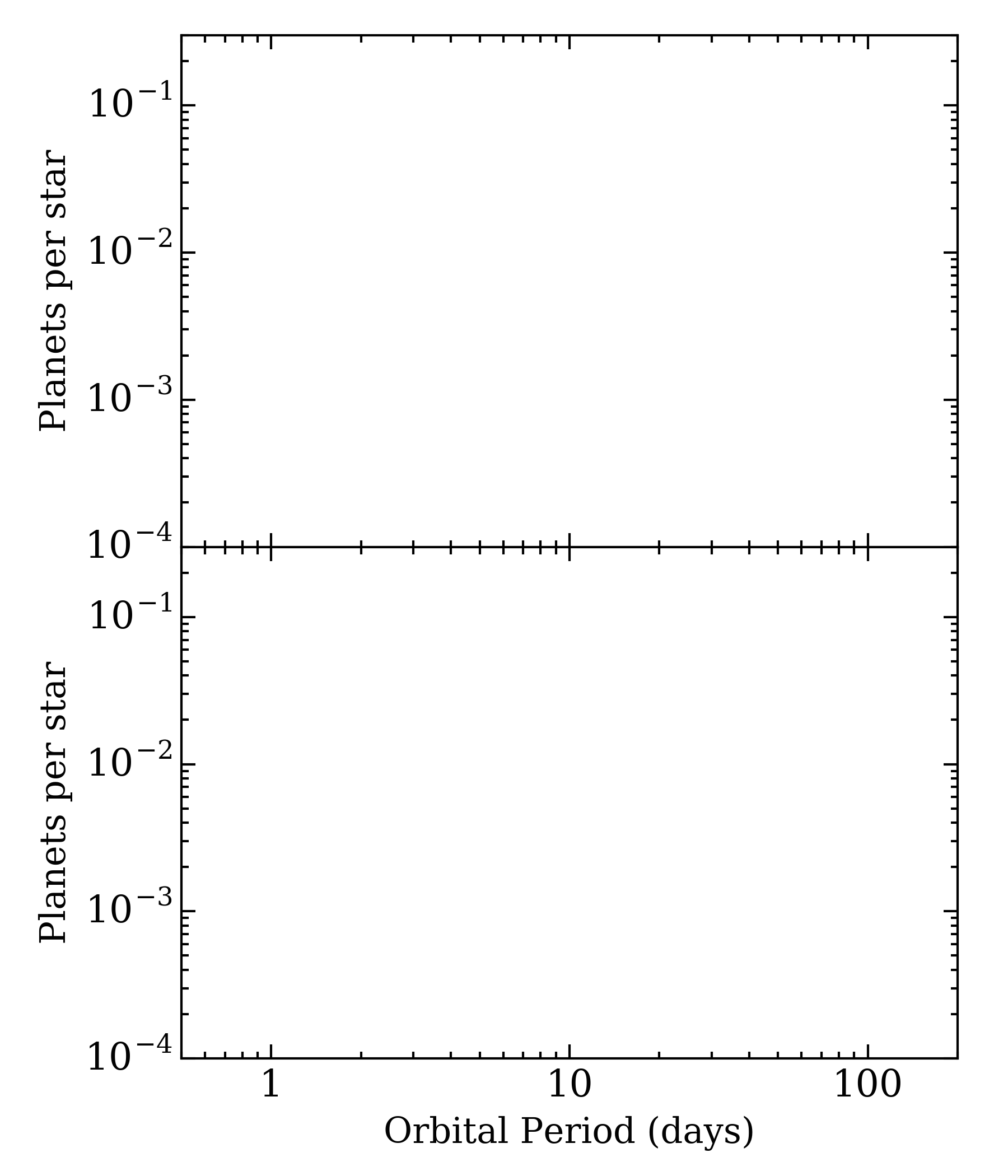}%
  \hspace{-\hsize}%
  \begin{ocg}{fig:0off}{fig:0off}{0}%
  \end{ocg}%
  \begin{ocg}{fig:0on}{fig:0on}{1}%
    \includegraphics[width=\hsize]{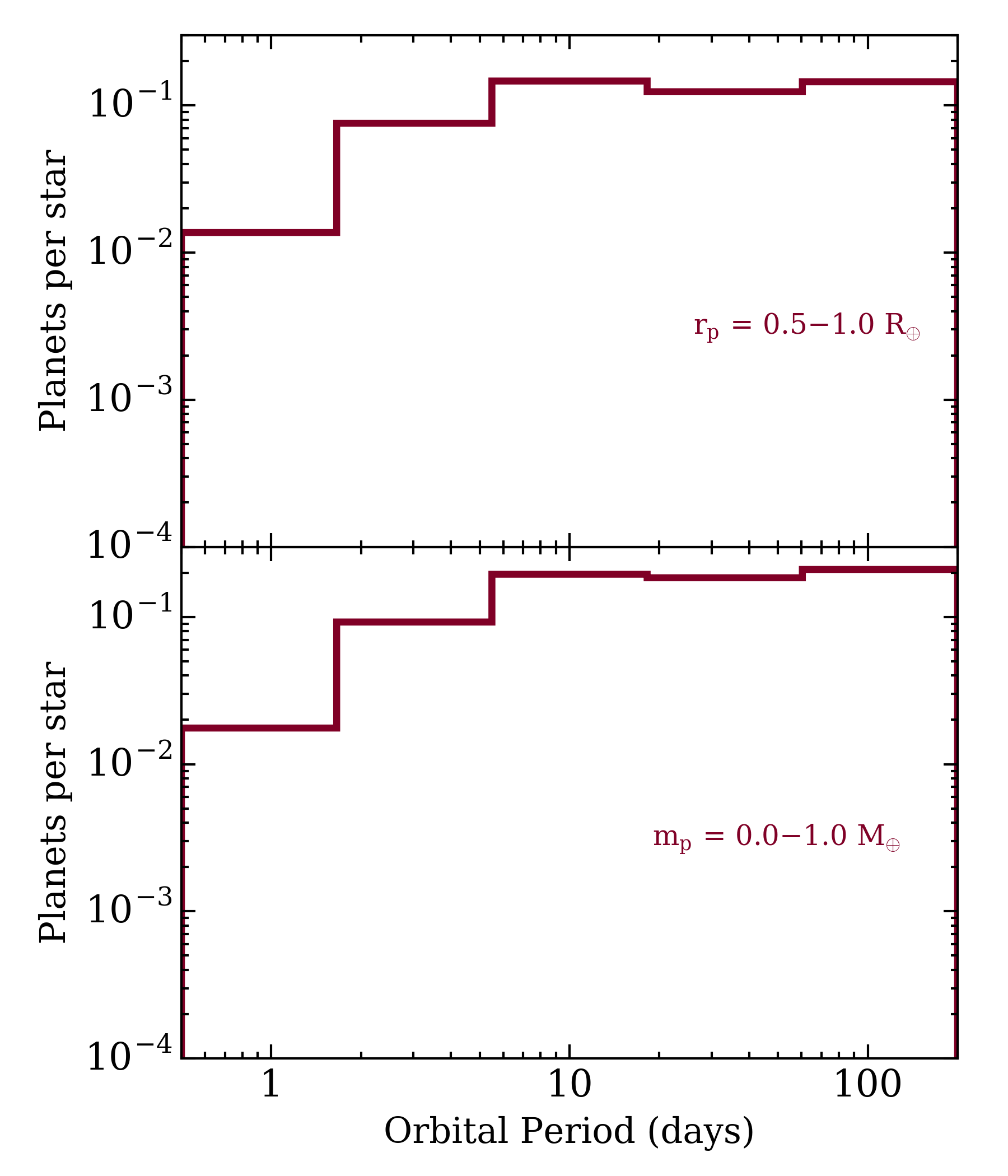}%
  \end{ocg}
  \hspace{-\hsize}%
  \begin{ocg}{fig:1off}{fig:1off}{0}%
  \end{ocg}%
  \begin{ocg}{fig:1on}{fig:1on}{1}%
    \includegraphics[width=\hsize]{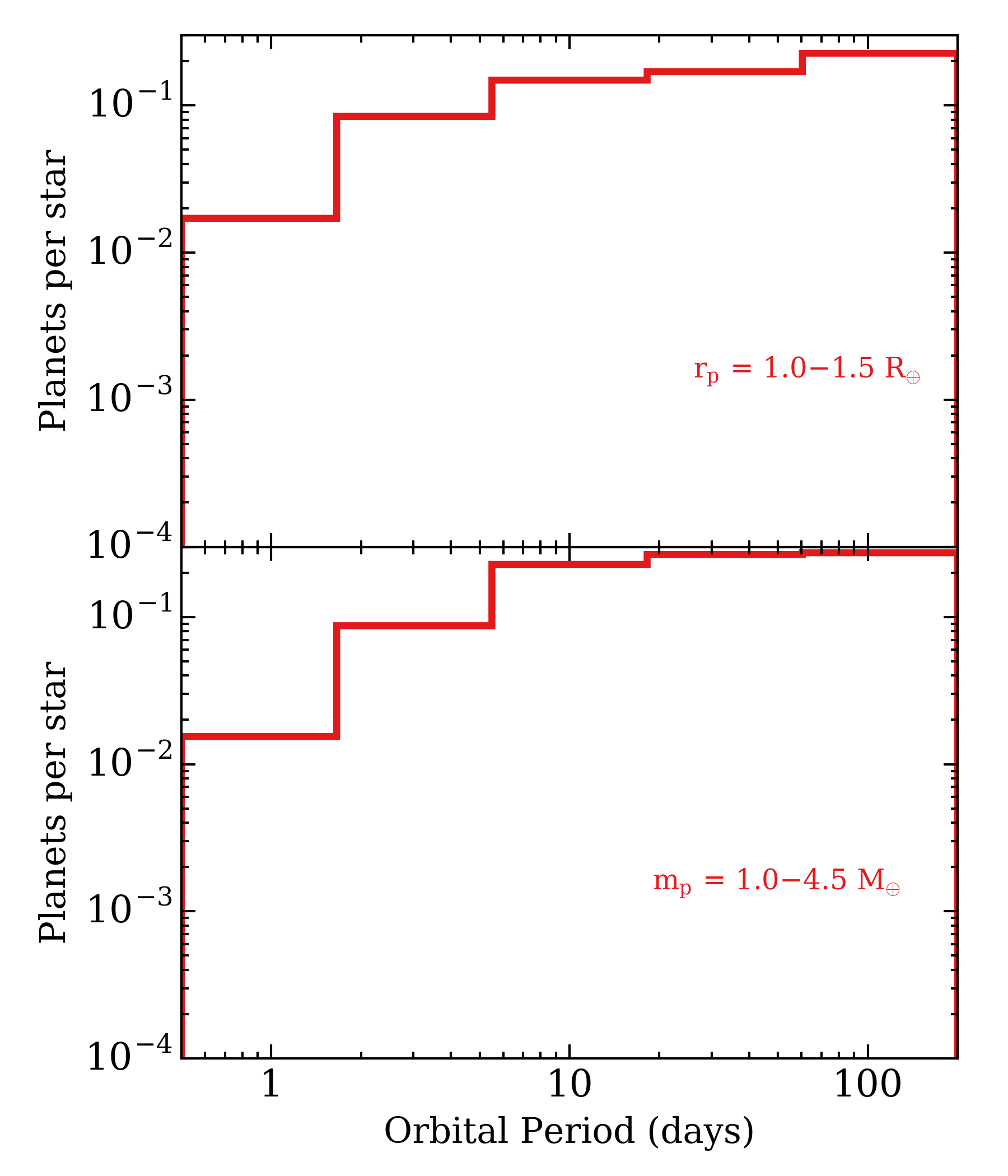}%
  \end{ocg}
  \hspace{-\hsize}%
  \begin{ocg}{fig:2off}{fig:2off}{0}%
  \end{ocg}%
  \begin{ocg}{fig:2on}{fig:2on}{1}%
    \includegraphics[width=\hsize]{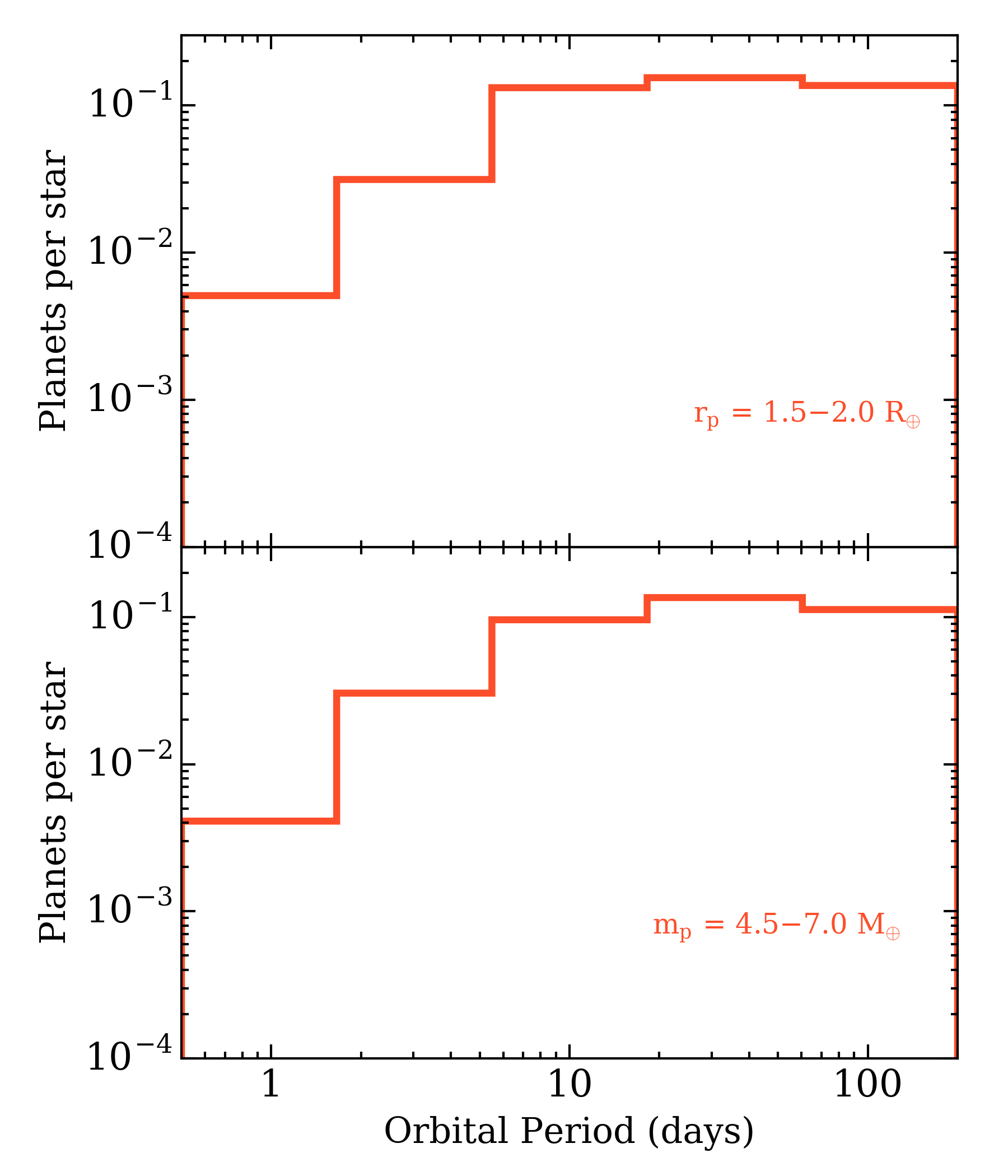}%
  \end{ocg}
  \hspace{-\hsize}%
    \begin{ocg}{fig:3off}{fig:3off}{0}%
  \end{ocg}%
  \begin{ocg}{fig:3on}{fig:3on}{1}%
    \includegraphics[width=\hsize]{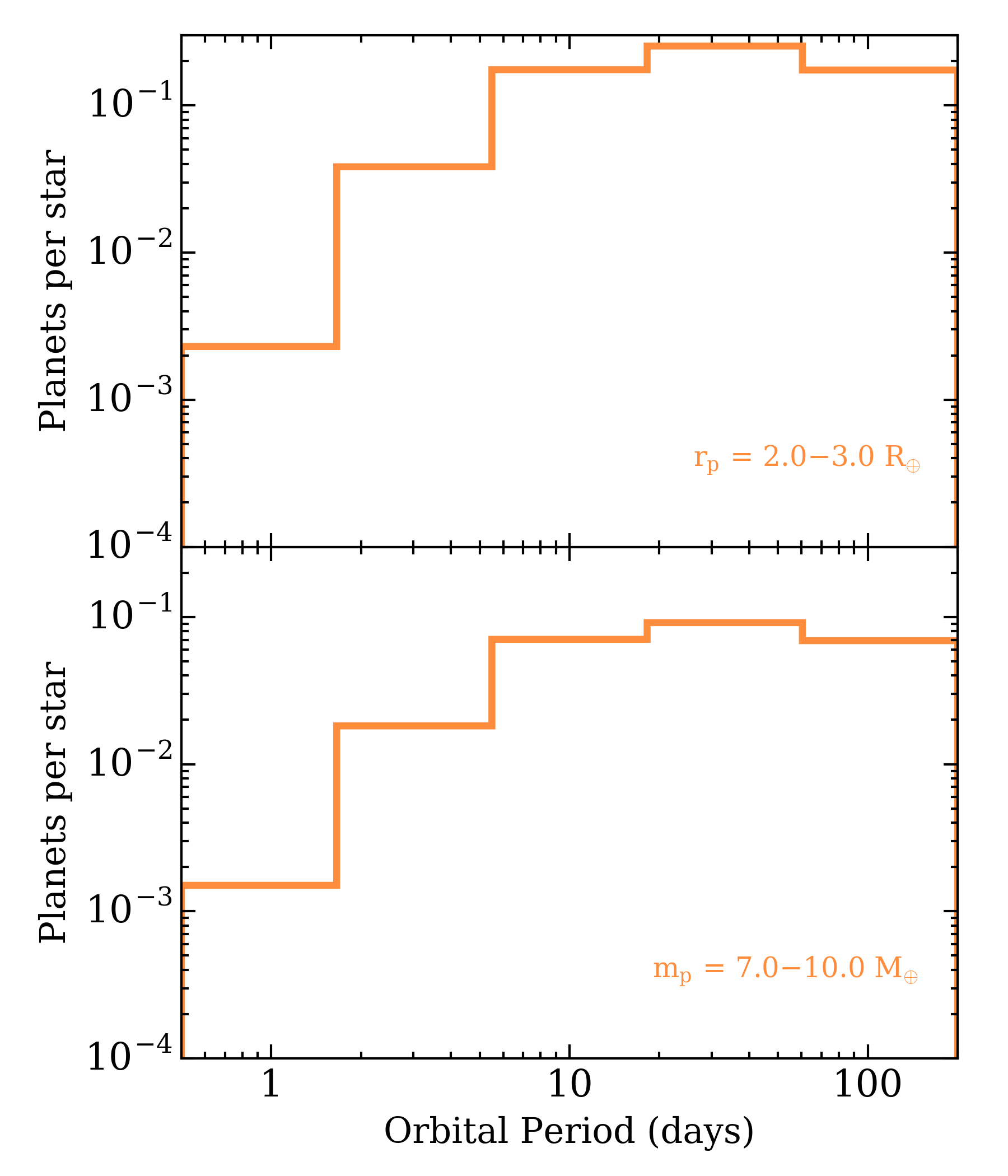}%
  \end{ocg}
  \hspace{-\hsize}%
    \begin{ocg}{fig:4off}{fig:4off}{0}%
  \end{ocg}%
  \begin{ocg}{fig:4on}{fig:4on}{1}%
    \includegraphics[width=\hsize]{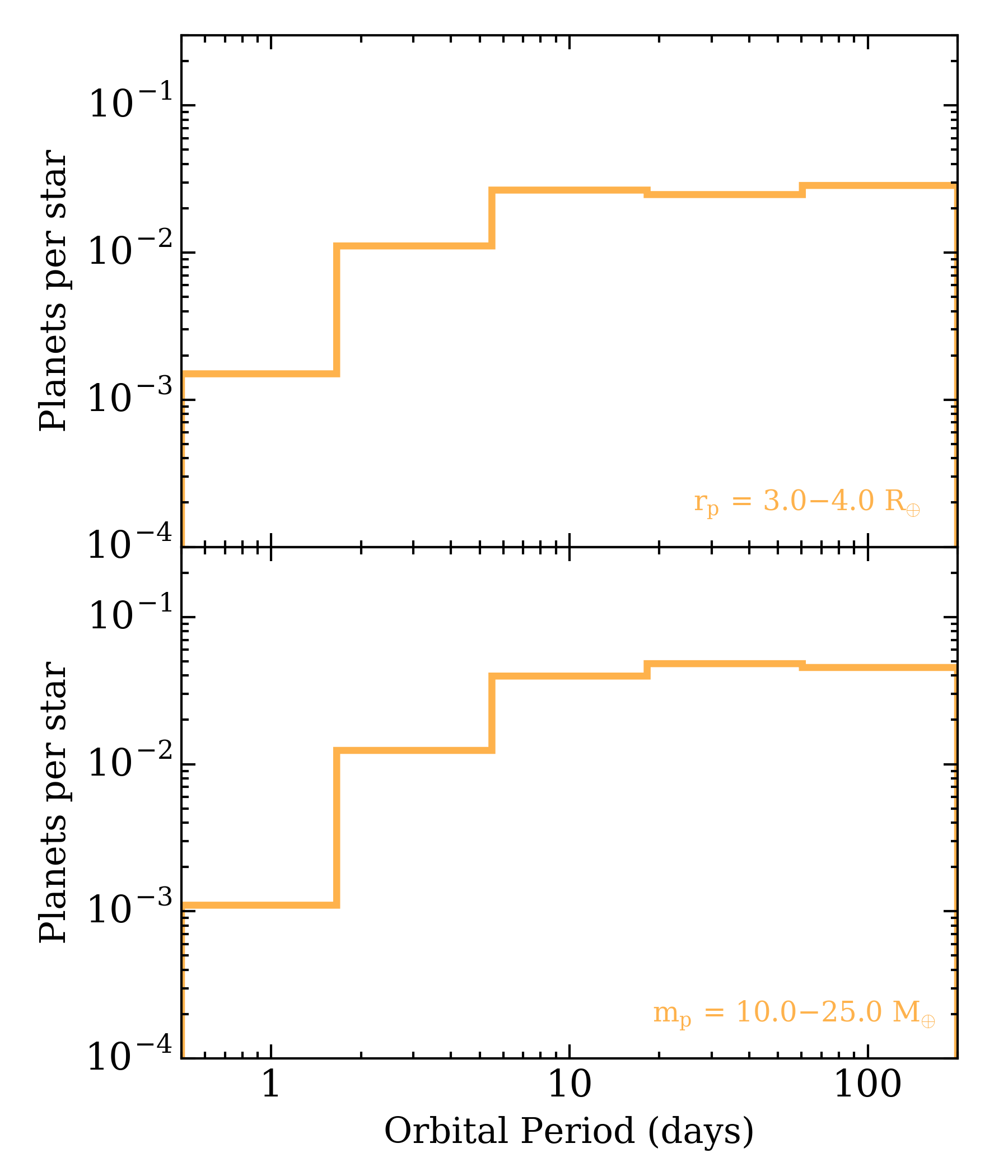}%
  \end{ocg}
  \hspace{-\hsize}%
  \caption{\emph{Top}: histograms of the injected small planet population in the simulated SLS-PS
    as a function of orbital period and planetary radius from \cite{dressing15a}. \emph{Bottom}:
    the same planet population as above converted to planetary mass using Eqs.~\ref{eq:mr}
    and~\ref{eq:mrscat}. For clarity each $r_p$ bin, and the approximately corresponding
    $m_p$ bin, can be viewed individually:
    \ToggleLayer{fig:0on,fig:0off}{\protect\cdbox{$r_{p,min}=0.5$ R$_{\oplus}$}},
    \ToggleLayer{fig:1on,fig:1off}{\protect\cdbox{$r_{p,min}=1$ R$_{\oplus}$}},
    \ToggleLayer{fig:2on,fig:2off}{\protect\cdbox{$r_{p,min}=1.5$ R$_{\oplus}$}},
    \ToggleLayer{fig:3on,fig:3off}{\protect\cdbox{$r_{p,min}=2$ R$_{\oplus}$}},
    \ToggleLayer{fig:4on,fig:4off}{\protect\cdbox{$r_{p,min}=3$ R$_{\oplus}$}}.}
  \label{fig:occurrence}
\end{figure}

As mentioned previously,
our dynamical considerations cause the injected planet population to not exactly match the adopted Kepler
occurrence rates. The modified Kepler planet occurrence
rates, or equivalently our injected planetary population, is shown in Fig.~\ref{fig:occurrence} as a function
of the input variables $P$ and $r_p$. We then report the same population after converting the planetary radii to
masses using Eqs.~\ref{eq:mr} and~\ref{eq:mrscat}.

\begin{figure}
  \centering
  \includegraphics[width=\hsize]{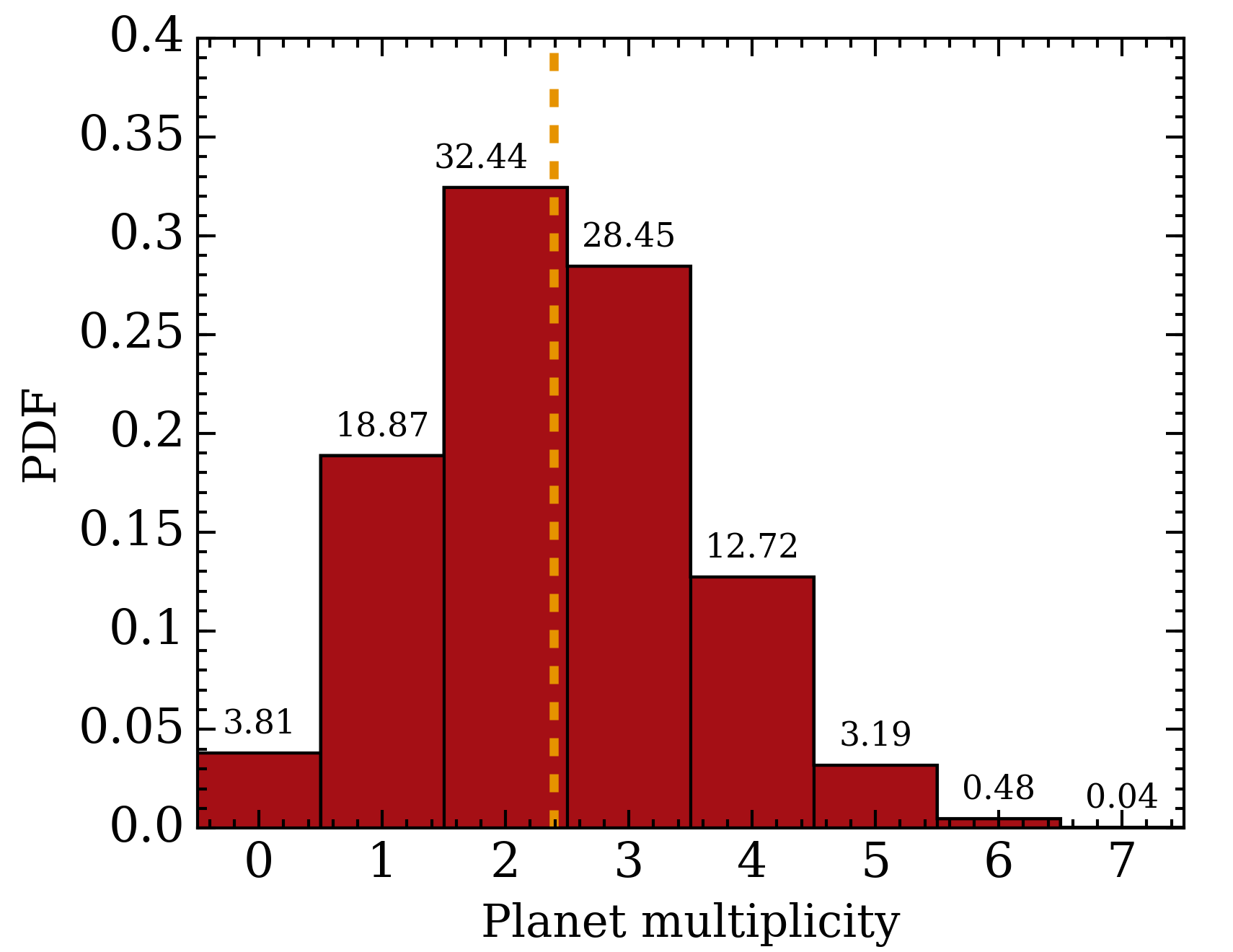}%
  \caption{The probability density function of planet multiplicities in the simulated SLS-PS.
    The fraction of simulated systems with a given multiplicity---in percentage---is annotated for
    each integer multiplicity. We find an average planet multiplicity of $\sim 2.4$ which is highlighted by the
    \emph{vertical dashed line} and is slightly less than the cumulative injected multiplicity of 2.5 planets
    with $P\leq 200$ days from \cite{dressing15a}.}
  \label{fig:mult}
\end{figure}

The resulting distribution of planetary system multiplicities is shown in Fig.~\ref{fig:mult}.
Simulated planetary systems that obey our dynamical stability criteria contain 0-7 planets although only 0.04\%
of simulated planetary systems can survive with 7 planets. Similarly, $\sim 4$\% of simulated planetary systems contain
no planets at all. 
The most common planet multiplicity is 2 with $\sim 32$\% of simulated planetary systems containing 2 planets. The resulting
average planet multiplicity is $\sim 2.4$ which is slightly less than the cumulative injected multiplicity of
$2.5 \pm 0.2$. When
recovering the planet yield of our simulated survey we will have to correct for this small discrepancy between the
cumulative planet multiplicity of our injected population and the \emph{true} multiplicity of 2.5 for planets
with $P \in [0.5,200]$ days and $r_p \in [0.5,4]$ R$_{\oplus}$ (see Sect.~\ref{sect:yield}).

\section{Activity Mitigation} \label{sect:GP}
\subsection{Overview of the Gaussian Process Formalism}
RV activity signals present in M dwarfs (e.g. Gl176, \citealt{forveille09}; Gl674, \citealt{bonfils07},
Proxima Centauri; \citealt{robertson16}) will deter our ability to detect new exoplanets in the SLS-PS.
A number of correction techniques have been developed by various groups
within the field to mitigate these effects and detect planets in RV \citep[see reviews in][]{fischer16, dumusque17}.
One particularly promising method is the use of Gaussian process regression to model correlated noise (i.e. activity)
in RV time-series.

Gaussian processes (GP) belong to a class of \emph{non-parametric} regression models from the field of machine-learning
\citep{rasmussen05}.
Being non-parametric, the functional form of the activity model is unspecified and instead is determined by the
data itself. In this way the GP activity model does not assume 
any physical model of the underlying processes responsible for the observed activity. Instead the GP is
used to model the covariance properties 
of an input time-series according to a user-defined covariance function which itself is described by a small number
of hyperparameters\footnote{`Small' compared to the size of the input time-series which would be required to fully
  describe the covariance properties of the input dataset.}.
By its non-parametric nature, GP regression is an attractive method for modelling the stochastic processes that give
rise to observable stellar activity.

The GP prior distribution, which is described by the aforementioned covariance function,
is a multi-variate Gaussian distribution of functions specified by a mean function
$\boldsymbol{\mu}(\mathbf{t})$, evaluated at the epochs contained in the window function $\mathbf{t}$,
and a covariance matrix

\begin{equation}
  K_{ij} = \sigma_i^2 \delta_{ij} + k(t_i,t_j). \label{eq:K}
\end{equation}

\noindent which
is computed from the vector of measurement uncertainties $\boldsymbol{\sigma}(\mathbf{t})$,
the Kronecker delta function $\delta_{ij}$, and
the user-defined covariance function $k(t_i,t_j)$ describing the covariance between two measurements taken at
times $t_i$ and $t_j$for $i,j \in [1,n_{\text{obs}}]$.
Obtaining the GP model of an arbitrary input time-series $\mathbf{y}$ and uncertainty vector
$\boldsymbol{\sigma}$ is done by maximizing the Gaussian logarithmic likelihood function

\begin{equation}
  \ln{\mathcal{L}} = -\frac{1}{2} \left( (\mathbf{y}-\boldsymbol{\mu})^T K^{-1}
  (\mathbf{y}-\boldsymbol{\mu}) + \ln{\mathrm{det} K} + n_{\text{obs}} \ln{2 \pi} \right), \label{eq:like}
\end{equation}
  
\noindent to obtain the `best-fit' values of the GP hyperparameters describing the covariance of the observations
$\mathbf{y}$ through the covariance function $k(t_i,t_j)$.
In the case of M dwarfs, activity predominantly arises from rotationally modulated ARs which also
evolve in time due to their varying lifetimes, spatial distribution, and contrast.
Because of this temporal evolution we adopt a quasi-periodic covariance function
of the form

\begin{equation}
k(t_i,t_j) = a^2 \exp{\left[ - \frac{|t_i-t_j|^2}{2\lambda^2} -\Gamma^2
    \sin^2{\left(\frac{\pi|t_i-t_j|}{P_{\text{GP}}} \right)} \right]}, \label{cov}
\end{equation}

\noindent described by four hyperparameters: $a$ the amplitude of the covariance in the units of $\mathbf{y}$,
$\lambda$ the exponential decay of correlations, $\Gamma$ the coherence scale of correlations,
and $P_{\text{GP}}$ the periodic timescale. 
From identical reasoning to our own, quasi-periodic covariance functions are commonly adopted in a
number of related astrophysical applications such as the recovery of stellar rotation periods (i.e.
setting $P_{\text{GP}}=P_{\text{rot}}$) \citep[e.g.][]{angus17} and to the modelling of RV activity
from spectroscopic activity diagnostics \citep[e.g.][]{haywood14, rajpaul15, cloutier17b}, 
photometry \citep[e.g.][]{cloutier17a, cloutier17b}, or the raw RVs themselves
\citep[e.g.][]{faria16, donati17, yu17} and is applicable when the baseline of the
observations spans at least a few rotation periods \citep{pont13}.

Following the optimization of the GP hyperparameters one obtains a unique GP prior distribution. Conditioning
the GP prior on the dataset $\mathbf{y}(\mathbf{t})$ results in the GP predictive distribution whose mean function
and posterior variance can be evaluated at previously unseen epochs $\mathbf{t}^*$ via

\begin{equation}
  \boldsymbol{\mu}(\mathbf{t}^*) = K(\mathbf{t}^*,\mathbf{t}) \cdot K(\mathbf{t},\mathbf{t})^{-1} \cdot \mathbf{y}(\mathbf{t})
  \label{eq:GPmean}
\end{equation}

\noindent and 

\begin{equation}
  C(\mathbf{t}^*) = K(\mathbf{t}^*,\mathbf{t}^*) - K(\mathbf{t}^*,\mathbf{t}) \cdot K(\mathbf{t},\mathbf{t})^{-1} \cdot
  K(\mathbf{t}^*,\mathbf{t})^T.
  \label{eq:GPvar}
\end{equation}

\noindent Here the covariance matrix $K(\mathbf{t}^*,\mathbf{t}^*)$ is evaluated at unseen epochs such that
$\boldsymbol{\sigma}(\mathbf{t}^*)$ must be set to zero in Eq.~\ref{eq:K}.

\subsection{Modelling RV Activity with GP Regression} \label{sect:regression}
As in nature, the RV time-series from our MC simulations contain contributions from both planetary companions
as well as activity from ARs. These sources of the RV activity signals also have manifestations in other
spectroscopic time-series, albeit not strictly at the same rotation period \citep{hebrard16}. One example is the
full width at half maximum (FWHM) of the CCF which we simulate along with the RV signals (see Sect.~\ref{sect:activity}).
As is often the case when searching for small planets
in RV, one cannot distinguish a-priori an RV signal from a planet or from activity. However the covariance
properties in the FWHM time-series will be related to the covariance properties in RV residuals after the removal
of planetary sources. We therefore use the FWHM time-series to train the hyperparameters of our quasi-periodic
GP prior distribution. In principle one could have chosen an alternative activity indicator such as the BIS or
contrast of the CCF however recent HARPS observations have demonstrated that the strongest correlations between the RVs and
an activity indicator often exist between the RV and FWHM time-series 
(e.g. \citealt{astudillodefru17}, Bonfils et al. 2017b in prep). However
this is not universal. Similarly, the contrast or depth of the CCF is not used in place
of the FWHM because although the two time-series are highly correlated, the CCF contrast is known to be more strongly
affected by instrumental noise in practice.
Using a spectroscopic training set like the FWHM also has the benefit of being obtained contemporaneously with the
RV measurements therefore probing the star's activity at the same epochs in which we are searching for planets.
In this way our GP activity model is also sensitive to variations in the RV activity occurring on long timescales such as
from magnetic activity cycles.

To reduce computational wall time of our full MC simulation
we do not compute a GP activity model in each MC realization of our simulated
SLS-PS. The GP activity model is only computed when the rms of the injected RV activity exceeds the median RV measurement
uncertainty. In such cases we run a Markov Chain Monte-Carlo
(MCMC) on the FWHM time-series to obtain the marginalized posterior probability density functions (PDF)
of the four GP hyperparameters: $a,\lambda,\Gamma$, and $P_{\text{GP}}$.
To sample the posterior PDFs we use the \texttt{emcee} affine invariant
MCMC ensemble sampler \citep{foremanmackey13} coupled with the fast GP package \texttt{george} 
\citep{ambikasaran15} to evaluate the likelihood function in Eq.~\ref{eq:like}.
Specifically we initialize 100 walkers with an effective chain length
$\sim 10$ autocorrelation times to ensure convergence of the chains.
Samples of the posterior PDFs are only saved following a burn-in phase of $\sim 10$ autocorrelation times.
From a set of supervised preliminary tests of the MCMC procedure, walkers are initialized in the
parameter space within Gaussian balls whose variance is chosen such that the mean acceptance fraction among
the walkers is $20-60$\%.

\begin{figure*}
  \centering
  \includegraphics[width=\hsize]{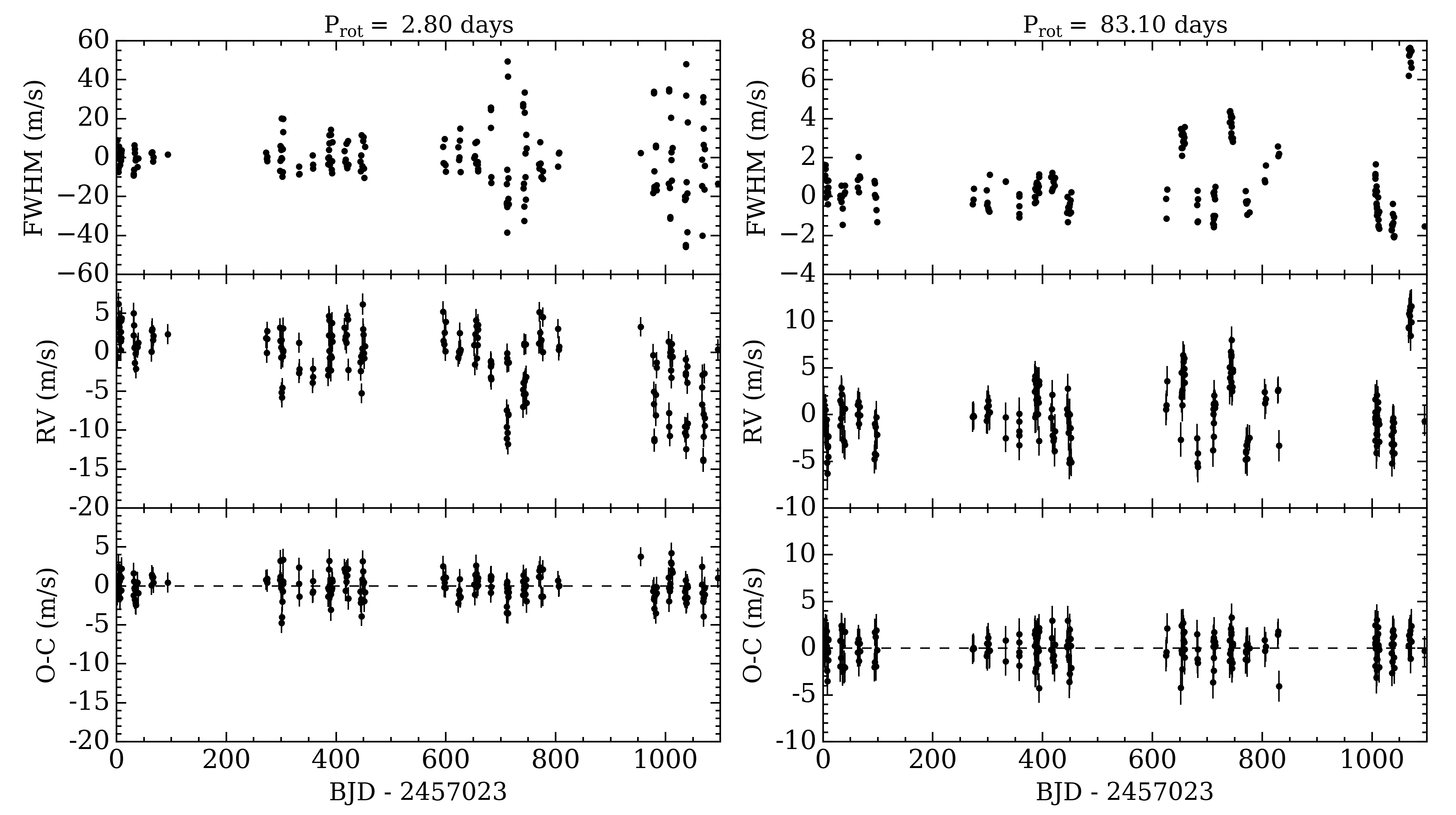}%
  \hspace{-\hsize}%
  \begin{ocg}{fig:GPoff}{fig:GPoff}{0}%
  \end{ocg}%
  \begin{ocg}{fig:GPon}{fig:GPon}{1}%
    \includegraphics[width=\hsize]{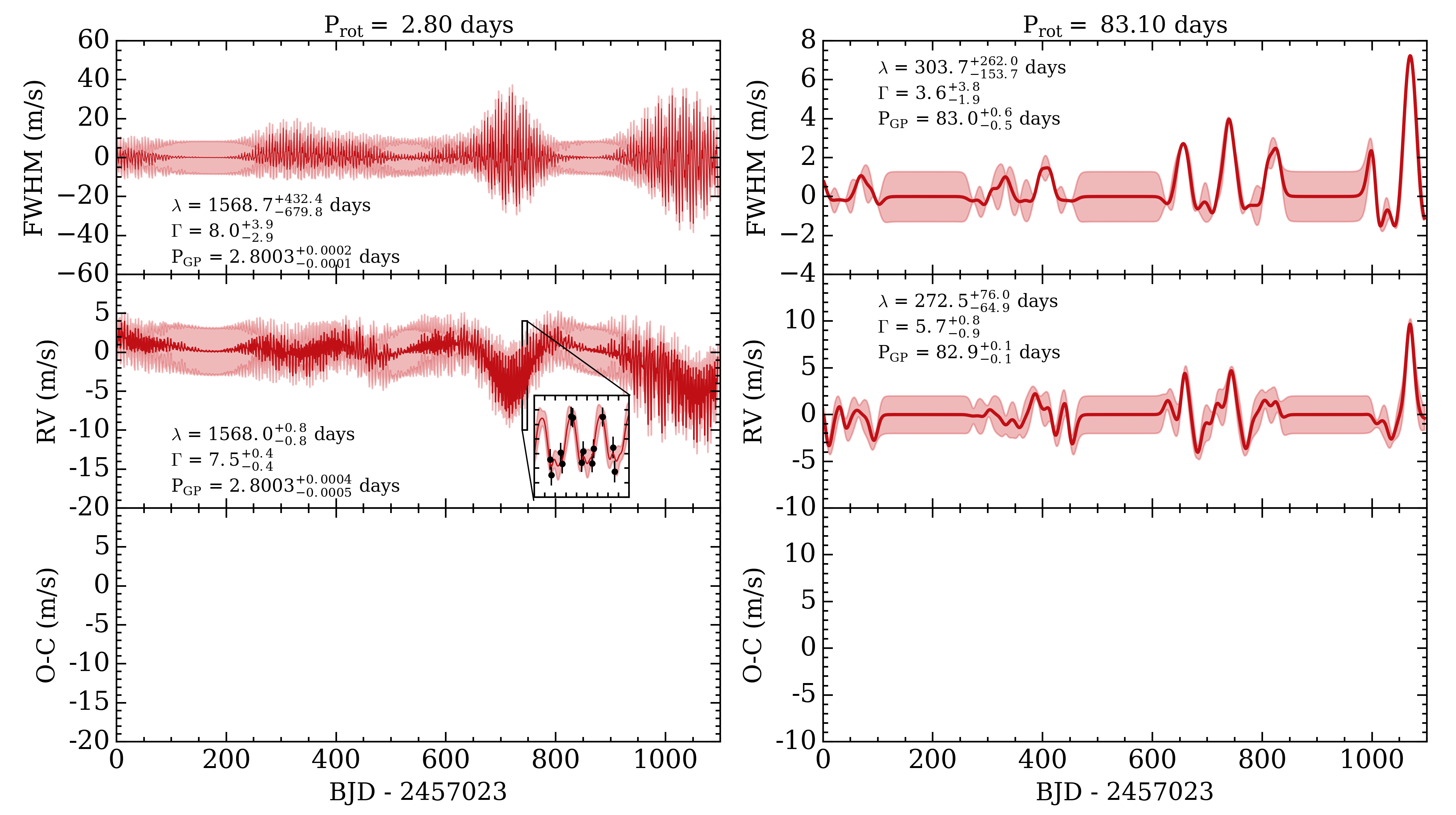}%
  \end{ocg}
  \hspace{-\hsize}%
  \caption{Examples of our GP formalism used to mitigate RV activity and detect underlying planetary signals for two
    simulated systems with either a rapidly (\prot{} $=2.8$ days; \emph{left column}) or slowly (\prot{} $=83.1$ days;
    \emph{right column}) rotating host star. \emph{Top panels}: the FWHM time-series used for training along with the
    \ToggleLayer{fig:GPon,fig:GPoff}{\protect\cdbox{mean GP regression model and its $1\sigma$ confidence interval}}
    shown in red. \emph{Middle panels}: the raw
    RV time-series along with the mean GP activity model (in the absence of planets) and its $1\sigma$ confidence interval.
    \emph{Bottom panels}: the RV residuals after removal of the mean GP activity model shown in the middle panels.
    Each system contains 3 planets which contribute to the $O-C$ residuals albeit with semi-amplitudes which are all
    $\lesssim \sigma_{\text{RV}}$. The two upper rows report the MAP values of GP hyperparameters along with their
    $16^{\text{th}}$ and $84^{\text{th}}$ percentiles.}
  \label{fig:GPexample}
\end{figure*}

We select broad non-informative priors for all GP hyperparameters with the
exception of $P_{\text{GP}}$ which is constrained to the narrow uniform range of
$P_{\text{GP}}/P_{\text{rot}} \in \mathcal{U}(0.9,1.1)$ either when the photometric rotation period is known for a
particular star or if \prot{} is detected in a LS periodogram
of the FWHM with a false-alarm probability (FAP) $\le 1$\%. Prescription of the GP hyperparameter priors are
reported in Table.~\ref{table:gppriors}. Throughout this paper we calculate FAPs via
bootstrapping with replacement using $10^4$ iterations and each LS periodogram is normalized by its standard
deviation. From the marginalized posterior PDFs of the GP hyperparameters, we adopt the \emph{maximum a-posteriori}
(MAP) values of each hyperparameter to construct a unique covariance matrix $K$ and thus our mean GP model
and its $1\sigma$ confidence interval from Eqs.~\ref{eq:GPmean} and~\ref{eq:GPvar}. Examples of resulting
GP models of the FWHM time-series for a rapidly and a slowly rotating star are shown in the upper panels in
Fig.~\ref{fig:GPexample}.

\begin{deluxetable*}{cc}
\tabletypesize{\scriptsize}
\tablecaption{Gaussian Process Hyperparameter Priors Used in Training\label{table:gppriors}}
\tablewidth{0pt}
\tablehead{Hyperparameter & Prior}
\startdata
Covariance amplitude, $a$ [\mps{]} & $\mathcal{J}(10^{-2}, 10^2) \cdot \max{|FWHM-\langle FWHM \rangle|})$ \\
Exponential decay timescale, $\lambda$ [days] & $\mathcal{J}(1, 10\cdot (\max{\mathbf{t}}-\min{\mathbf{t}}))$  \\
Coherence scale, $\Gamma$ & $\mathcal{J}(10^{-2}, 10^2)$ \\
Periodic timescale, $P_{\text{GP}}$ [days] & $\mathcal{J}(0.1,300)$\tablenotemark{a} \\
& $\mathcal{U}(0.9, 1.1) \cdot P_{\text{rot}}$\tablenotemark{b}
\enddata
\tablenotetext{a}{If \prot{} \emph{is not} detected in the FWHM time-series.}
\tablenotetext{b}{If \prot{} \emph{is} detected in the FWHM time-series.}
\end{deluxetable*}

Following the training phase on the FWHM time-series
we proceed with modelling the RVs simultaneously with a trained GP activity model plus keplerian planetary
signals. The marginalized posterior PDFs of the GP hyperparameters $\lambda$, $\Gamma$, and $P_{\text{GP}}$
from training are used as informative priors in the joint RV analysis which treats the GP amplitude
$a$ as a free parameter. For each assumed mean function $\boldsymbol{\mu}$, containing between zero and three
keplerian solutions, we compute the MAP GP activity model from the hyperparameter values sampled using MCMC.
Assuming a zero mean function, the resulting mean GP activity models
for the two stars shown in Fig.~\ref{fig:GPexample} are shown in the middle panels of the figure. The residuals
following the removal of the mean GP activity model is also shown. In each case the stellar rotation period is
detected in the LS periodogram of the FWHM time-series and therefore is used to constrain $P_{\text{GP}}$ during
training. For the rapid rotator with \prot{} $=2.8$ days, the rms of the injected activity is 4.43 \mps{} and
is reduced to 1.59 \mps{} after removing the mean GP activity model ($\chi_{\text{red}}^2 =1.4$ for four GP hyperparameters).
This resulting RV rms is more comparable
to the median RV measurement uncertainty $\sigma_{\text{RV}}=1.35$ \mps{.} In the slow rotator case (\prot{} $=83.1$ days)
the rms of the injected activity is reduced from 3.60 \mps{} to 1.53 \mps{} compared to
$\sigma_{\text{RV}}=1.63$ \mps{} ($\chi_{\text{red}}^2 =0.9$). In both test cases considered in
Fig.~\ref{fig:GPexample}, a planet is detected in the RV residuals and fit simultaneously with the activity assuming
a new mean model containing a single keplerian solution. Details of our planet detection algorithm are discussed in
Sect.~\ref{sect:detection}.

\subsection{GP Activity Model Performance}
As an overview of the performance of our GP activity modelling, we can compare the rms of the known
injected RV activity with the rms of the residuals following the removal of our mean GP activity model. The residual
rms should never exceed the rms of the injected activity otherwise our GP formalism would be adding additional noise
into the RV time-series rather than modelling and reducing it as is its intended purpose. Similarly, optimal GP fits
will result in a residual rms that is close to the median RV measurement uncertainty of the time-series.

\begin{figure}
  \centering
  \includegraphics[width=\hsize]{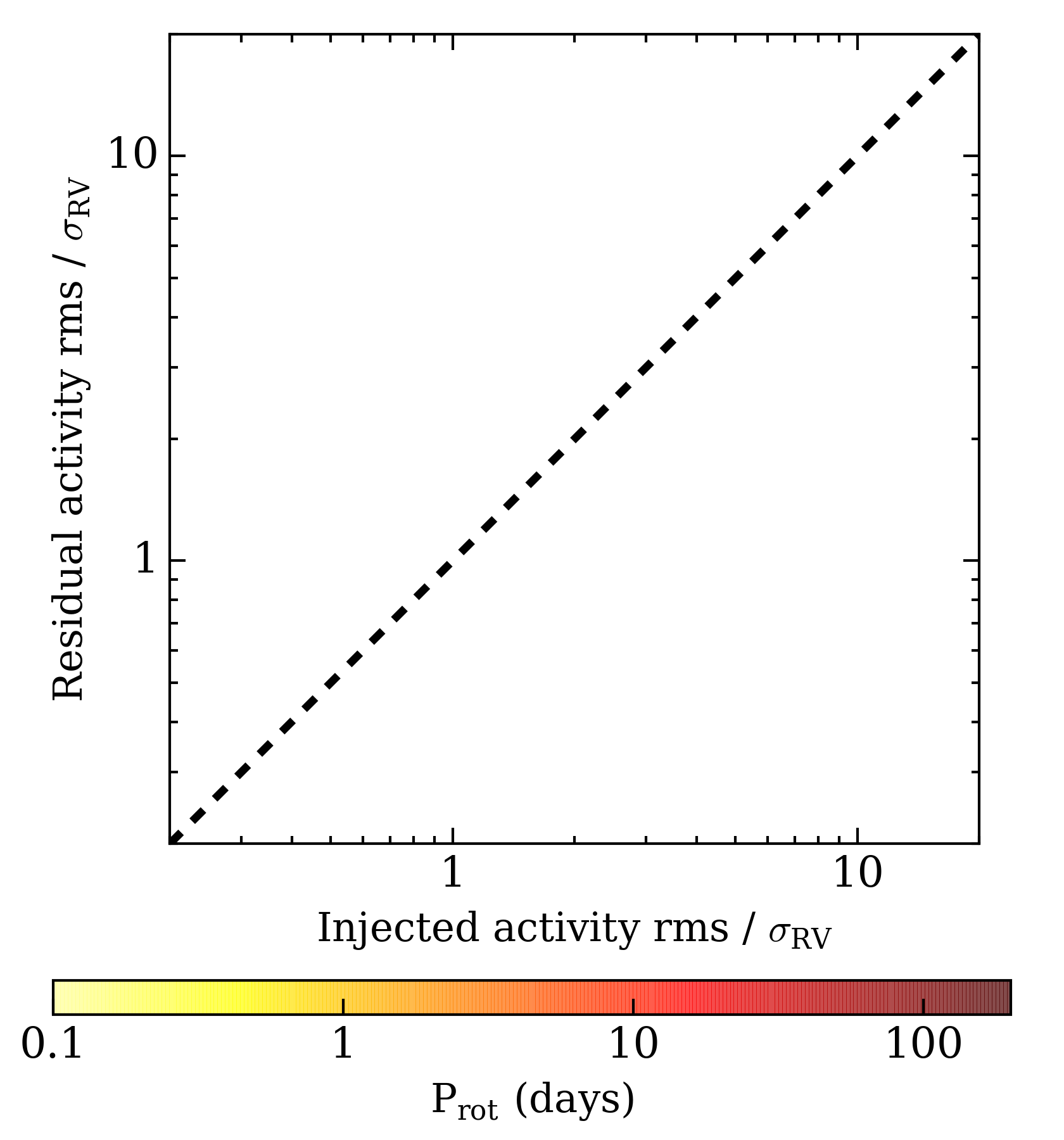}%
  \hspace{-\hsize}%
  \begin{ocg}{fig:detoff}{fig:detoff}{0}%
  \end{ocg}%
  \begin{ocg}{fig:deton}{fig:deton}{1}%
   \includegraphics[width=\hsize]{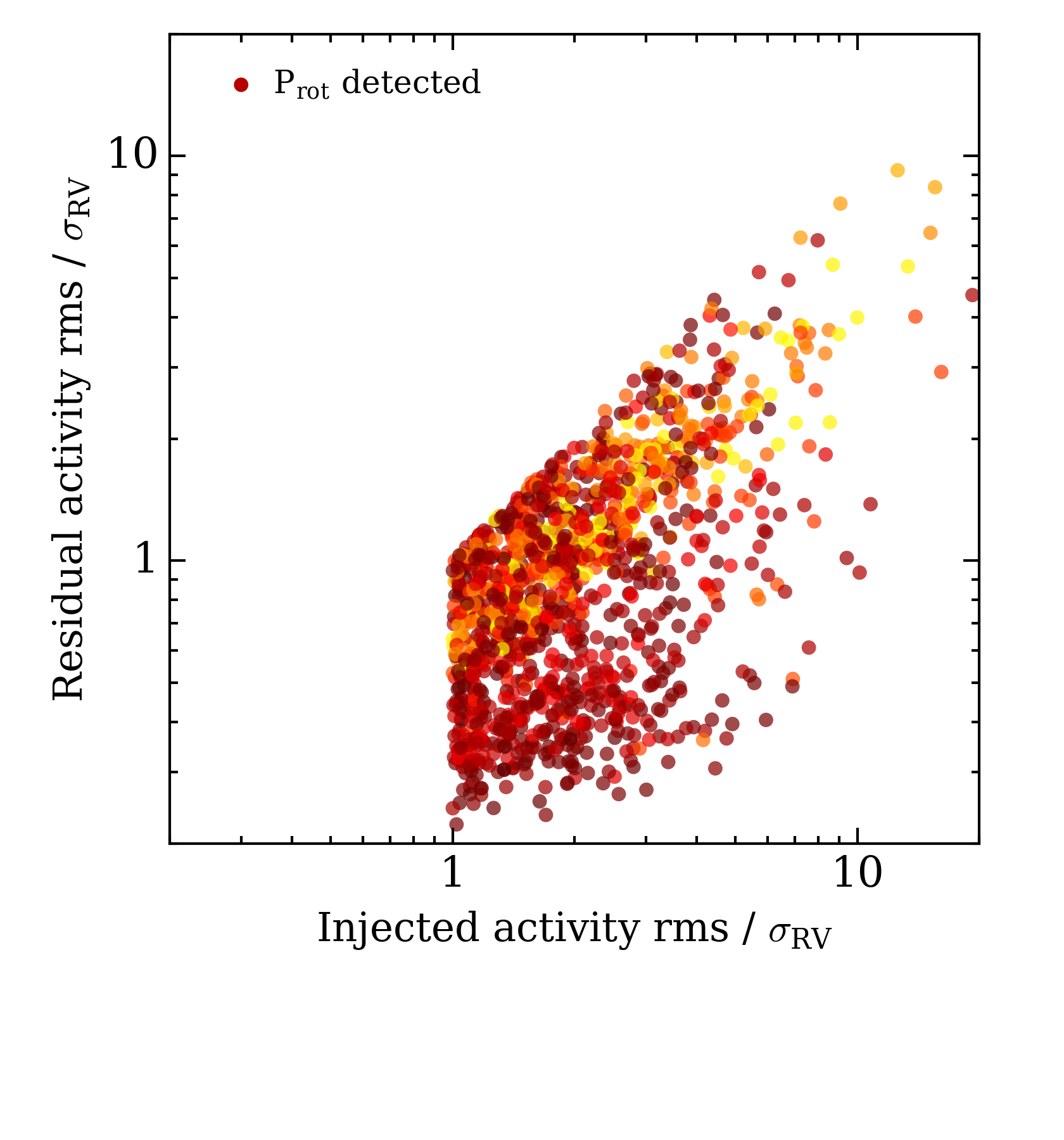}%
  \end{ocg}
  \hspace{-\hsize}%
  \begin{ocg}{fig:nondetoff}{fig:nondetoff}{0}%
  \end{ocg}%
  \begin{ocg}{fig:nondeton}{fig:nondeton}{1}%
   \includegraphics[width=\hsize]{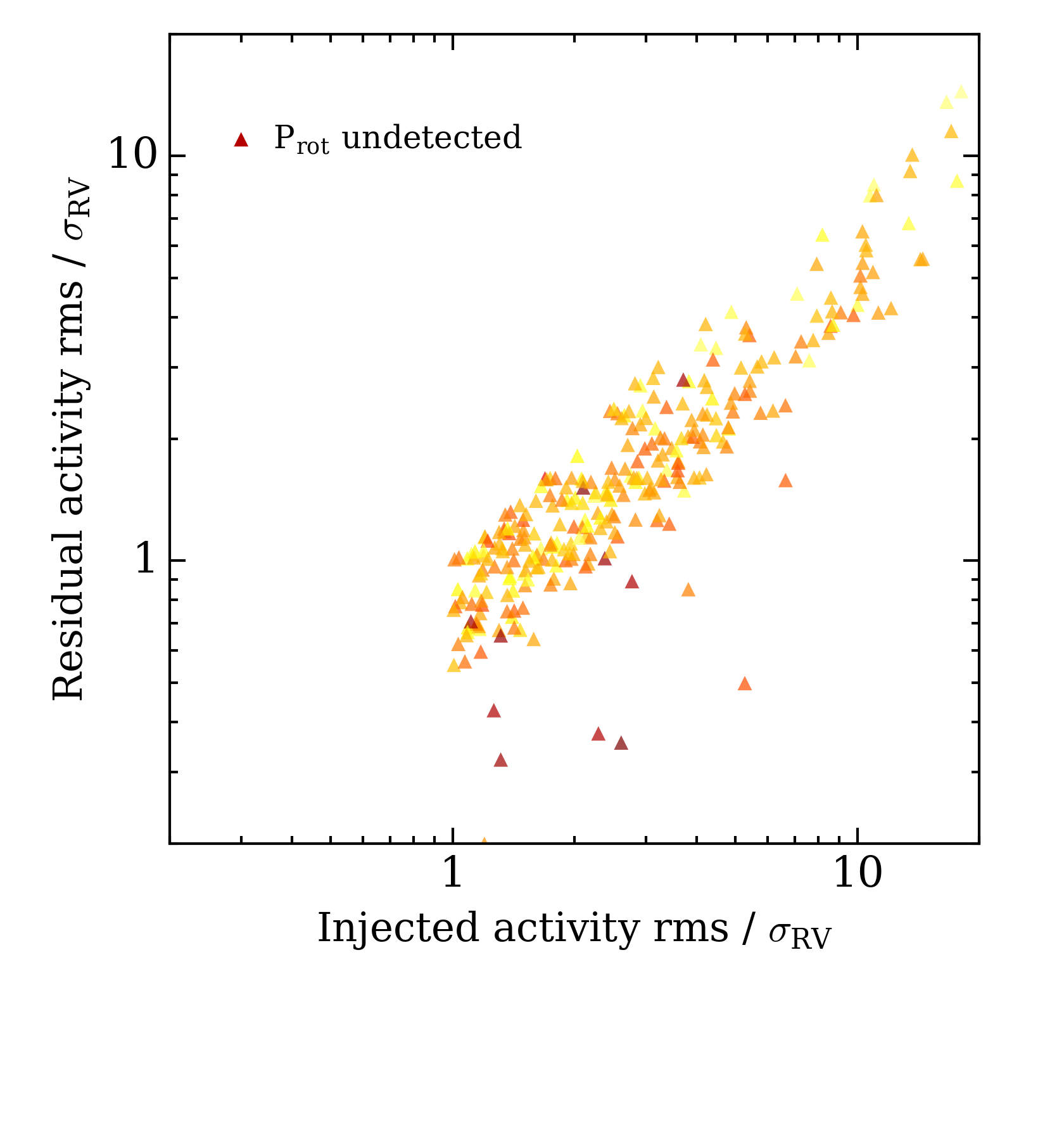}%
  \end{ocg}
  \hspace{-\hsize}%
  \caption{Comparison of the rms of the \emph{injected} RV activity to the rms of the \emph{residual} RV activity following
    the removal of the mean GP activity model which is
    computed in the absence of planets. Each rms value is normalized by the
    RV measurement uncertainty $\sigma_{\text{RV}}$ of its time-series. The \emph{dashed line} $y=x$ is indicative of cases
    wherein the GP activity model does not reduce rms due to activity. Time-series in which \prot{} is detected in the
    FWHM time-series are depicted as 
    \ToggleLayer{fig:deton,fig:detoff}{\protect\cdbox{\emph{circles}}}. Whereas time-series in which \prot{} is not detected in
    the FWHM times-series are depicted as \ToggleLayer{fig:nondeton,fig:nondetoff}{\protect\cdbox{\emph{triangles}}}.}
  \label{fig:compareGPres}
\end{figure}

Fig.~\ref{fig:compareGPres} compares the rms of the \emph{injected} RV activity to the rms of the \emph{residual} RV activity
after removing the mean GP activity model computing assuming a zero planet model. Recall that we only compute a GP activity model
when the injected rms is $> \sigma_{\text{RV}}$ such that the injected activity rms in units of $\sigma_{\text{RV}}$
is always greater than or equal to unity in Fig.~\ref{fig:compareGPres}.
We note that the residual activity rms never exceeds the injected activity rms as expected; i.e. the residual rms
always lies beneath the line $y=x$ in Fig.~\ref{fig:compareGPres}. Secondly,
there appears to be a positive correlation between \prot{} and the relative reduction of the activity rms,
which is analogous to GP performance. That is that within our sample of RV time-series which are modelled with a GP activity
model, the activity rms is maximally reduced in systems
with the longest rotation periods. Conversely, the GP performance when applied to rapid rotators (\prot{} $\gtrsim 2$ days)
is often marginal in comparison. The exact cause of these effects above may be related to the poor time-sampling of our
observations compared to the short stellar rotation period but is ultimately beyond the scope of this paper and is reserved
for a future study (Cloutier et al. in prep).

We note that in many slow rotator cases, the dimensionless residual rms is often less than unity. This suggests that in such
cases the non-parametric GP is actually modelling the noise and not just the stellar activity signal.
Unfortunately when the GP activity model is over-fitting the RV noise, planetary signals can be absorbed into the mean GP
model and thus avoid detection. Furthermore,
this apparent over-fitting is a common feature in many of our simulations and does not appear to depend on
whether or not the stellar rotation period is detected a-priori in the FWHM time-series.
However it is true that on average,
the GP modelling out-performs cases in which \prot{} is known from the FWHM compared to cases in which \prot{} remains
undetected thus providing very weak constraints on $P_{\text{GP}}$.
This highlights the importance having a-priori knowledge of \prot{} from any of the
ancillary spectroscopic times-series, polarimetric time-series \citep{hebrard16},
or from previously obtained long-baseline photometry \citep[e.g.][]{newton16a}.

\section{Automated Planet Detection} \label{sect:detection}
Due to the large number of planetary systems in our simulated SLS-PS, we must detect planets in an automated
way. The steps in our automated planet detection algorithm
represent computationally tractable calculations given the large number of planetary systems
for which each step must be performed. We note however that other---potentially more robust---planet
detection algorithms may be adopted in the real SLS-PS which are likely to include more
human intervention than the automated techniques described in the following subsections.

\subsection{Establishing Putative Planetary Detections} \label{sect:det}
\begin{figure}
  \centering
  \includegraphics[width=\hsize]{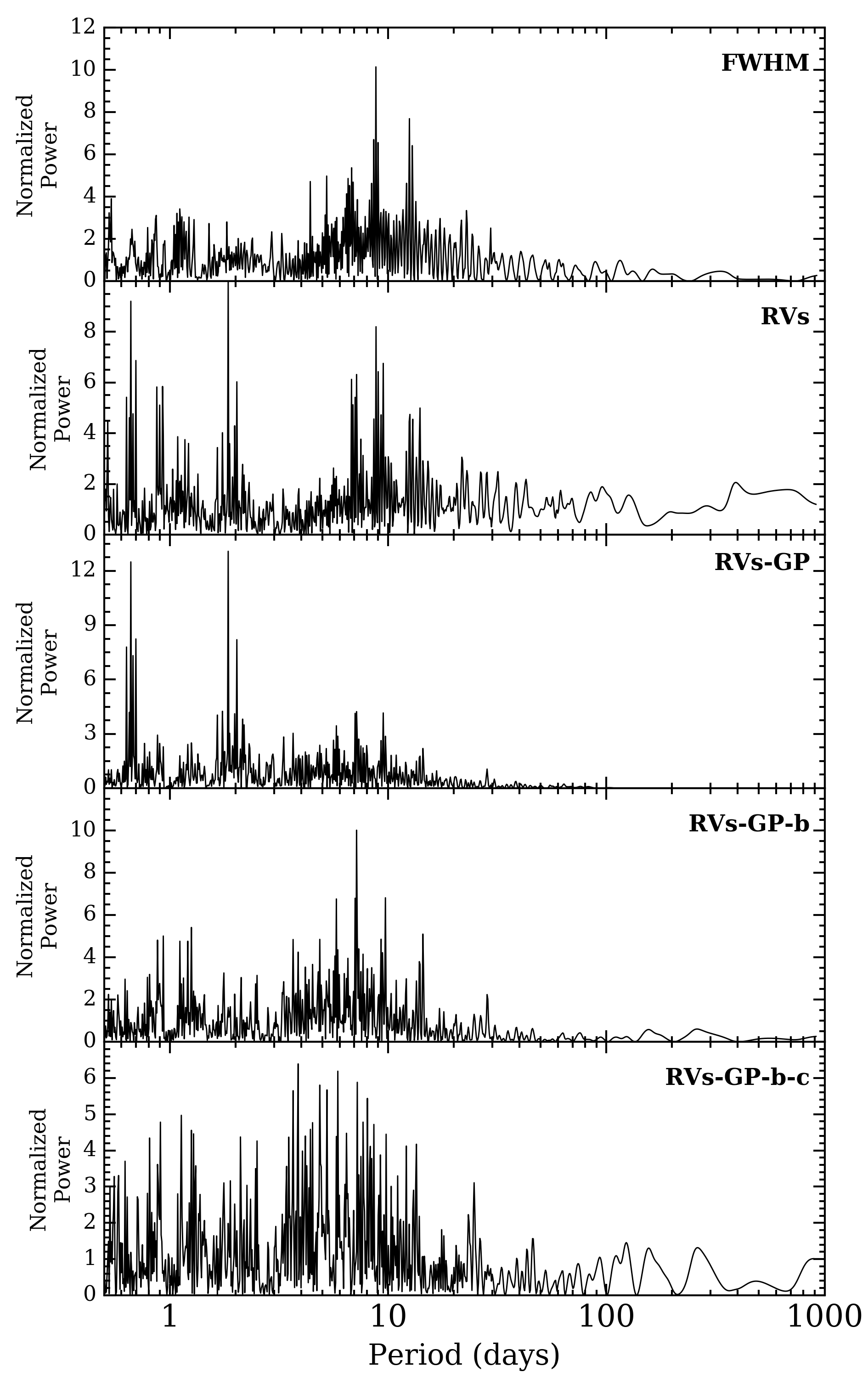}%
  \hspace{-\hsize}%
  \begin{ocg}{fig:Poff}{fig:Poff}{0}%
  \end{ocg}%
  \begin{ocg}{fig:Pon}{fig:Pon}{1}%
    \includegraphics[width=\hsize]{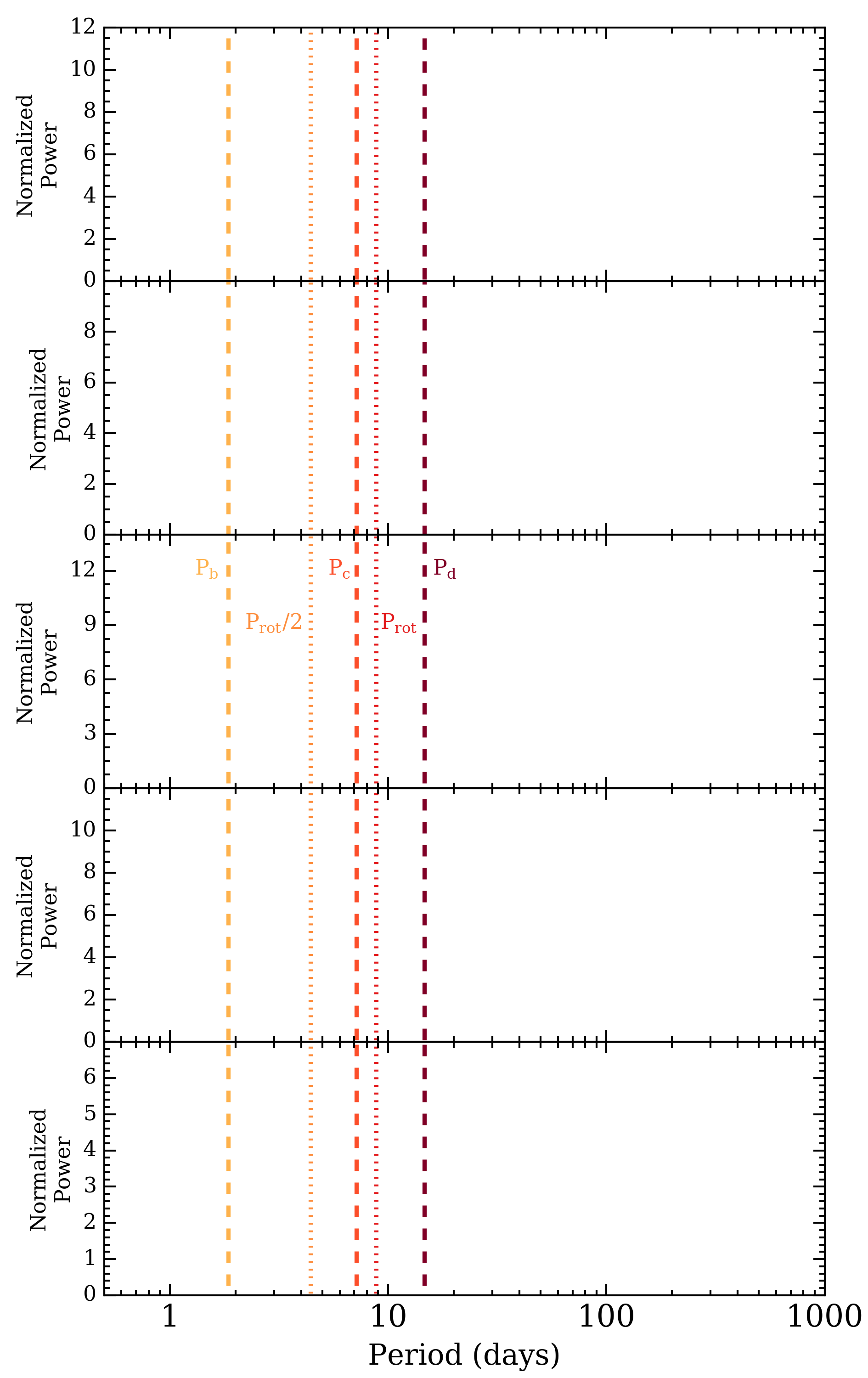}%
  \end{ocg}
  \hspace{-\hsize}%
  \begin{ocg}{fig:FAPoff}{fig:FAPoff}{0}%
  \end{ocg}%
  \begin{ocg}{fig:FAPon}{fig:FAPon}{1}%
    \includegraphics[width=\hsize]{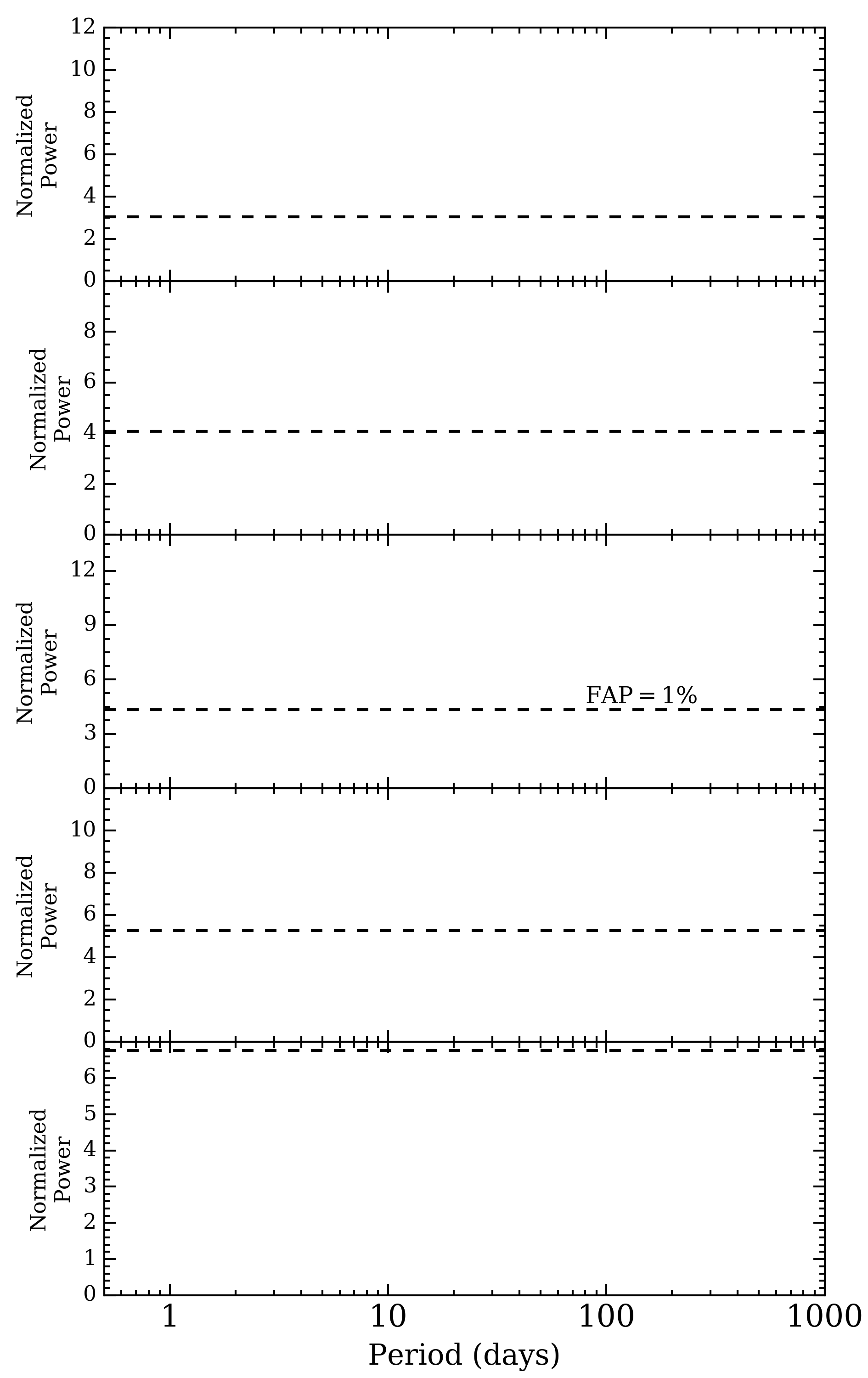}%
  \end{ocg}
  \hspace{-\hsize}%
  \caption{\emph{Top to bottom}: Lomb-Scargle periodograms of the full width at half maximum (FWHM)
    of the cross-correlation function, raw radial velocities (RVs), RVs corrected for activity (RVs-GP),
    RVs corrected for activity and a planet (RVs-GP-b), and the RVs corrected for activity and two planets
    (RVs-GP-b-c). Periodicities equal to the stellar rotation period, its first harmonic, and the
    orbital periods of the three planets in the system are highlighted with
    \ToggleLayer{fig:Pon,fig:Poff}{\protect\cdbox{\emph{vertical lines}}}. The power
    corresponding to a 1\% false alarm probability in each periodogram is highlighted by a  
    \ToggleLayer{fig:FAPon,fig:FAPoff}{\protect\cdbox{\emph{horizontal dashed line}}} in each panel.}
  \label{fig:periodograms}
\end{figure}

We proceed by searching for planetary periodicities in an iterative manner using the LS periodograms
of the RVs following the removal of various periodic signals.
An example of this iterative process is visualized in Fig.~\ref{fig:periodograms} for a 3 planet system
($P_b,P_c,P_d=1.85,7.17,14.66$ days, $K_b,K_c,K_d=4.7,2.3,0.07$ \mps{} respectively)
around a moderately active star with a measured photometric rotation period \prot{} $\sim 8.8$ days and
an RV rms in the absence of planets of $\sim 4.2$ \mps{.}

For MC realizations in which the rms of the injected RV activity exceeds the median RV measurement
uncertainty---as is the case for the system shown in Fig.~\ref{fig:periodograms}---we compute two
versions of our initial periodogram: one of the raw RVs \emph{only} and a second
of the RV residuals after fitting the data with the trained GP activity model and zero mean function
(i.e. no planet model). This GP model predicts the RV activity in the absence of planetary signals.
For the remaining MC realizations containing quiet stars
we only compute the LS periodogram of the raw RVs thus neglecting any
modelling of correlated RV residuals. For cases in which we use a trained GP to model RV activity,
it is beneficial for
the periodic term of the assumed quasi-periodic covariance kernel to be constrained by our training set.
For the example shown in Fig.~\ref{fig:periodograms}, the top panel shows the LS periodogram of the FWHM
in which \prot{} is detected and is subsequently used in our RV modelling (see Sect.~\ref{sect:regression}).

The second panel in Fig.~\ref{fig:periodograms} depicts the LS periodogram of the raw RVs only whereas
the third panel depicts the raw RVs corrected by the trained GP activity model with
$P_{\text{GP}}=P_{\text{rot}}=8.8$ days. Comparing these two periodograms it is apparent
that some power at \prot{} is diminished when removing the mean GP activity model
along with power at long periods due to the large exponential
timescale found during training. More importantly, the strongest periodic signal in each periodogram is
at the orbital period of the innermost planet at $\sim 1.85$ days suggesting the presence of a
planet. In these periodograms, and in all subsequent iterations, we claim a putative planet detection
for periodicities with i) FAP $\leq 1$\% ii) is within 2\% of an injected planet's orbital period iii)
is not associated with significant
periodic signals seen in the FWHM time-series iv) is not an alias of the time-sampling (see example in
Fig.~\ref{fig:wfs})
and v) is $>2$\% away from the stellar rotation period and any of its first four harmonics. The signals
at $\sim 1.85$ days in the second and third panels of Fig.~\ref{fig:periodograms} obey these criteria and
therefore constitute a putative planet detection. We proceed by referring to this putatively
detected planet as `b'.

We note that the second condition for a putative planet detection above cannot be utilized in a real survey
because any planetary periodicities are not known a-priori. Instead we invoke this condition to
accelerate our automated planet detection algorithm relative to the steps that must be taken in a real survey
to secure planet detections \citep[e.g. Bayesian model comparison;][]{ford07}.
Without a-priori knowledge of planet orbital periods, LS periodicities at high
significance that obey all of the remaining aforementioned putative planet criteria, may represent false positive
signals if not carefully modelled.
In our simulations we find that on average our time-series generate $\sim 0.5$ such false positives in their LS
periodograms following the removal of planet models. Although
a fraction of these false positive periodicities are likely to be aliases of each other making the above
estimate an upper limit. Ideally the determination of these signals as false positives or as true planetary
signals would be solved via a formal model comparison in a subsequent analysis. However, such calculations are
not guaranteed to converge to the correct solution \citep[see][]{dumusque17}.

The second periodogram iteration requires that we fit the putative planetary signal and search for
additional signals in the LS periodogram of the residuals. To model planet `b'
we adopt the maximum-likelihood keplerian model parameters at the detected periodicity
and re-compute the GP activity model which is modified due to the new mean function (i.e. a one planet model
rather than the previously assumed no planet model). Note that the GP activity model is re-computed after each
iteration due to the changing mean function. The fourth panel in Fig.~\ref{fig:periodograms} shows
the resulting periodogram after correcting the raw RVs with the new GP activity model and a keplerian solution
for planet `b'. The strongest residual
periodic signal is at the orbital period of the middle planet at $\sim 7.17$ days which
we claim as a second putative planet detection because the periodicity obeys the aforementioned criteria.
This second putative planet is referred to as `c'.

The bottom panel of Fig.~\ref{fig:periodograms} depicts the LS periodogram of the RVs after being
corrected for activity and the superposition of the two maximum-likelihood planet models. Now the strongest
residual periodicity has a FAP $>1$\% implying that our automated planet detection algorithm has ceased to
detect planetary signals. Therefore in the example shown in Fig.~\ref{fig:periodograms}, the
third injected planet at $\sim 14.66$ days remains undetected. An unsurprising result given the small RV
semi-amplitude of the planet ($K_d=7$ cm s$^{-1}$) compared to the time-series' median RV measurement
uncertainty $\sigma_{\text{RV}}=1.8$ \mps{.} In each MC realization we perform this iterative procedure
until no putative planets are detected and up to
a maximum of three planets despite many planetary systems having $>3$ injected planets
(see Fig.~\ref{fig:mult}). In this way we are at least
sensitive to the expected number of planets per M dwarf \citep[$2.5 \pm 0.2$;][]{dressing15a} 
and limit the computational expense of detecting planets dominated by repeatedly computing GP activity
models and LS periodogram FAPs.

\subsection{Model Selection} \label{sect:CV}
In Sect.~\ref{sect:det} we established putative planet detections based on low FAP LS periodogram
periodicities. However the robust detection of a planet with a particular set of keplerian model parameters
must be favoured over competing models that lack such a planet. The proper diagnostic for model
selection is the ratio of Bayesian model evidences which are
notoriously difficult and time-consuming to calculate \citep{ford07}. 
As an alternative model selection technique we turn to time-series
cross-validation (CV). This technique is a specialized version of general K-fold CV
and is suitable to data featuring strong correlations in time as is the case with RV time-series
\citep{arlot10}.

For MC realizations featuring at least one putative planetary detection
we perform time-series CV 
on models that contain an increasing number of planets, including the null hypothesis i.e. no planets.
The latter model has zero keplerian parameters whereas a model containing $N_p \ge 1$ planets
contains $3N_p$ model parameters where the three parameters per planet are its orbital period,
time of inferior conjunction, and RV semi-amplitude. For the purpose of model selection we
will assume circular orbits for all planets to limit the size of the parameter space.

The CV algorithm proceeds by first splitting the RV time-series $y_1,\dots,y_{n_{\text{obs}}}$ into training and
testing sets. For some $t>1$, each competing model is fit to the training set
$y_1,\dots,y_t$ using a Levenberg-Marquardt optimization routine. The optimized model is then evaluated
at the next epoch $t+1$ ($\mu_{t+1}$) and the lnlikelihood of the testing set $y_{t+1}$ given the optimized
model is computed
using Eq.~\ref{eq:like}. When computing the lnlikelihood, we adopt a white covariance matrix for systems
wherein the GP analysis is not used but otherwise assume the MAP GP hyperparameters from the
iterative procedure in Sect.~\ref{sect:det}. These steps are repeated for
$t=N_{\text{min}},\dots,N-1$ where the minimum size of the training set $N_{\text{min}}$ is set to 20. 
The favoured model is determined by which of the competing models has a largest median lnlikelihood per
measurement among the $N-N_{\text{min}}$ CV iterations. In cases wherein two models are consistent
within their median absolute deviations, the model containing less planets is accepted as an imposition of
Occam's razor.

\subsection{Vetting of Planet Detections} \label{sect:vett}
A consequence of our automated planet detection methodology is
various non-deterministic effects which can result in planet detections that are highly unlikely to be
favored by model comparison in the real SLS-PS, yet are marginally detected in our simulations.
Such planets are commonly those whose RV semi-amplitude is
close to the rms of the RV time-series. These planets would likely be rejected by any human
vetting which we do not conduct in our simulations. We therefore undergo a vetting procedure in an
attempt to restrict the detected planet population to be maximally realistic. Our adopted vetting procedure
is based on the methods of \cite{cumming08} from the Keck Planet Search. For vetting we define 
the condition that a bona fide planet detection must satisfy $K/\sigma_{\text{K}} \ge 3$. \emph{That is
  that a true planet detection is one in which the planet's semi-amplitude $K$ is detected with a minimum
  expected detection significance of $3\sigma$}.

To estimate the expected uncertainty in the RV semi-amplitude $\sigma_{\text{K}}$ we compute the
\emph{Fisher information matrix} $B$ which quantifies the information content in an observable time-series
$\mathbf{y}(\mathbf{t})$ regarding unknown model parameters $\boldsymbol{\theta}$. The model parameter
covariance matrix is related to the Fisher information via $C=B^{-1}$. We can therefore
use the Fisher information matrix to analytically predict the measurement uncertainty of the RV model
parameters of interest given an input time-series with \nobs{} measurements $\mathbf{y}=(y_1,\dots,y_{n_{\text{obs}}})$
obtained at the epochs $\mathbf{t}=(t_1,\dots,t_N)$ and with measurement uncertainties
$\boldsymbol{\sigma}=(\sigma_1,\dots,\sigma_N)$.

The Fisher information matrix is a Hessian matrix of the lnlikelihood of a single keplerian model 
with respect to its model parameters $\boldsymbol{\theta}=\{P,T_0,K\}$:

\begin{equation}
  B_{i,j} = -\frac{\partial^2 \ln{\mathcal{L}}}{\partial \theta_i \partial \theta_j}.
\end{equation}

\noindent The Fisher information matrix is symmetric and is $3 \times 3$ in our case as
we assume circular orbits.

In order to simplify the calculation of 
$B$ we consider planets individually and account for the residual RV signal from additional planets
through an ``effective'' RV uncertainty $\sigma_{\text{eff}}$
in place of the RV measurement uncertainties $\sigma_i$
in Eq.~\ref{eq:K}. The effective RV uncertainty is the rms of the RVs
after removal of the keplerian signal from the planet being considered. It therefore contains
contributions from any additional planets, stellar activity, and systematic errors.
Because we do not fit for each planet's orbital eccentricity the
keplerian model simplifies to $\mu(t_k) = -K \sin{\phi_k}$ where $\phi_k = \frac{2\pi}{P} (t_k-T_0)$.
Using this mean model in the lnlikelihood (Eq.~\ref{eq:like}) along with $K_{ij}$ approximated by a
white covariance matrix $K_{ij} = \sigma_{\text{eff}} \delta_{ij}$, we can compute each element of $B$ analytically
(see Appendix~\ref{app:fisher}). The Fisher information matrix
is then inverted to obtain the covariance matrix of the model
parameters $C$. The measurement uncertainty of the semi-amplitude is then $\sigma_{\text{K}} = \sqrt{C_{K,K}}$.

\subsection{Summary of the Automated Planet Detection Algorithm}
To recapitulate our process of claiming planet detections in our simulated SLS-PS we
recall the three steps discussed throughout this section. Firstly we search for putative
planetary signals in the LS periodogram of either the raw RVs or the RVs corrected for activity using a
zero-mean GP activity model. We are careful to ensure that significant periodicities have a
planetary origin and are not associated with stellar activity signals or our time sampling. Secondly we
compute model lnlikelihoods using time-series cross-validation and compare models with and without the
putative planet. We only retain planets which are favored by this model selection technique. The third and
final step consists of vetting our planet detections by insisting that they must have an RV semi-amplitude
detection significance greater than $3\sigma$ where the detection significance is estimated from the
planet's known semi-amplitude $K$ and an analytical estimate of the measurement uncertainty on $K$ from
the Fisher information. Planet detections which pass our vetting procedure are treated as bona fide
detections.

\begin{figure}
  \centering
  \includegraphics[width=\hsize]{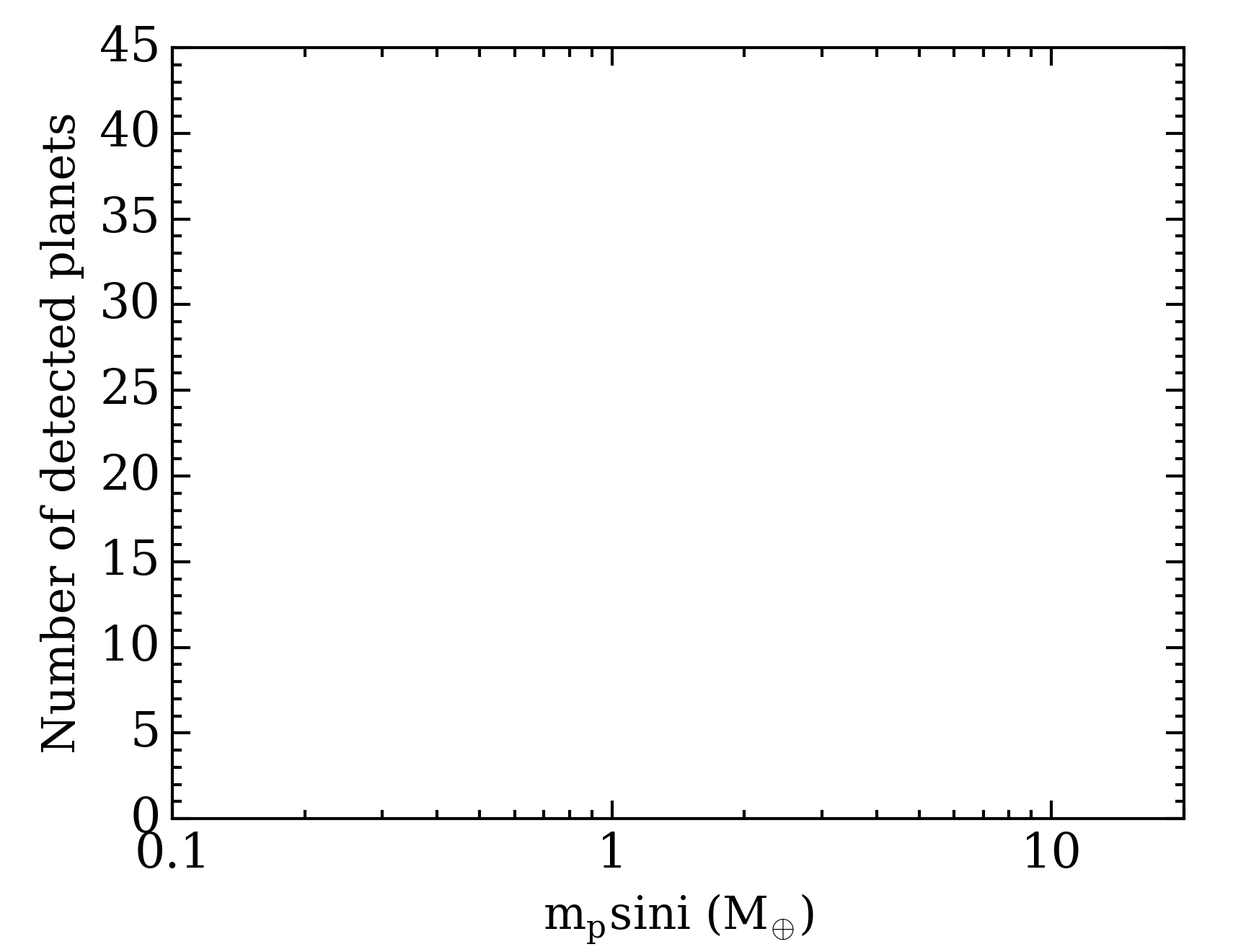}%
  \hspace{-\hsize}%
  \begin{ocg}{fig:1off}{fig:1off}{0}%
  \end{ocg}%
  \begin{ocg}{fig:1on}{fig:1on}{1}%
    \includegraphics[width=\hsize]{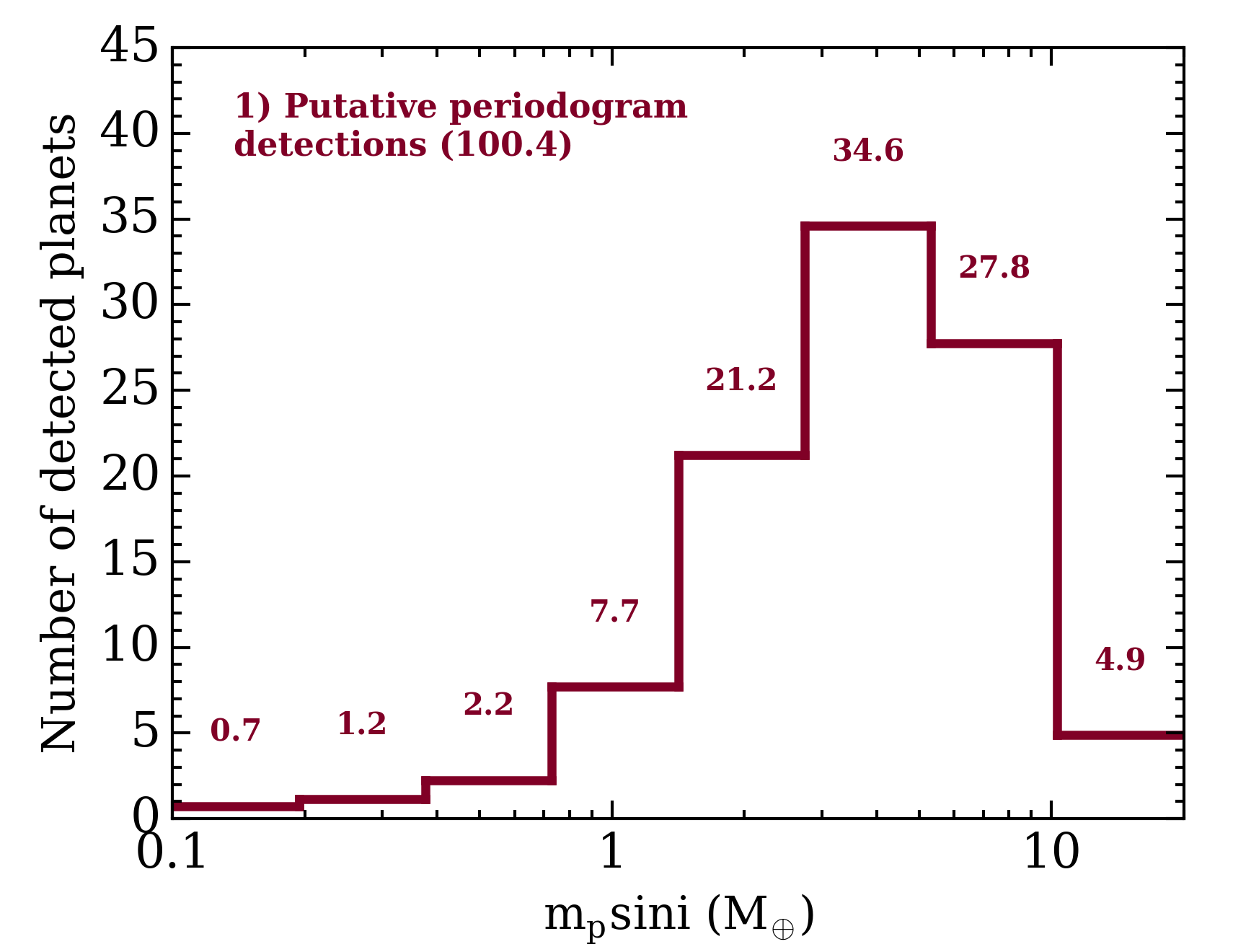}%
  \end{ocg}
  \hspace{-\hsize}%
  \begin{ocg}{fig:2off}{fig:2off}{0}%
  \end{ocg}%
  \begin{ocg}{fig:2on}{fig:2on}{1}%
    \includegraphics[width=\hsize]{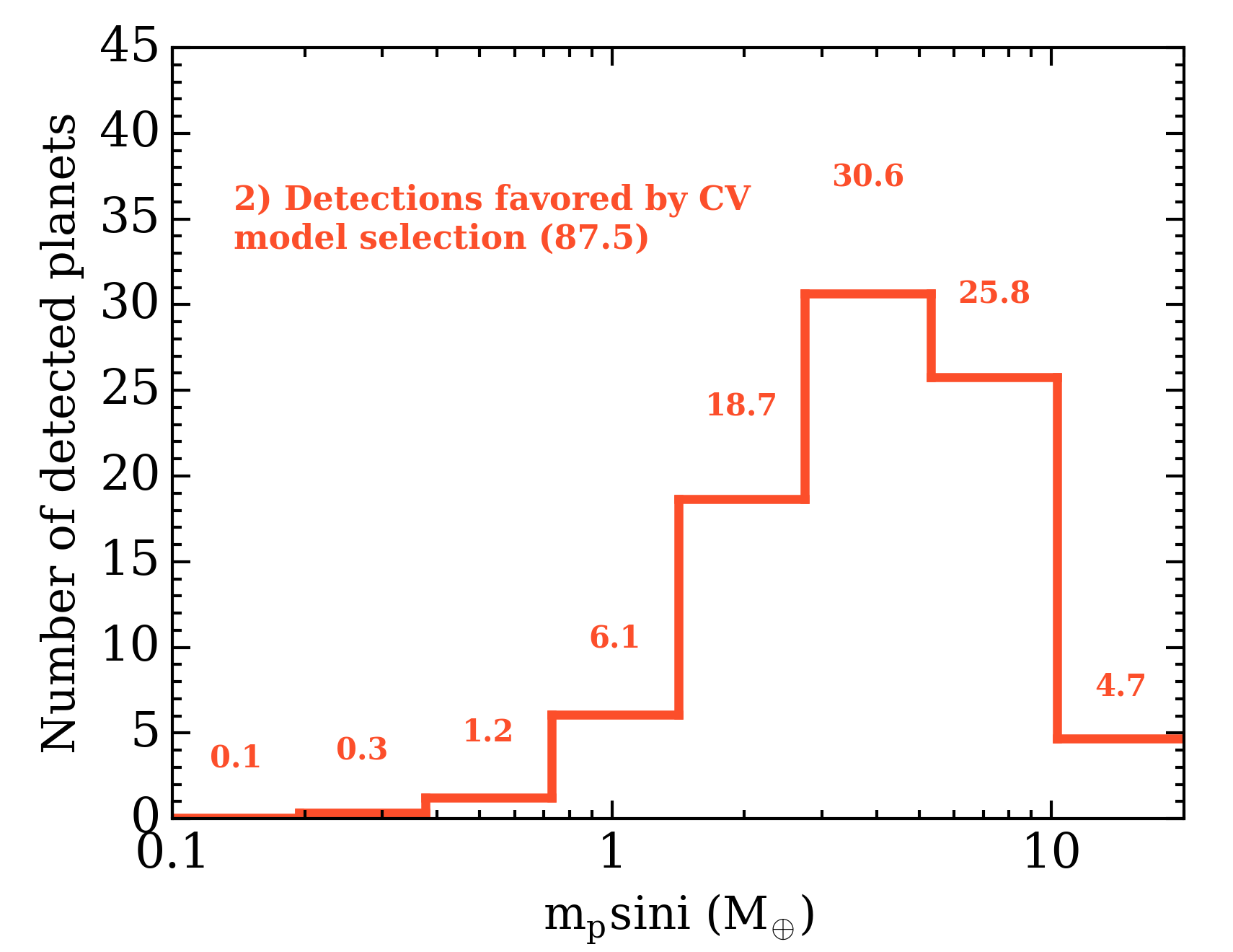}%
  \end{ocg}
  \hspace{-\hsize}%
  \begin{ocg}{fig:3off}{fig:3off}{0}%
  \end{ocg}%
  \begin{ocg}{fig:3on}{fig:3on}{1}%
    \includegraphics[width=\hsize]{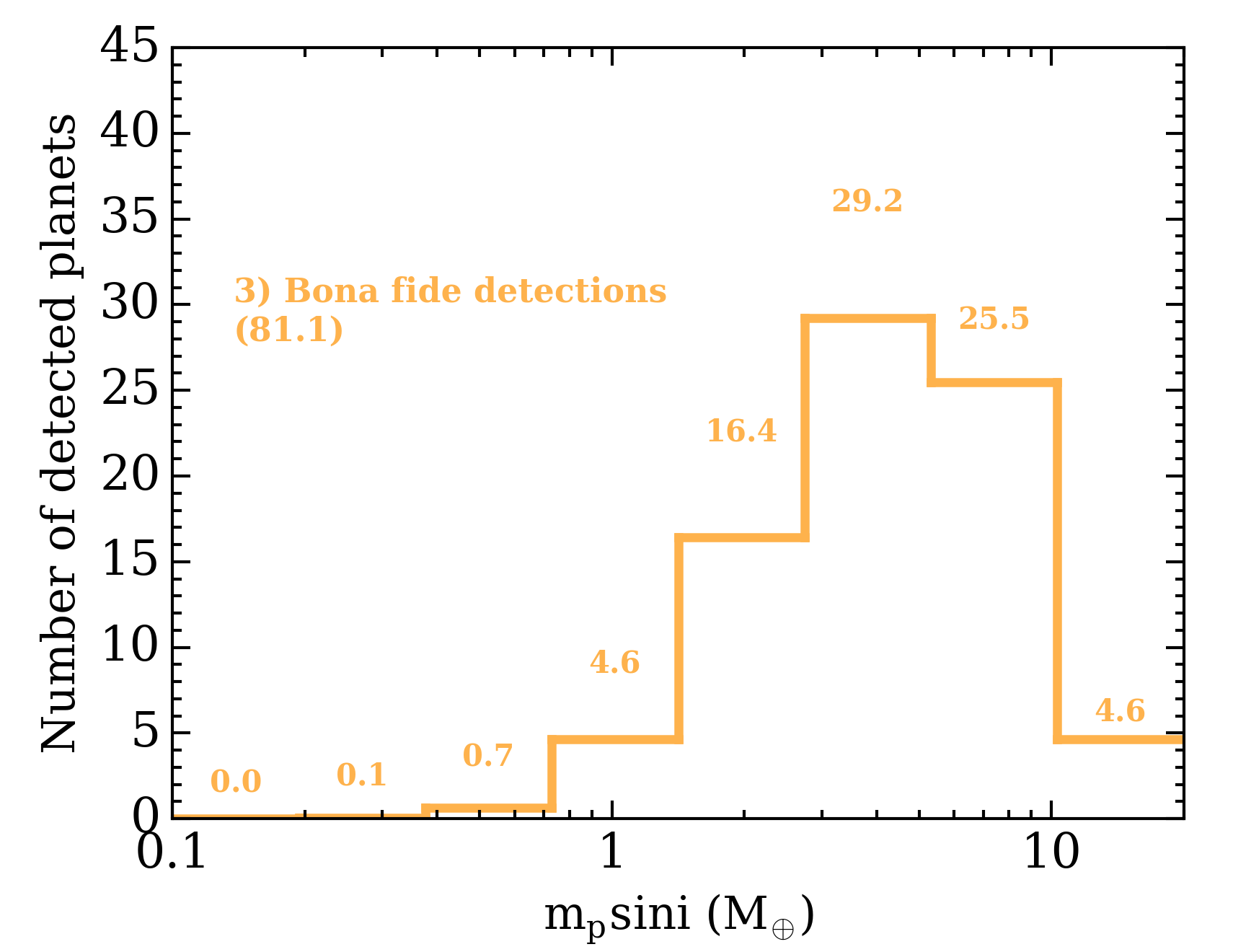}%
  \end{ocg}
  \hspace{-\hsize}%
  \caption{The number of planets detected as a function of \msini{} after each of the three steps in
    our automated planet detection algorithm. The number of detected planets in each
    \msini{} bin is annotated above the bin to help quantify the decrease in planet detections
    following each step. The distribution of detected planet minimum masses
    following each step can be viewed independently for clarity
    (\ToggleLayer{fig:1off,fig:1on}{\protect\cdbox{1}},
    \ToggleLayer{fig:2off,fig:2on}{\protect\cdbox{2}},
    \ToggleLayer{fig:3off,fig:3on}{\protect\cdbox{3}}).}
    \label{fig:detections}
\end{figure}

The distributions of detected planet minimum masses after each step in our automated planet detection
algorithm are shown in Fig.~\ref{fig:detections}. Here we only include planet detections with
$0.1 \leq m_p\sin{i}/\text{M}_{\oplus} \leq 20$. In this SLS-PS there are a small number of giant planets
with \msini{} $> 20$ M$_{\oplus}$ detected thus resulting in an underestimated total planet yield annotated
in Fig.~\ref{fig:detections} (see Sect.~\ref{sect:yield} for a full description of the detected planet
population).

It is clear from Fig.~\ref{fig:detections} that each step in our automated planet detection algorithm
reduces the number of detected planets somewhat.
In each \msini{} bin other than the most massive bin ($9.4 \leq m_p\sin{i}/\text{M}_{\oplus} \leq 20$),
our CV model selection technique rejects between $\sim 1-3$ planets or
$\sim 13$\% of all putative planet detections in the range of minimum planet masses considered. Similarly
our vetting procedure reduces the number of detected planets in each \msini{} bin by $\sim 1-2$ planets
for intermediate minimum masses; $0.7 \lesssim m_p\sin{i}/\text{M}_{\oplus} \lesssim 6$.
Our vetting procedure therefore does not reject a significant number of detected
planets at the lowest masses (\msini{} $\lesssim 0.7$ M$_{\oplus}$) nor at the highest
(\msini{} $\gtrsim 6$ M$_{\oplus}$). The former being the result of the small number of putative low mass planets
detected and the latter being due to the large RV semi-amplitude of the most massive planets thus resulting
in a typically large detection significance. Vetting rejects $\sim 7$\% of planet detections favored by
CV. Therefore $\sim 80$\% of putative planet detections from the periodogram analysis materialize
into bona fide planet detections.

\section{SLS-PS Sensitivity} \label{sect:sensitivity}
The detection sensitivity is defined as the recovery fraction of injected planets in our simulated SLS-PS.
Because we have a-priori knowledge of the injected planet population we can compute the detection sensitivity
for each star in our sample by simply dividing the number of detected planets by the number of injected
planets over any desired range of planet properties. We perform this calculation over the discretized
parameter space in $P$, \msini{} and in $S$, \msini{.}
Here $S=L_s / 4\pi a^2$ is the insolation received by the planet where $L_s$ is the stellar
luminosity\footnote{The stellar luminosity is computed from the evolutionary models of \cite{baraffe98} 
based on the stellar mass on the main sequence at 2 Gyrs.} and $a$ is the planet's semi-major axis. We 
focus on the
following ranges of parameter values which encompass the vast majority of the injected planet population:
$P \in [0.5,200]$ days, $S \in [0.01,100]$ S$_{\oplus}$, and \msini{} $\in [0.4,15]$ M$_{\oplus}$. 

We note that in this
study the recovery fraction is uniquely determined by the performance of our automated planet detection  
algorithm (see Sect.~\ref{sect:detection}). Conversely, the actual SLS-PS will have a much higher degree of
human intervention on the data analysis effort. This is afforded by the relatively small size of the RV
datasets compared to large surveys (e.g. Kepler and TESS)
which benefit greatly from automated detection algorithms. Therefore the detection sensitivity in the actual
SLS-PS may not correspond exactly to what is presented here although the automated algorithm used
in this study is designed to closely mimic the analysis that will be conducted on the actual SLS-PS data.

\begin{figure*}
  \centering
  \includegraphics[width=0.5\hsize]{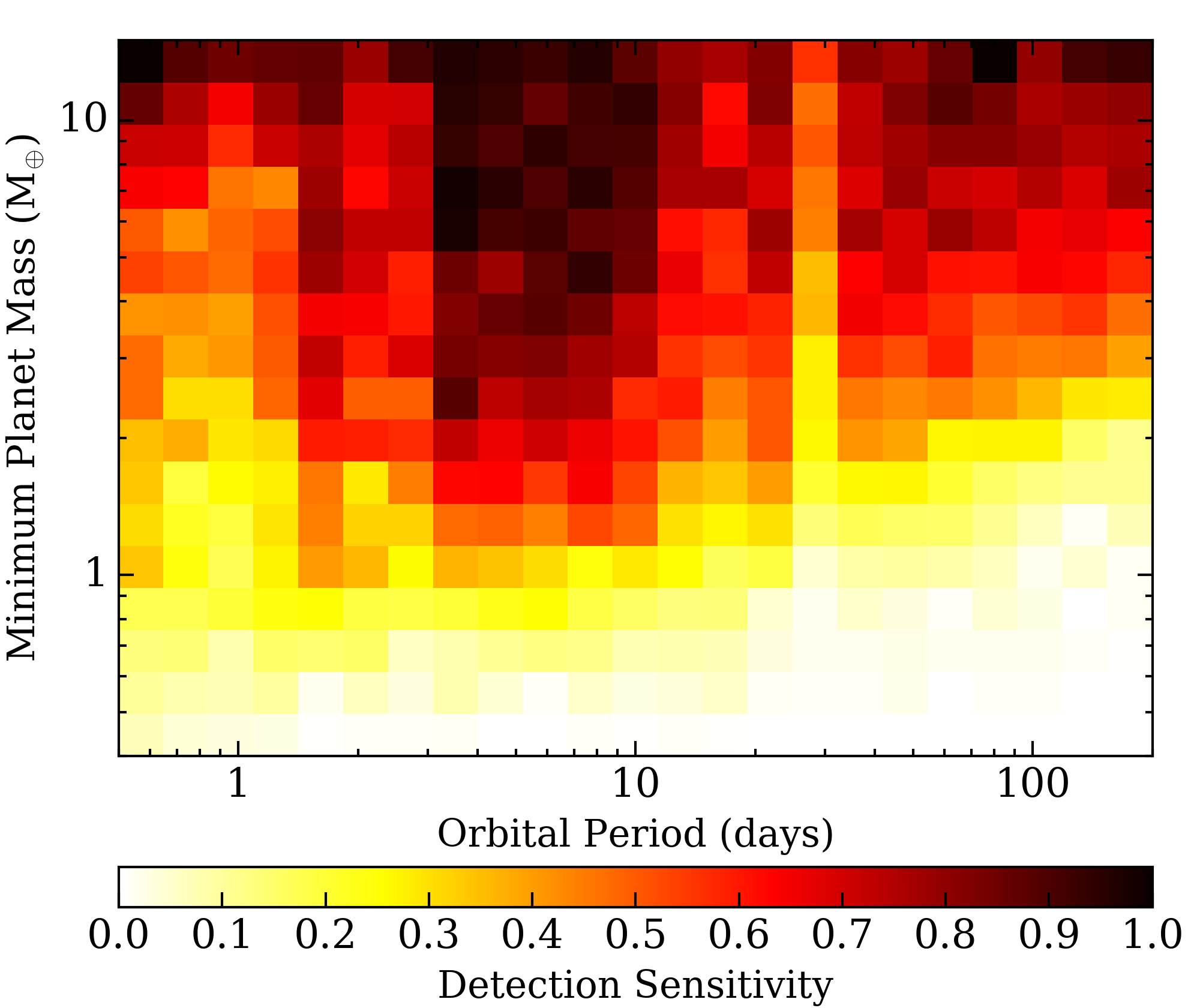}%
  \includegraphics[width=0.5\hsize]{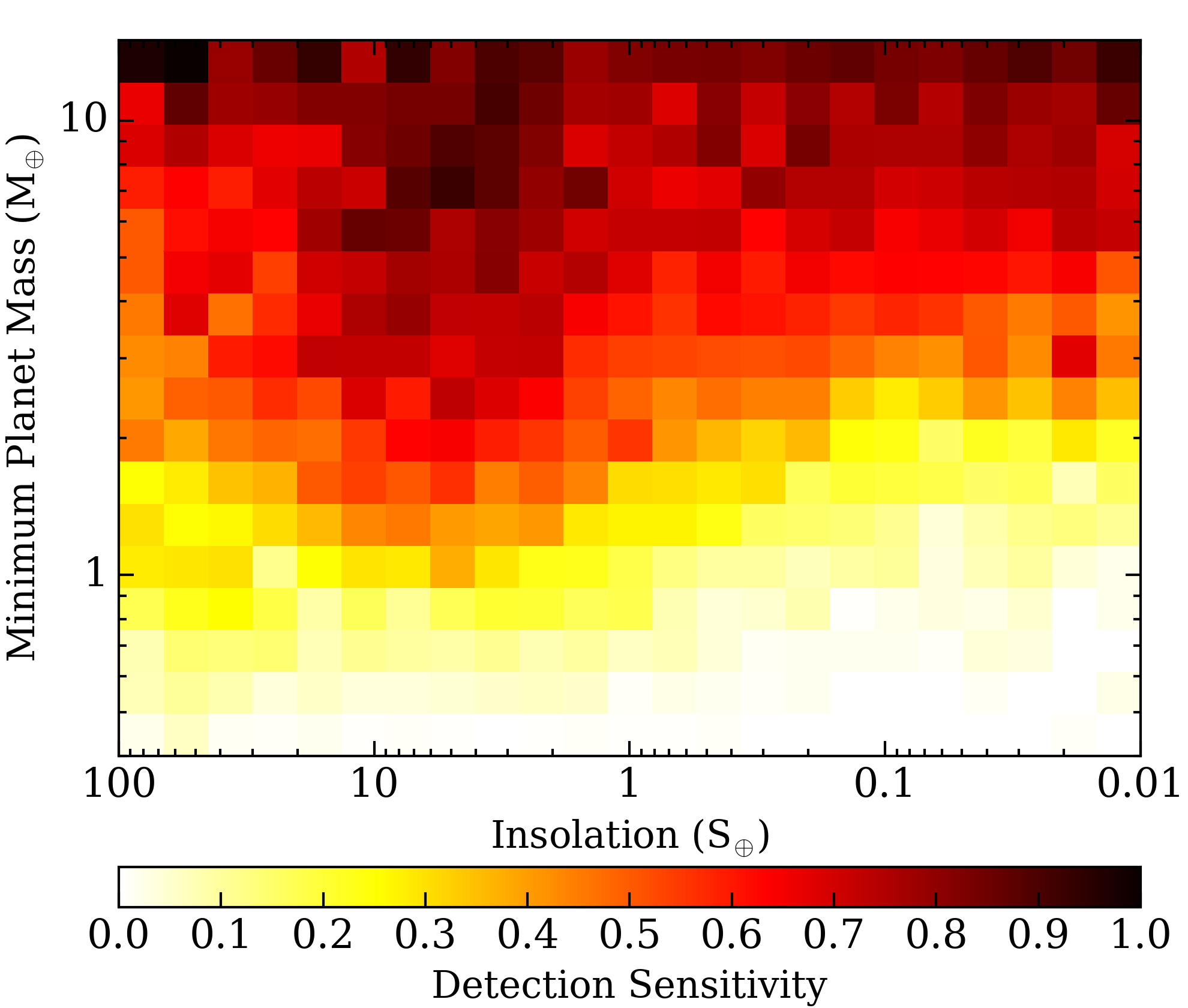}%
  \hspace{-\hsize}%
  \begin{ocg}{fig:shadeoff}{fig:shadeoff}{0}%
  \end{ocg}%
  \begin{ocg}{fig:shadeon}{fig:shadeon}{1}%
    \includegraphics[width=0.5\hsize]{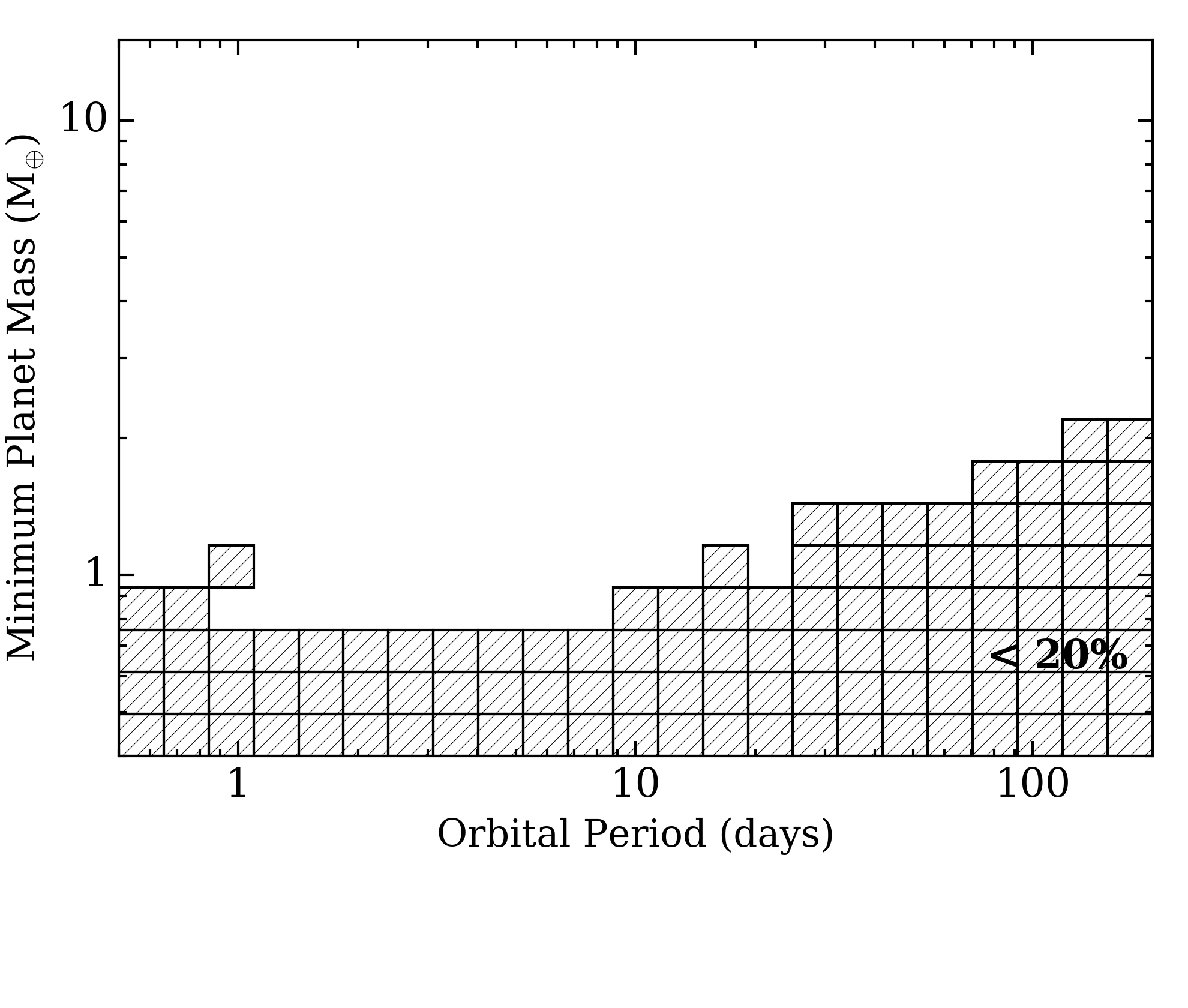}%
    \includegraphics[width=0.5\hsize]{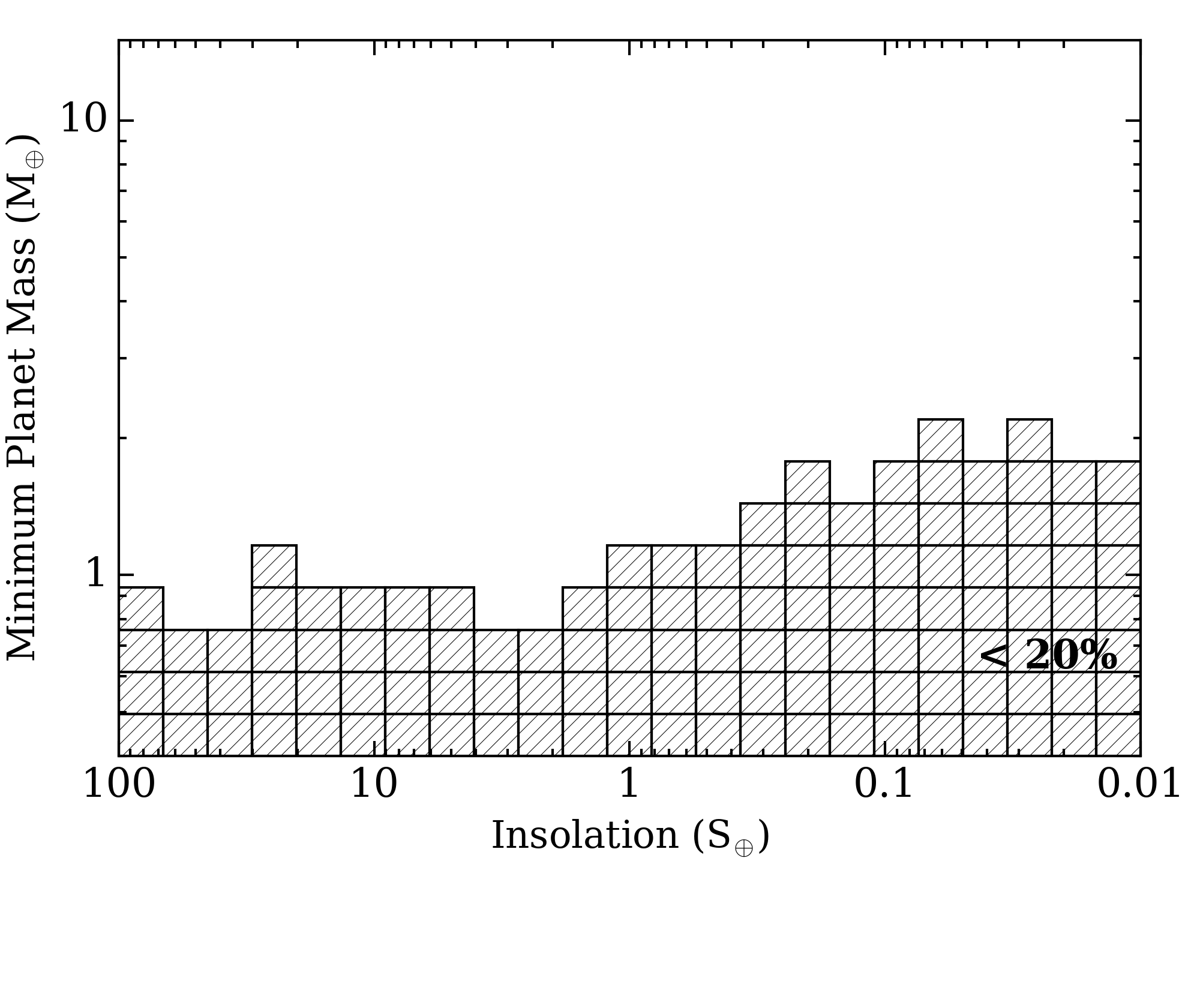}%
  \end{ocg}
  \hspace{-\hsize}%
  \begin{ocg}{fig:Koff}{fig:Koff}{0}%
  \end{ocg}%
  \begin{ocg}{fig:Kon}{fig:Kon}{1}%
    \includegraphics[width=0.5\hsize]{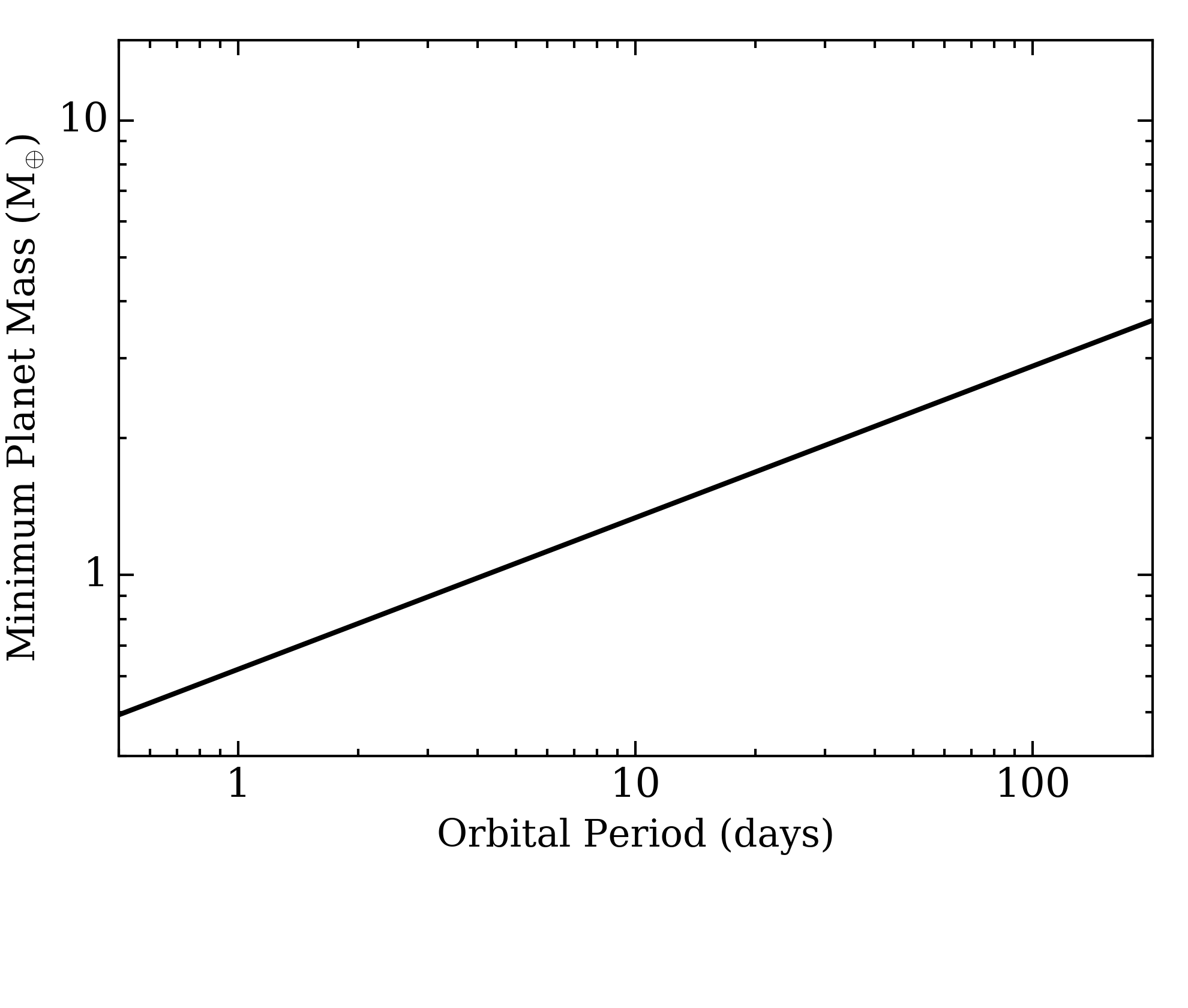}%
    \includegraphics[width=0.5\hsize]{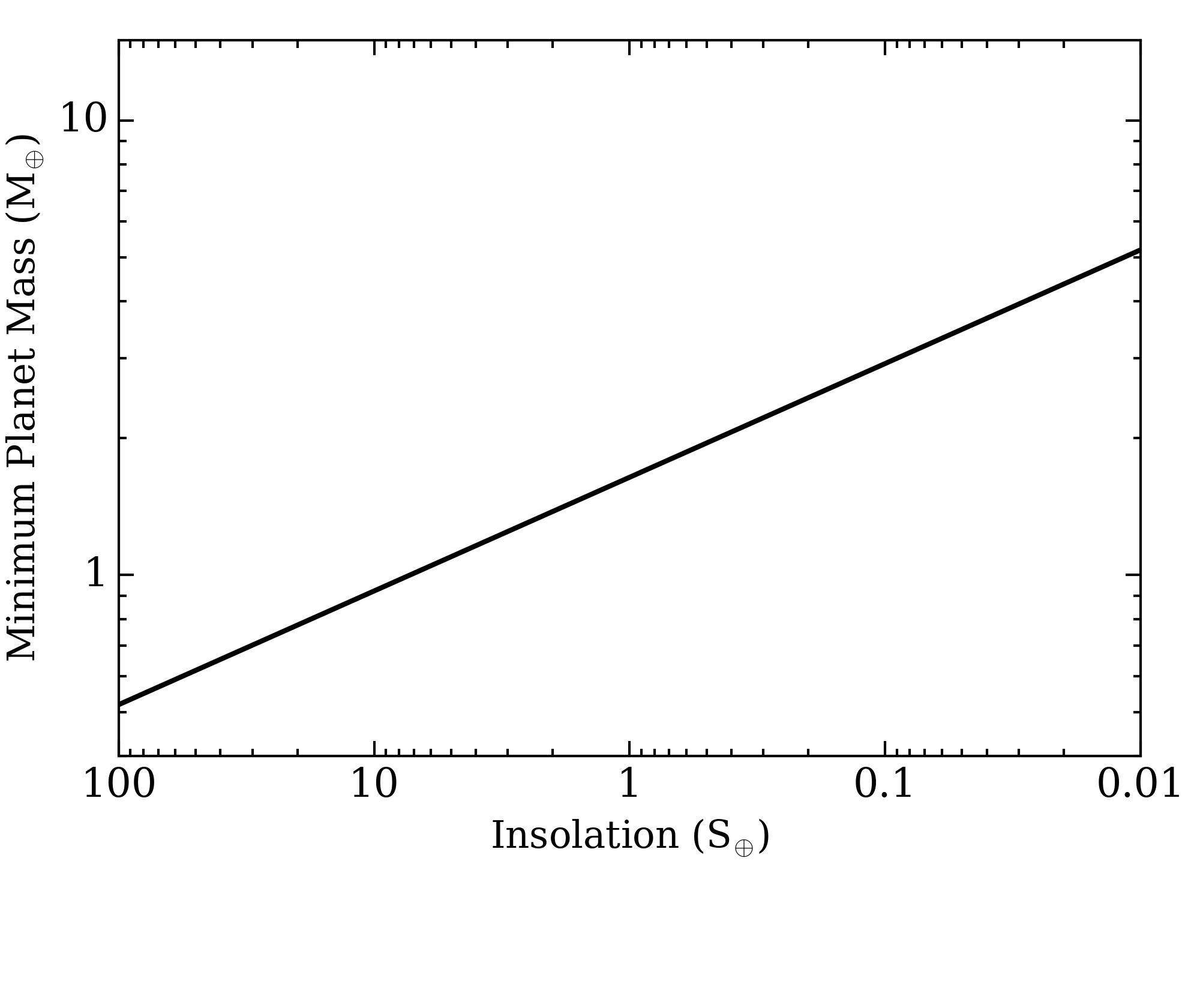}%
  \end{ocg}
  \hspace{-\hsize}%
  \begin{ocg}{fig:HZoff}{fig:HZoff}{0}%
  \end{ocg}%
  \begin{ocg}{fig:HZon}{fig:HZon}{1}%
    \includegraphics[width=0.5\hsize]{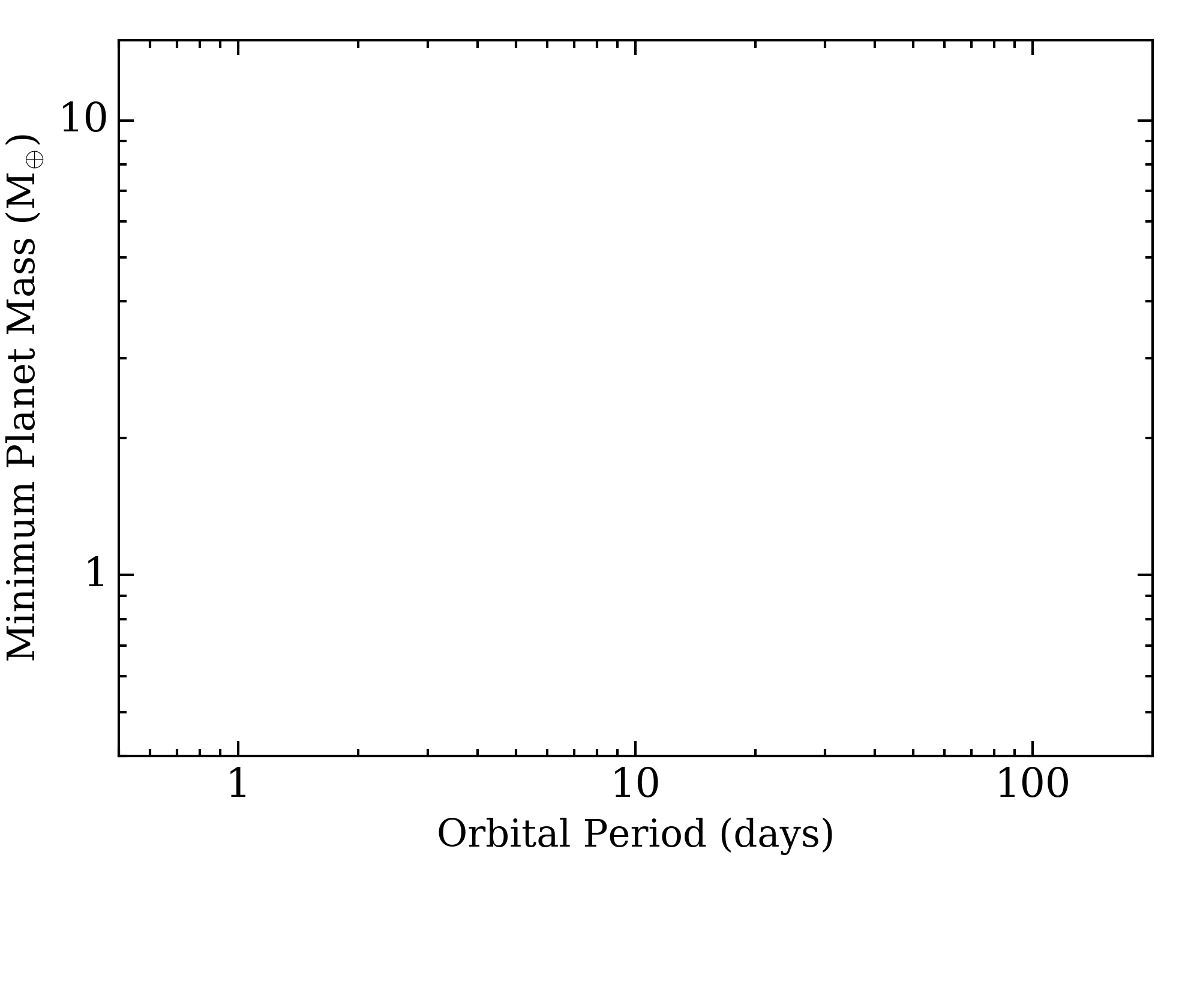}%
    \includegraphics[width=0.5\hsize]{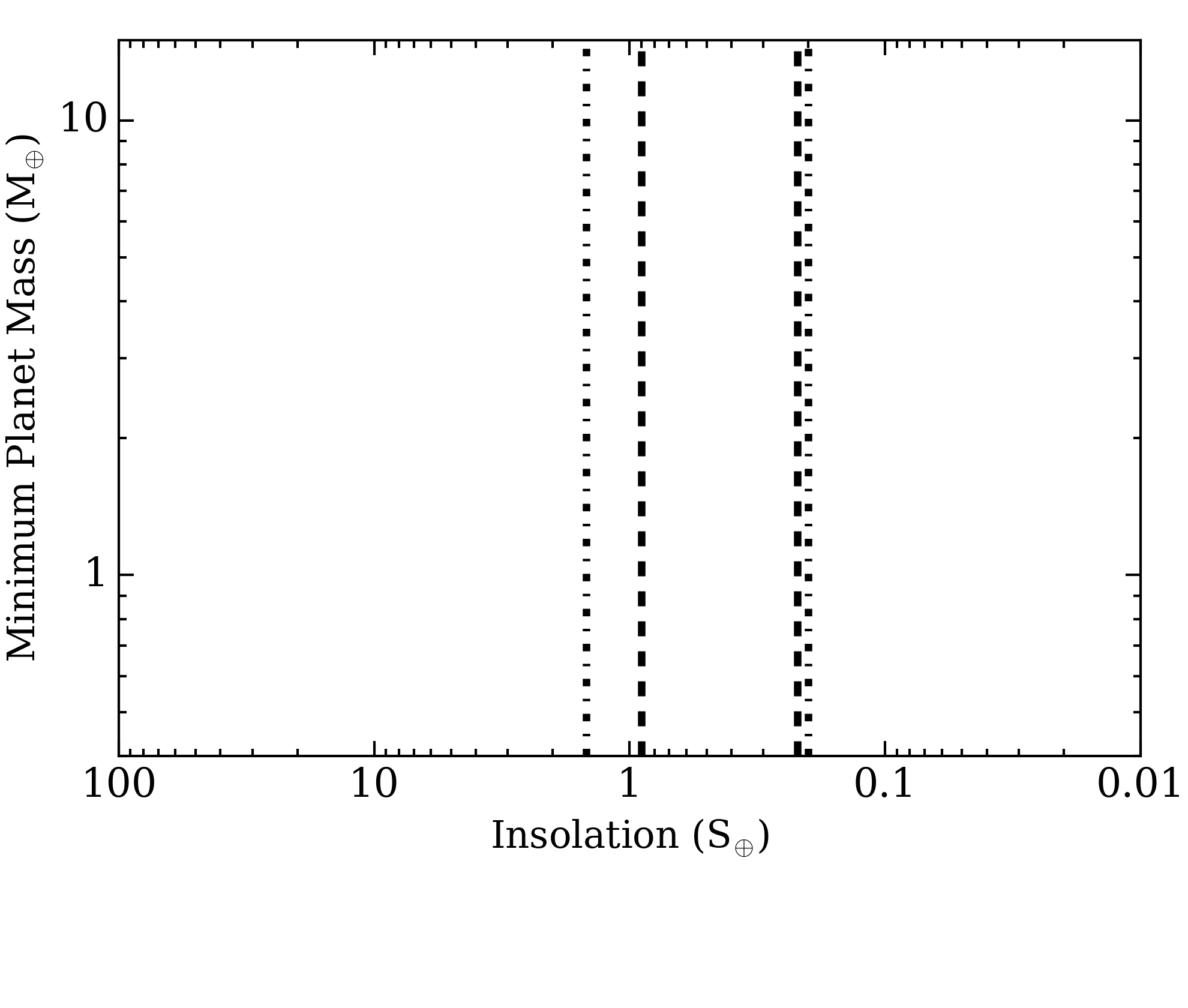}%
  \end{ocg}
  \hspace{-\hsize}%
  \caption{Binned maps of the detection sensitivity of the full SPIRou input catalog derived using our
    automated planet detection algorithm as a function of minimum planet mass and orbital period 
    (\emph{left}) or insolation (\emph{right}). The \emph{dashed vertical lines} in the
    insolation panel indicate the approximate `water-loss' and `maximum-greenhouse' insolation
    limits of the \ToggleLayer{fig:HZon,fig:HZoff}{\protect\cdbox{HZ}} from \cite{kopparapu13}.
    The \emph{dashed-dotted vertical lines}
    indicate the less conservative `recent-Venus' and `early-Mars' HZ limits \citep{kopparapu13}
    \ToggleLayer{fig:shadeon,fig:shadeoff}{\protect\cdbox{\emph{Shaded}}} bins highlight
    regions of the parameter space wherein our detection sensitivity is $<20$\%. The
    \ToggleLayer{fig:Kon,fig:Koff}{\protect\cdbox{\emph{solid lines}}} highlight the curve with $K=1$
    \mps{} for a star with mass equal to our sample's median stellar mass of 0.25 M$_{\odot}$.}
  \label{fig:sensitivity}
\end{figure*}

The detection sensitivity to planets varies from star-to-star due to their changing stellar
properties which can affect our ability to detect planets in radial velocity (e.g. apparent magnitude,
stellar mass, stellar rotation, etc).
Computing the detection sensitivity for each star individually is necessary for calculating planet
occurrence rates (see Sect.~\ref{sect:measurements}). To improve the detection statistics across the
full range of planetary parameters considered we augment the MC realizations for each star with an additional
set of planetary systems with logarithmic $P$, $S$, and \msini{} sampled uniformly rather than from the
planet occurrence rates. The individual detection sensitivity maps for each star can then be combined
to obtain the average sensitivity maps for the full
SLS-PS as a function of $P$, $S$ and \msini{} as shown in Fig.~\ref{fig:sensitivity}.
In this way we marginalize over the stellar properties of our sample stars including
the aforementioned parameters which are known to influence the detection sensitivity for each individual
star. Hence our sensitivity results might be scaled to various stellar samples provided that its global
properties are consistent with our current sample. In Fig.~\ref{fig:sensitivity}
the uncertainties in the detection sensitivity within each grid cell come from counting or Poisson
statistics and therefore benefit from a large number of simulated planetary systems.

Given the binning in Fig.~\ref{fig:sensitivity}, nowhere do we achieve a 100\%
detection sensitivity. This is true even for the most massive close-in planets whose RV
semi-amplitudes are typically much greater than the characteristic RV measurement uncertainty so
long as the system is not orientated close to face-on (i.e. $\sin{i} \sim 0^{\circ}$). The geometric effect
as well as potential aliasing of periodic signals arising from stellar rotation or the window
function, also prevent the detection of certain types of planets. For example, in Fig.~\ref{fig:protcdf}
we see that in our stellar sample there is a dearth of \prot{} $\sim 3-10$ days. Thus
we run into minimal aliasing from \prot{} at those orbital periods and achieve an increased detection
sensitivity relative to planets with equivalent \msini{} but with smaller $P$
(\emph{left panel} Fig.~\ref{fig:sensitivity}).

Similarly we can see a steep decrease in detection sensitivity at periods of $\sim 30$ days. Recall that
in our window functions we are restricted by the telescope's observing schedule to only observe outside
of dark-time whose cycle follows the lunar cycle with a cadence of  $\sim 30$ days.
Consequently the origin of the
decrease in detection sensitivity at $\sim 30$ days can be traced back to aliasing from our time
sampling at that period; an unfortunate circumstance as the range of orbital periods corresponding to the
HZ around M2-4 dwarfs, spans 30 days. 

Due to the
aforementioned aliasing effects from stellar rotation and our window functions, the SPIRou detection
sensitivity is maximized for the most massive planets with $3 \lesssim P \lesssim 10$ days. 
Another pertinent effect at orbital periods close to one day is the potential for signal aliasing due to the
Earth's rotation. This---in part---is responsible for the low detection sensitivity at orbital periods close
to a day in Fig.~\ref{fig:sensitivity}; another unfortunate circumstance given the general interest in
quantifying the occurrence rate of close-in planets \citep{mulders15}.

The average detection sensitivity across the full range of $P$ and
\msini{} in Fig.~\ref{fig:sensitivity} is $44.8 \pm 0.5$\%.
This average detection sensitivity is sufficiently high such that the
resulting planet detections from the SLS-PS will be able to place strong constraints on the cumulative
occurrence rate of planets around SPIRou stars (see Sect.~\ref{sect:measurements}).

When considering the detection sensitivity as a function of insolation, we see the same
qualitative structure as is seen as a function of orbital period (\emph{right panel}
Fig.~\ref{fig:sensitivity}). For example, the increased
sensitivity between $\sim 3-10$ days has a broad manifestation at $S \in [1,10]$ S$_{\oplus}$.
Similarly, in both cases it is unsurprising to see that the detection sensitivity increases
towards more massive planets. In both cases the SPIRou
detection sensitivity reaches its lowest values for the least massive planets on wide-orbits.
Notably, we achieve a detection sensitivity of $\lesssim 20$\% for all planets with
\msini{} $\lesssim 1$ M$_{\oplus}$ thus making it difficult to detect Earth-mass planets and
smaller in the SLS-PS. The average detection sensitivity across the full range of $S$ and
\msini{} considered here is $47.6 \pm 0.5$\%.

\subsection{Detection Sensitivity to HZ Planets}
It is also important to consider our detection sensitivity to HZ planets as these targets are often
flagged for various observational follow-up campaigns. In this study we adopt the `water-loss' and
`maximum-greenhouse' definitions
as our fiducial HZ limits from \cite{kopparapu13} which are derived from a 1D radiative-convective
climate model in the absence of clouds. Following \cite{kasting93}, the inner edge of the HZ is defined
by the `water-loss' limit which arises from the photolysis of water in the upper atmosphere and subsequent
hydrogen escape. The outer edge of the HZ is determined by the `maximum-greenhouse' limit wherein an
increase in atmospheric CO$_2$ levels will result in a net cooling as the increased albedo from Rayleigh
scattering begins to dominate over the increasing greenhouse effect. The insolation levels 
approximately corresponding to our adopted HZ definition are $S \in [0.22,0.90]$ S$_{\oplus}$ for our
stellar sample. For comparison we
also consider the less conservative ($S \in [0.20,1.48]$ S$_{\oplus}$; \citealt{kopparapu13})
`recent-Venus' and `early-Mars' HZ limits which assume that both Venus and Mars were habitable early-on
in the lifetime of the Solar System.

From the fiducial definition of the HZ we find an average
detection sensitivity to HZ planets of $47.3 \pm 1.0$\% which is consistent with the average
detection sensitivity over the full $S$,\msini{} grid in the right panel
of Fig.~\ref{fig:sensitivity}
($46.2 \pm 0.5$\%). Adopting the more generous HZ limits, we find a comparable average
detection sensitivity of $48.7 \pm 0.9$\%.
However, the average detection sensitivity to Earth-like planets ($m_p \in [1,5]$ M$_{\oplus}$) in the HZ
is significantly reduced to $36.7 \pm 1.2$\%. Here we have defined Earth-like planets in terms of their
absolute mass where the mass upper limit is approximately equal to 
the planet mass obtained when evaluating the mean mass-radius relation (Eq.~\ref{eq:mr}) at the
proposed maximum radius of a rocky planet; $\sim 1.5-1.8$ R$_{\oplus}$ \citep{weiss14, rogers15, fulton17}.

\section{SLS-PS Predicted Yield} \label{sect:yield}
\begin{figure*}
  \centering
  \includegraphics[width=0.5\hsize]{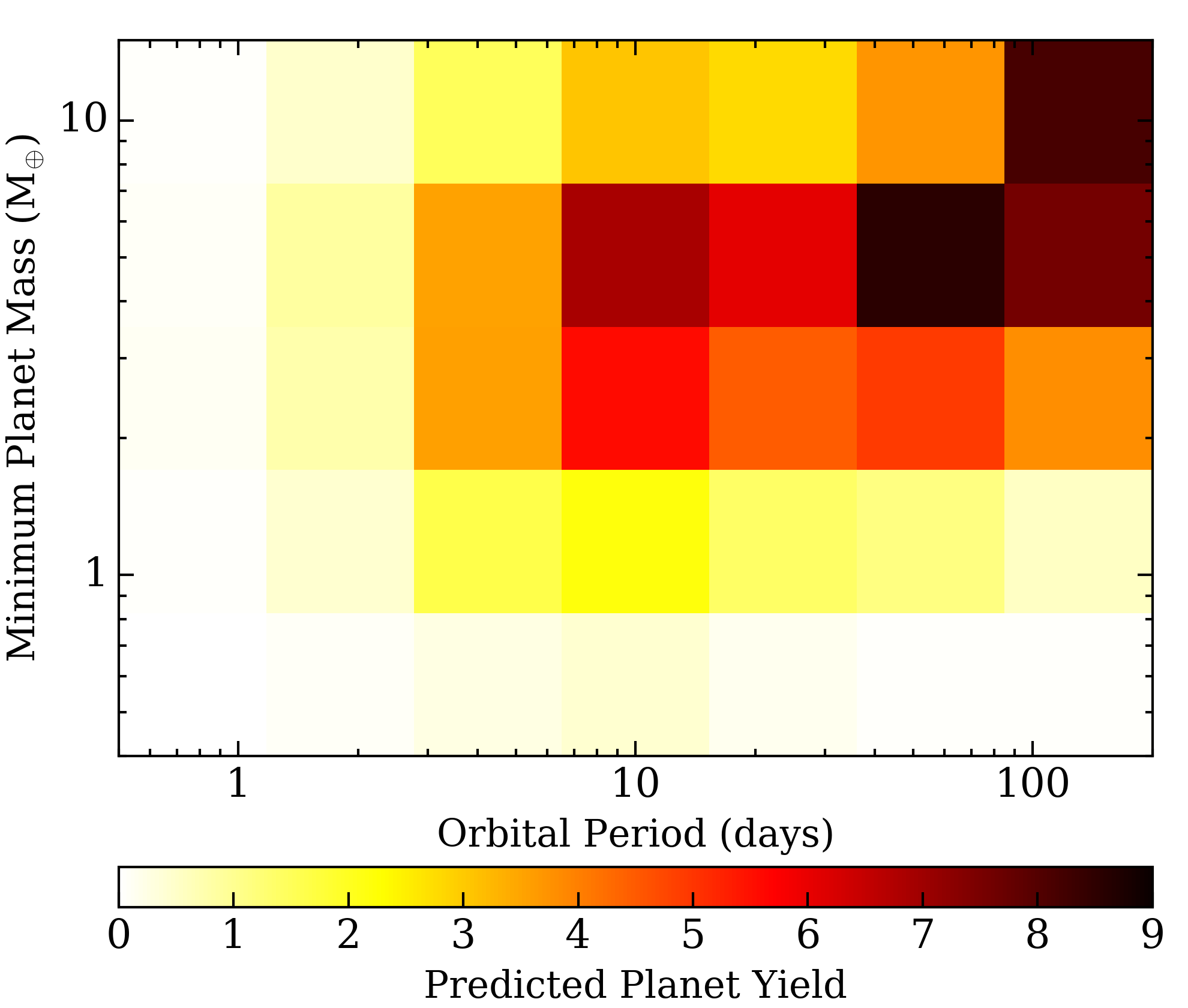}%
  \includegraphics[width=0.5\hsize]{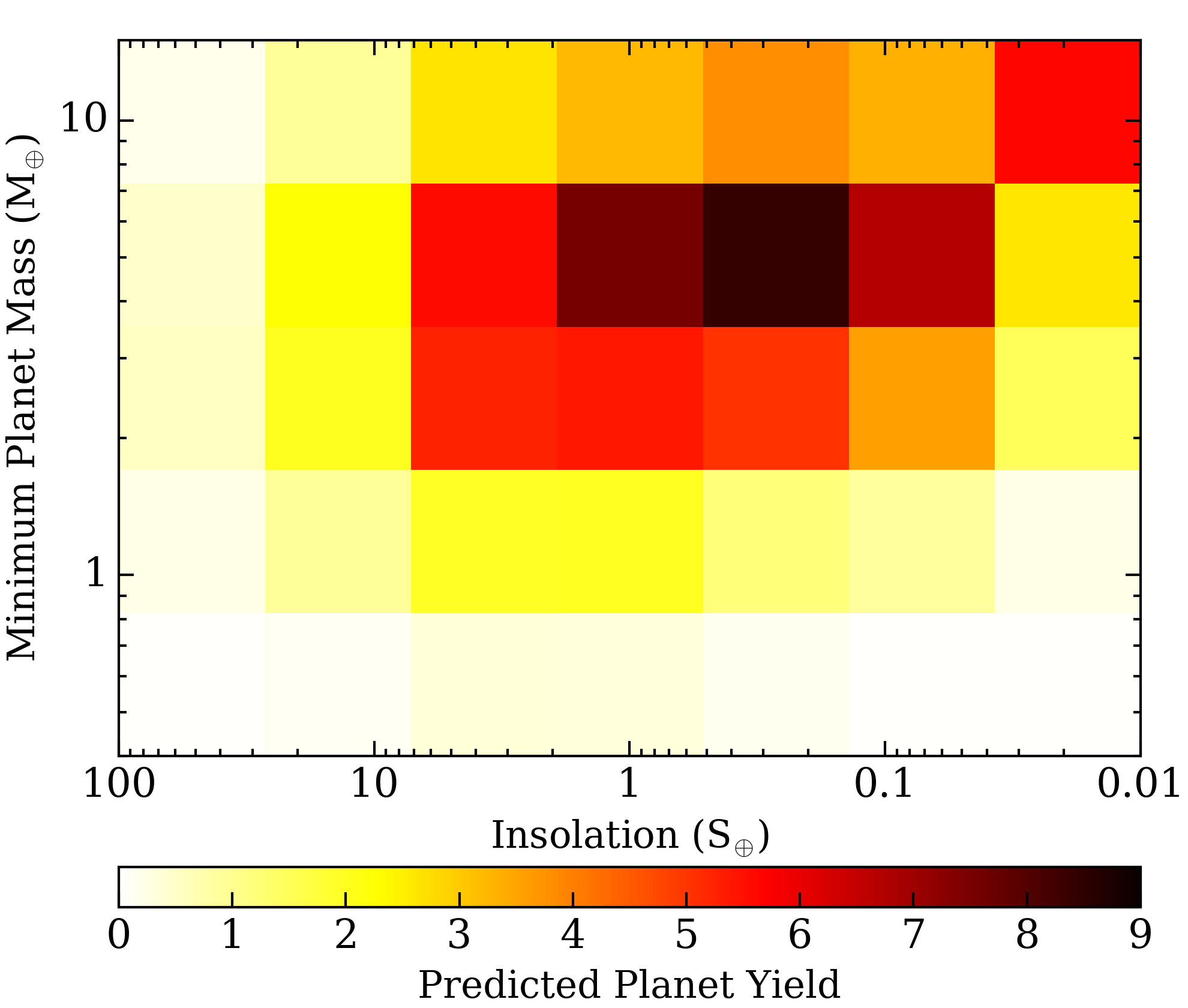}%
  \hspace{-\hsize}%
  \begin{ocg}{fig:shadeoff}{fig:shadeoff}{0}%
  \end{ocg}%
  \begin{ocg}{fig:shadeon}{fig:shadeon}{1}%
    \includegraphics[width=0.5\hsize]{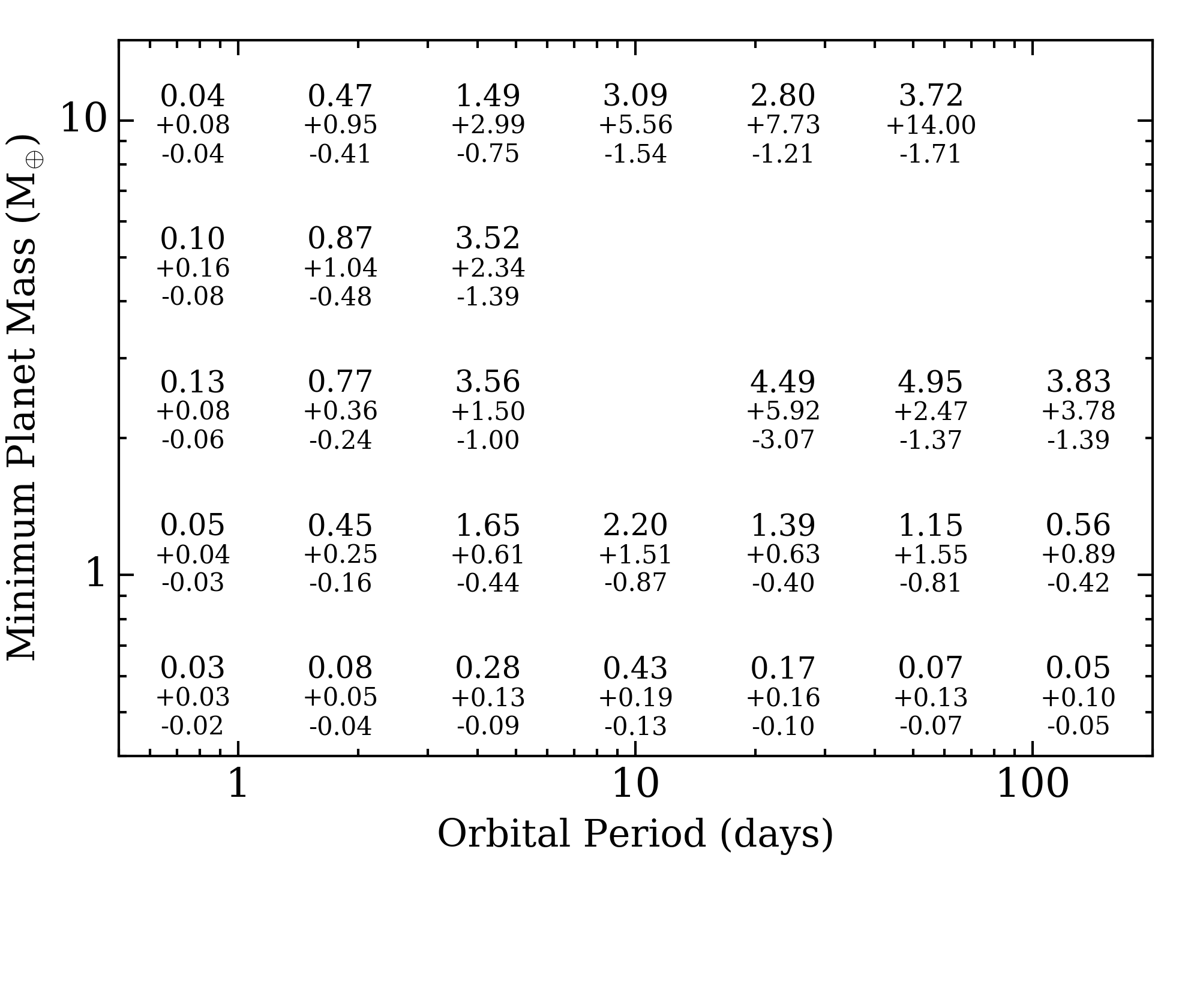}%
    \includegraphics[width=0.5\hsize]{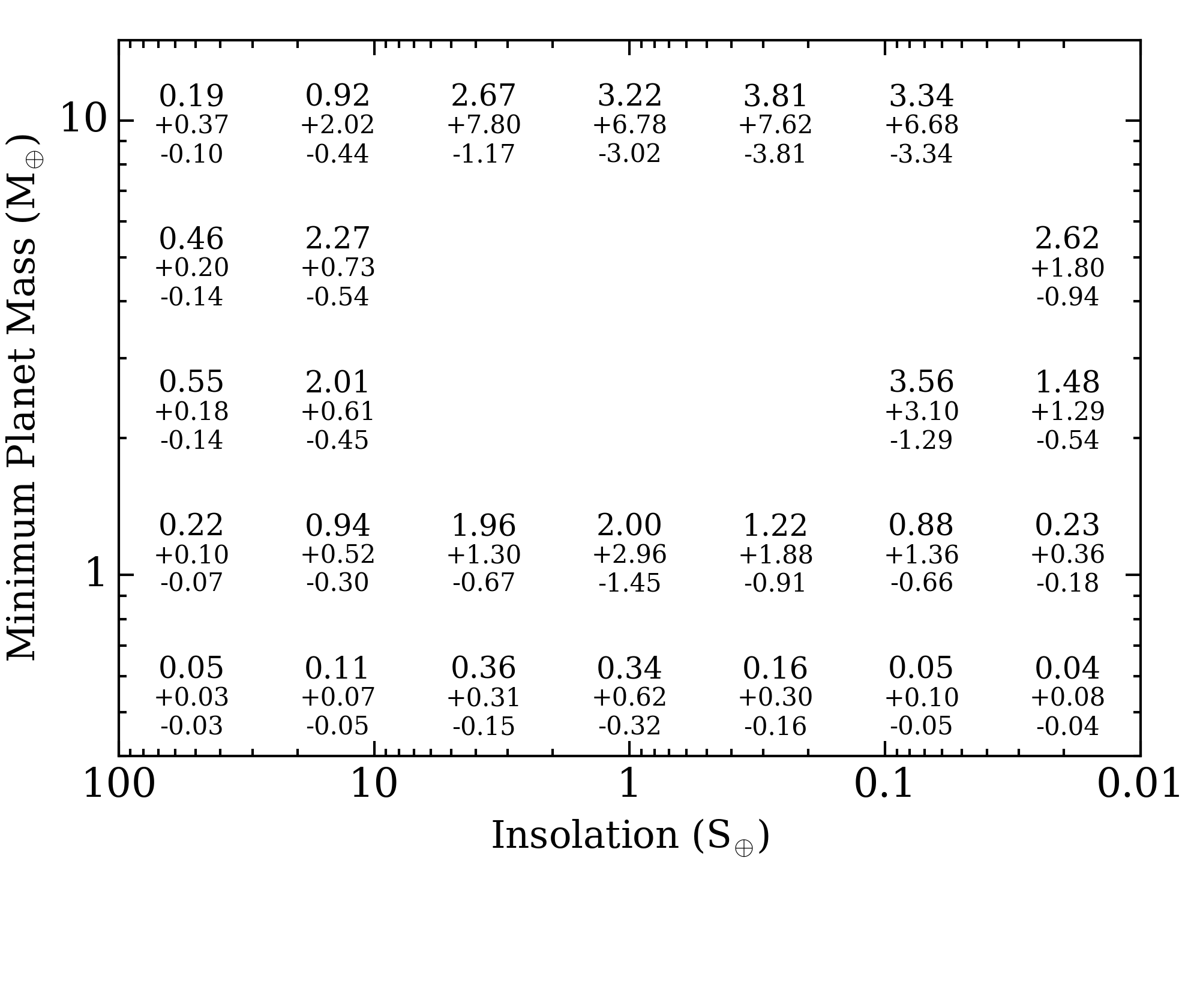}%
  \end{ocg}
  \hspace{-\hsize}%
  \begin{ocg}{fig:HZoff}{fig:HZoff}{0}%
  \end{ocg}%
  \begin{ocg}{fig:HZon}{fig:HZon}{1}%
    \includegraphics[width=0.5\hsize]{yieldPMpsini_HZ.png}%
    \includegraphics[width=0.5\hsize]{yieldFMpsini_HZ.png}%
  \end{ocg}
  \hspace{-\hsize}%
  \caption{Coarsely binned maps of the predicted planet yield from 
    the SLS-PS as a function of minimum planet mass and orbital period 
    (\emph{left}) or insolation (\emph{right}). The \emph{dashed vertical lines} in the
    insolation panel indicate the approximate `water-loss' and `maximum-greenhouse' insolation
    limits of the \ToggleLayer{fig:HZon,fig:HZoff}{\protect\cdbox{HZ}} from \cite{kopparapu13}.
    The \emph{dashed-dotted vertical lines}
    indicate the less conservative `recent-Venus' and `early-Mars' HZ limits \citep{kopparapu13}. The
    \ToggleLayer{fig:shadeon,fig:shadeoff}{\protect\cdbox{annotated numbers}} in each bin
    report the predicted number of detected planets and the uncertainties on the prediction
    which are dominated by uncertainties in the input planet occurrence rates.}
  \label{fig:yield}
\end{figure*}

For each of the 100 stars in the SPIRou input catalog,
we can compute the number of planets detected as a function of $P$, $S$,
and \msini{} given the input occurrence rates and calculations of each star's detection sensitivity.
Here we must correct the reduced injected cumulative planet occurrence rate of 2.4 planets per star
over our grid of $P \in [0.5,200]$ days and \msini{} $\in [0.4,15]$ M$_{\oplus}$ to be equal to the
intended 2.5 planets per star over the \cite{dressing15a} grid ($P \in [0.5,200]$ days, $r_p \in [0.5,4]$
R$_{\oplus}$). Then after dividing out the number of simulated planetary systems per star,
we compute the total planet detection yield predicted by our
simulated survey. The predictions are shown in Fig.~\ref{fig:yield} over a more coarsely
binned map than in Fig.~\ref{fig:sensitivity}, due to the small number of planets detected in each
bin.

The cumulative planet yield over the orbital period domain and minimum mass range considered in
Fig.~\ref{fig:yield} is $85.7^{+29.3}_{-12.5}$ out of $\sim 180$ injected planets.
Of these, $\sim 53.7$ planets (62.6\%) are
the only planet detected in the system while $\sim 26.8$ planets (31.3\%) are detected in a
2-planet system. The remaining
$\sim 5.2$ planets (6.1\%) are found in systems with 3 detected planets; i.e. we expect to detect
$1-2$ 3-planet systems in the SLS-PS.

The number of simulated planetary systems per SPIRou star is large in our simulations. The
result is that the uncertainties in the
predicted yield are dominated by uncertainties in the input planet occurrence
rates from Kepler. Due to Kepler's low detection sensitivity to small planets on
wide-orbits the planet occurrence rate and hence the predicted SPIRou yield is
poorly constrained at large orbital periods/low insolation levels.
Approximately half of our planet detections are super-Earths
with minimum masses \msini{} $\in [3,7]$ M$_{\oplus}$ owing to their assumed frequency,
which is intrinsically
high, and the moderately high detection sensitivity achieved across the
range of orbital periods considered ($\sim 30-85$\%). Considering the `water-loss' and
`maximum-greenhouse' definitions of the HZ limits from
\citep{kopparapu13} we detect $22.0^{+18.4}_{-7.9}$ out of $\sim 47$ injected
HZ planets in the SLS-PS over the range of minimum masses considered in Fig.~\ref{fig:yield}. These include
$9.0^{+8.5}_{-3.6}$ Earth-like HZ planets out of $\sim 25$ injected planets with $m_p \in [1,5]$ M$_{\oplus}$.
When adopting the more generous `recent-Venus' and `early-Mars' HZ limits, these numbers get inflated to
$31.5^{+26.8}_{-11.4}$ and $13.4^{+12.7}_{-5.3}$ respectively. 

\begin{figure*}
  \centering
  \includegraphics[width=\hsize]{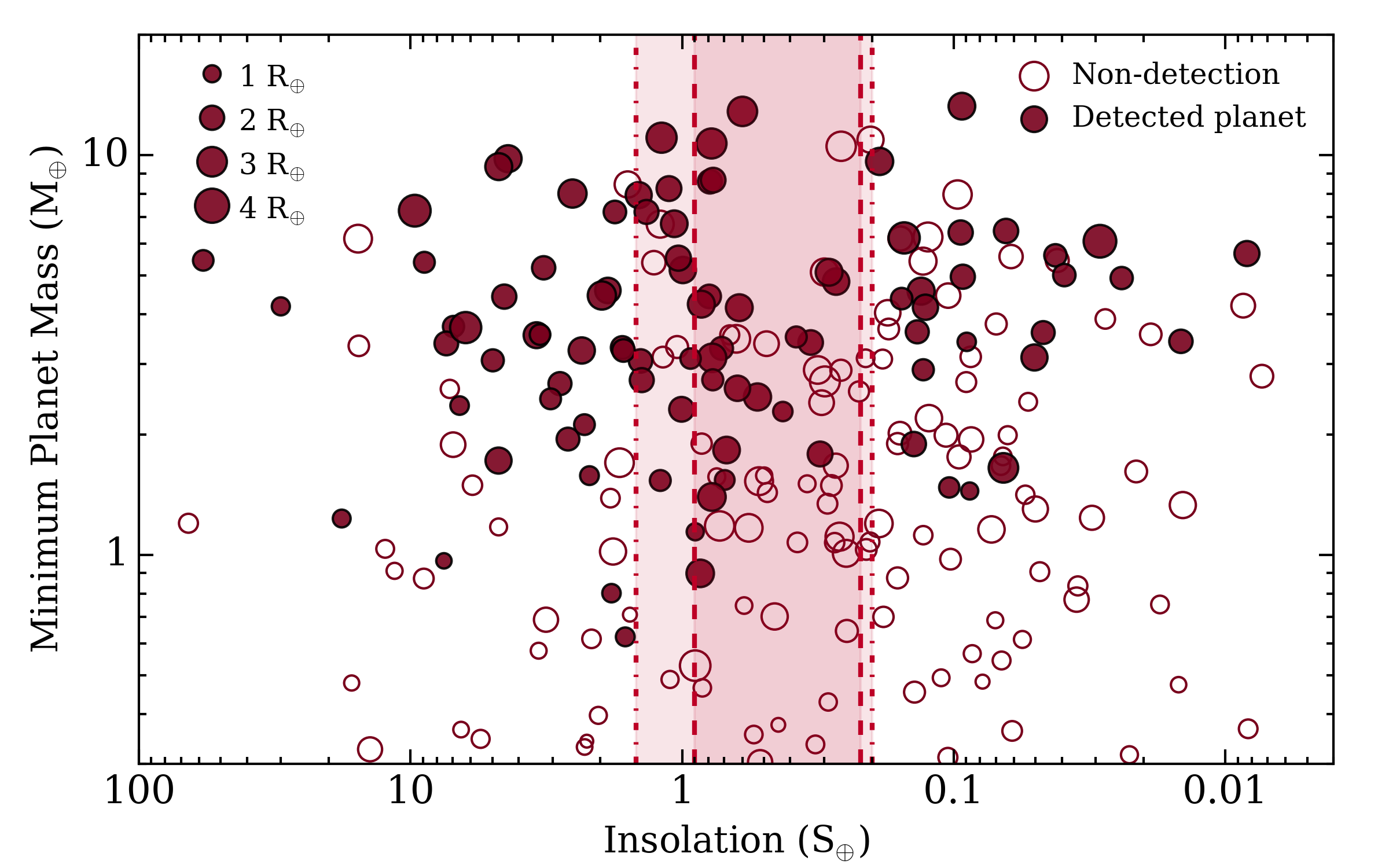}%
  \caption{A random subset of the simulated SPIRou planets representative of the underlying planet
    population investigated in the SLS-PS in the insolation/minimum planet mass plane.
    SPIRou planet detections are marked by
    \emph{solid circles} whereas \emph{open circles} represent injected planets that remain undetected by SPIRou.
    We detect 90 planets around $100$ stars in the subset of simulated planets shown here.
    The size of each planet's marker is proportional to its radius. The \emph{inner shaded region} highlights
    the approximate `water-loss' and `maximum-greenhouse' limits of the HZ whereas the
    \emph{outer shaded region} highlights the `recent-Venus' and `early-Mars' HZ limits \citep{kopparapu13}.}
  \label{fig:spiroudet}
\end{figure*}

The population of SPIRou planets can be visualized slightly differently in the
insolation/minimum planet mass plane as shown in Fig.~\ref{fig:spiroudet}. Here we
present a random subset of all simulated planetary systems. The size of this subset is
chosen such that the integer number of detected planets in the subset is consistent with
the predicted planet yield of $89.9$ planets. The resulting total number
of planets shown in Fig.~\ref{fig:spiroudet} is 250 with 90 planets detected. Because the
subset of planets shown in Fig.~\ref{fig:spiroudet} is random, it does not preserve the
number of detected planets in various subsets of the planet parameter space. For example,
in the full simulated SLS-PS we detected 9 Earth-like planets whereas the random subset
shown in Fig.~\ref{fig:spiroudet} only depicts 4 Earth-like planet detections.

The stellar parameters, planetary parameters, and simulated time-series for each MC realization are made
available to the community on
github\footnote{\href{https://github.com/r-cloutier/SLSPS\_Simulations}{https://github.com/r-cloutier/SLSPS\_Simulations}}
in the form of \texttt{python pickles}.
We also provide the combined results of all realizations for interested users to analyze the full SPIRou
input catalog and each star's suite of simulated planetary systems. These data may be used, for example,
to reconstruct most of the figures shown in this paper. Instructions and examples of how to read and
access these data are also provided along with all the required \texttt{python} scripts.

\subsection{Giant Planet Detections}
Although the vast majority of our predicted SPIRou planet population are derived from the Kepler occurrence
rates of small planets ($r_p \leq 4$ R$_{\oplus}$ or \msini{} $\lesssim 15$ M$_{\oplus}$), the non-zero frequency
of giant planets from RV surveys results in some giant planet detections. From our simulations we find that
4.1 SPIRou detections are giant planets with \msini{} $\geq 20$ M$_{\oplus}$. Thus the total SPIRou planet yield
including all planets becomes $89.9^{+30.7}_{-13.0}$. Furthermore, we find
that the average multiplicity of simulated planetary systems containing a giant planet is 2.2 despite our
imposed dynamical stability criteria which disfavor giant planets in multi-planet systems. This suggests that
M dwarf systems containing both small and giant planets could be discovered with SPIRou. Such systems---should
they exist---would be crucial for informing planet formation scenarios around M dwarfs.

\section{The effect of an increased planet frequency around late M dwarfs} \label{sect:2occ}
Recall that thus far we have assumed that the Kepler occurrence rates,
computed using stars with $T_{\text{eff}} \gtrsim 3200$ K \citep[i.e. spectral
  types of M4.5 and earlier;][]{luhman03}, are also applicable to the later
M dwarfs in our stellar sample. However there are some lines of evidence which
suggest that the cumulative planet occurrence rate will increase towards later M dwarfs.
For example, planet formation models around very
low-mass stars predict many small planets ($r_p \sim 1$ R$_{\oplus}$) at short
orbital periods \citep[e.g.][]{alibert13,alibert17} and the detection of seven
Earth-sized planets around the ultra-cool dwarf TRAPPIST-1 from a small sample
of ultra-cool dwarfs \citep{gillon17, luger17} are suggestive of an increased
cumulative planet occurrence rate around late M dwarfs compared to early
M dwarfs. If this is true then two possible outcomes on the SPIRou SLS-PS
planet yield are imaginable. Either the predicted planet yield will increase as a
result of the greater number of potential planets to detect, or the predicted
planet yield will decrease because adding additional planets will contribute to the
observed RV rms thus deterring our ability to detect individual planets.
Because the latter effect will modify our detection sensitivity we cannot simply estimate the
resulting planet yield from the product of a scaled-up planet occurrence rate with our nominal
detection sensitivity from Fig.~\ref{fig:sensitivity}. Instead, to address this
caveat in a simplified way, we simulate a new version of the SLS-PS 
in which we artificially \emph{increase} the planet occurrence rates by a scaling factor
and recompute the SLS-PS detection sensitivity and planet yield.
All inputs in this simulation other than the planet occurrence
rates are identical to the fiducial survey presented throughout this paper. 

To increase the planet occurrence rates we will use a simple scaling factor.
That is because in practice, the planet occurrence rates around late M dwarfs are not
well-characterized so we adopt the same Kepler occurrence rates $f(P,r_p)$
but increase it by a factor of 2 (i.e. a revised cumulative planet occurrence
rate of $5 \pm 0.4$ planets per M dwarf). We then conduct the simulated
SLS-PS identically as before. However, the dynamical stability considerations (see 
Sect.~\ref{sect:dynam}) that restrict the types of simulated multi-planet systems
can have a more pertinent effect when the number of sampled planets per planetary
system is doubled. We find that in practice these considerations cause the resulting planet
occurrence rates to only be increased by a factor of $\sim 1.5$ instead of 2
such that the resulting cumulative planet occurrence rate is 3.56 rather than 5
planets per star. 

\begin{figure}
  \centering
  \includegraphics[width=\hsize]{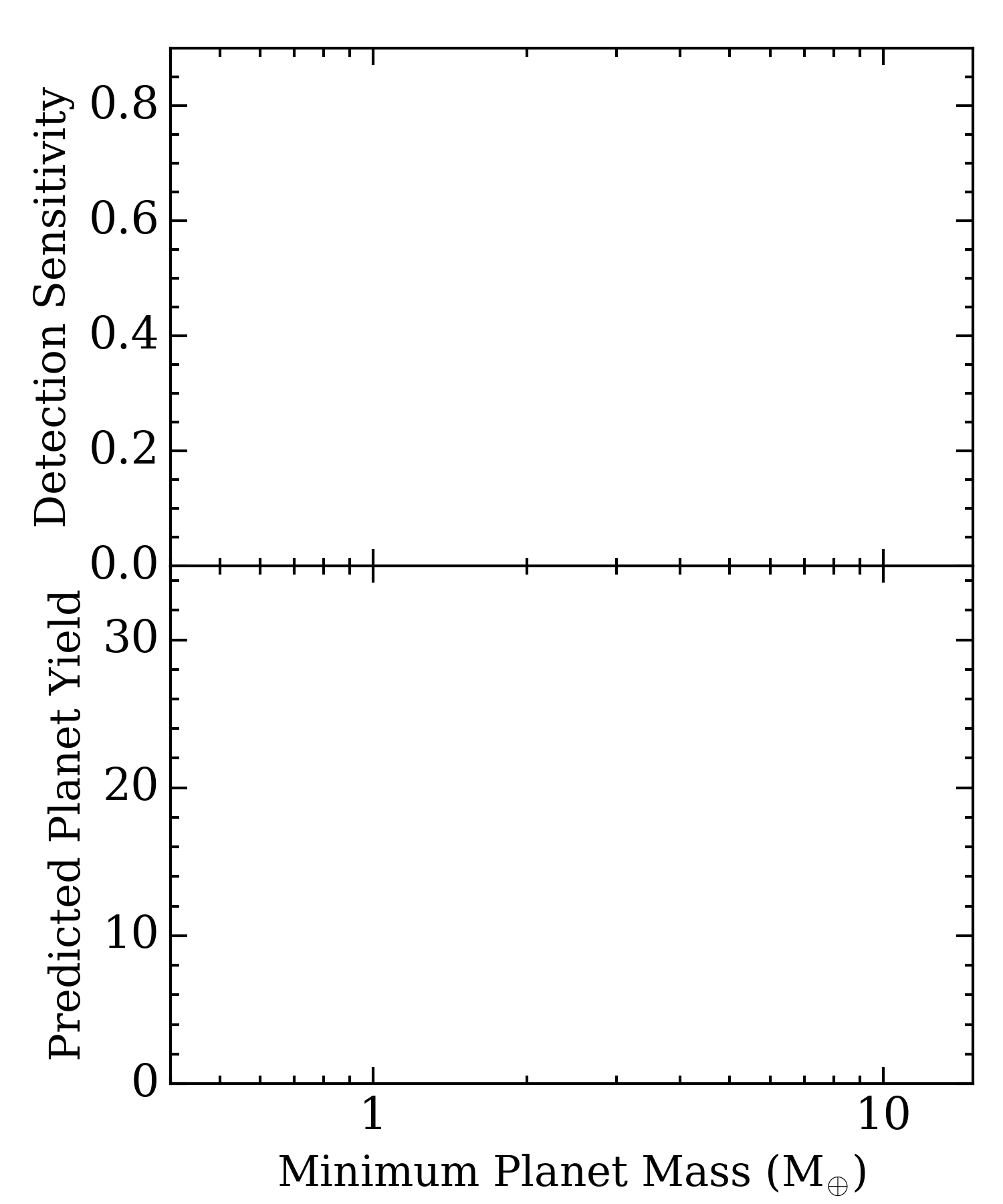}%
  \hspace{-\hsize}%
  \begin{ocg}{fig:1off}{fig:1off}{0}%
  \end{ocg}%
  \begin{ocg}{fig:1on}{fig:1on}{1}%
    \includegraphics[width=\hsize]{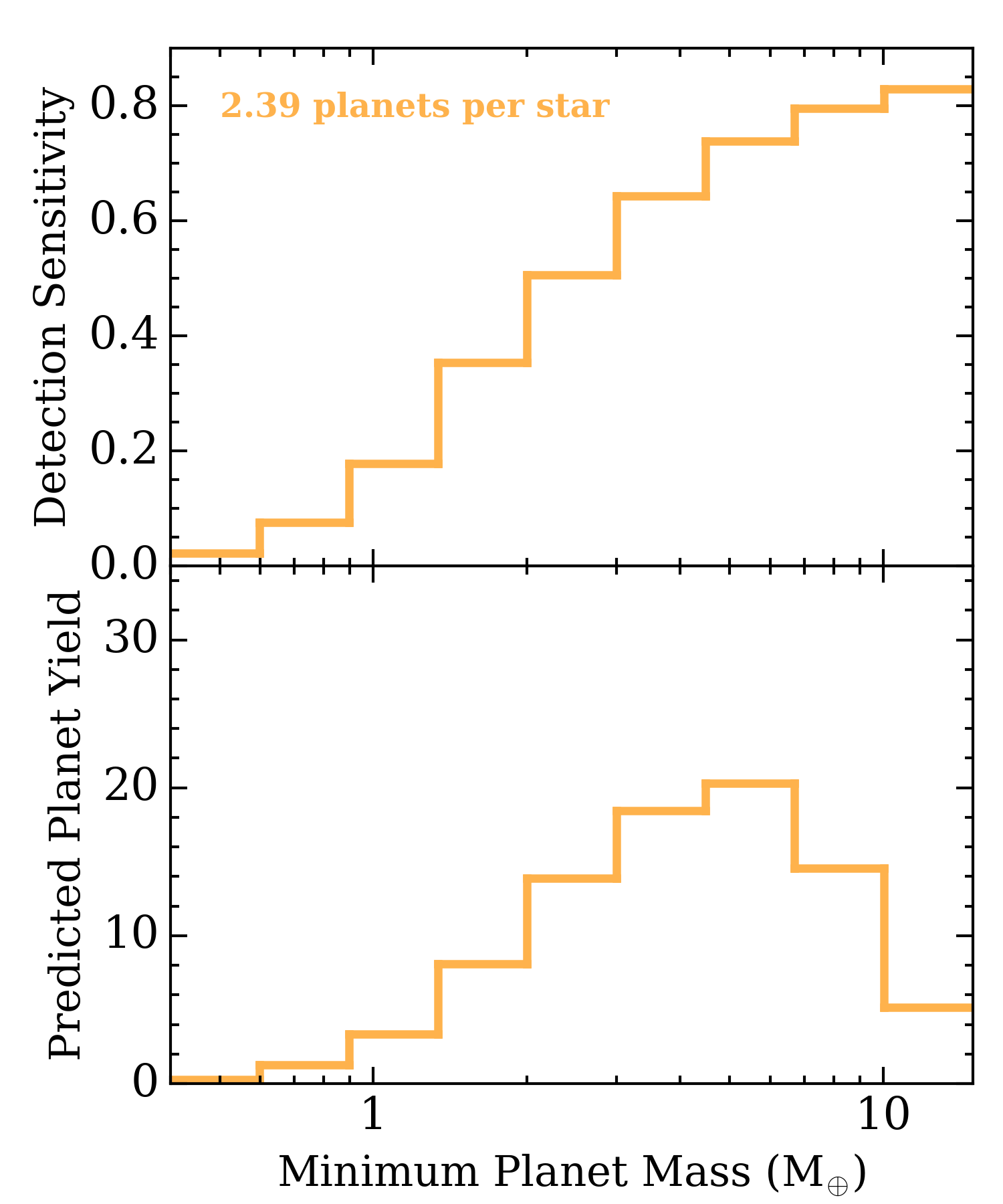}%
  \end{ocg}
  \hspace{-\hsize}%
  \begin{ocg}{fig:2off}{fig:2off}{0}%
  \end{ocg}%
  \begin{ocg}{fig:2on}{fig:2on}{1}%
    \includegraphics[width=\hsize]{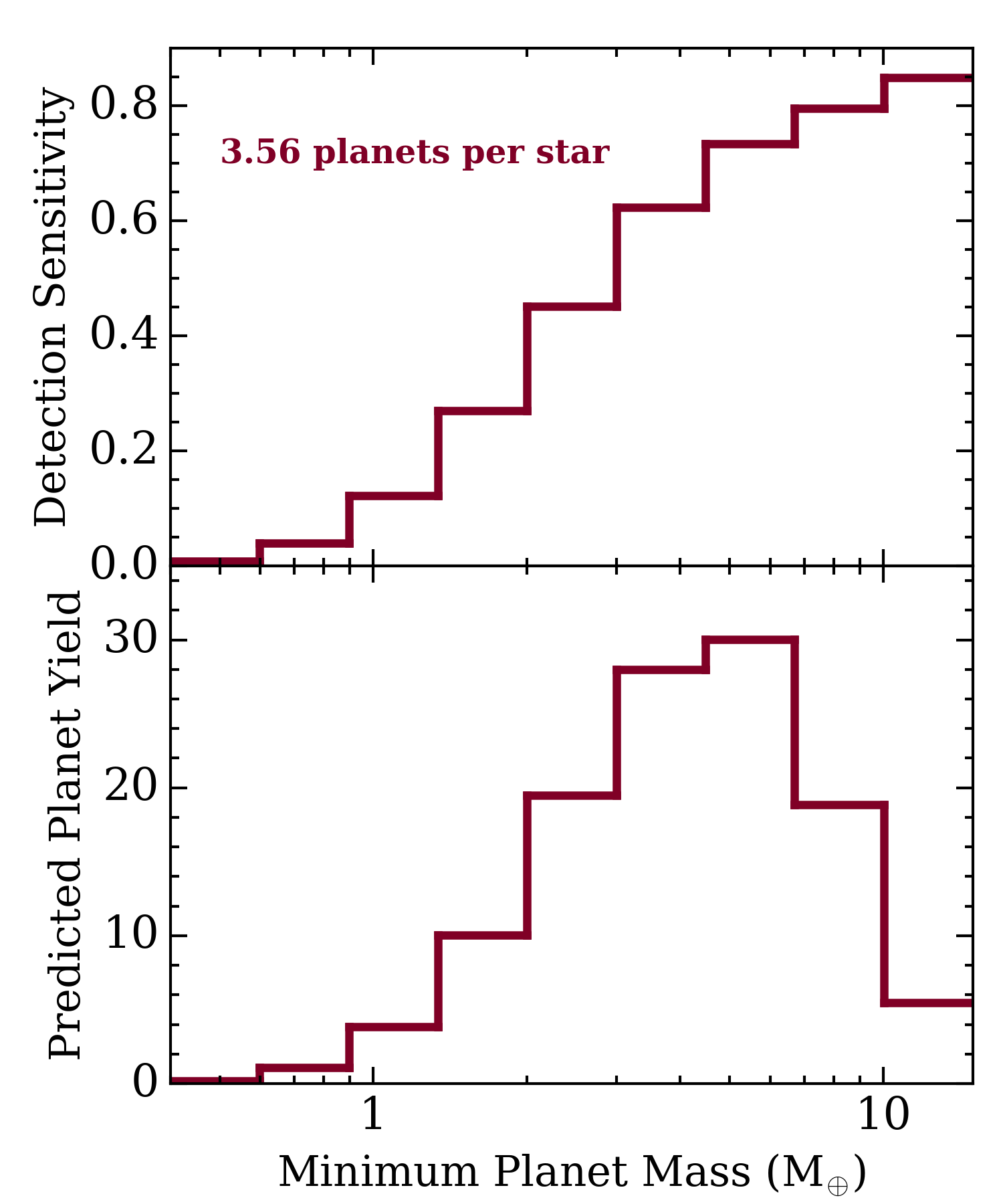}%
  \end{ocg}
  \hspace{-\hsize}%
  \caption{\emph{Top}: the detection sensitivity as a function of minimum planet mass
    for the nominal fiducial version of the SLS-PS containing 
    \ToggleLayer{fig:1off,fig:1on}{\protect\cdbox{2.39 planets per star}} and for a
    modified version of the SLS-PS that is nearly identical except for the increased cumulative
    planet occurrence rate of
    \ToggleLayer{fig:2off,fig:2on}{\protect\cdbox{3.56 planets per star}}.
    \emph{Bottom}: the predicted planet yield as a function of minimum planet mass
    for the two aforementioned survey versions. Over the range of \msini{} $\in [0.4,15]$
    M$_{\oplus}$, a factor of 1.3 more planets are detected in the survey with
    the increased cumulative planet occurrence rate.}
  \label{fig:2occ}
\end{figure}

Fig.~\ref{fig:2occ} compares the detection sensitivity and planet yield recovered by the
fiducial version of the SLS-PS---with 2.4 planets per star---with those obtained after increasing
the cumulative planet occurrence rate to 3.56 planets per star. The detection sensitivity as
a function of \msini{} is
only slightly modified by the addition of, on average, 1.2 planets per planetary system.
That is that, the fiducial version of the survey presented throughout this paper
has only a slightly greater detection sensitivity to planets with \msini{} $\in [0.4,20]$ M$_{\oplus}$
because the fewer planets on average per planetary system contribute to a
lower RV rms than when the planet occurrence is increased. Explicitly, the average detection sensitivity
over the range of \msini{} considered in Fig.~\ref{fig:2occ} is 47.5\% for our fiducial
survey version compared to 41.9\% when the cumulative planet occurrence rate is increased.

The net effect in the new version of the SLS-PS of having more planets per
planetary system and a comparable average detection sensitivity is that more planets are
detected overall. In each \msini{} bin shown in Fig.~\ref{fig:2occ}, more planets are detected
when the cumulative planet occurrence rate is increased resulting in
1.3 times more SPIRou planet detections ($110.7^{+28.5}_{-12.6}$ planets) than when assuming the nominal Kepler
planet occurrence rates.

Despite there being evidence for an increased planet occurrence rate $f$ around late M dwarfs
than around the early-to-mid M dwarfs observed by Kepler, consideration of an SLS-PS
in which the cumulative planet occurrence rate is increased is not intended to be
interpreted as a new estimate of the SLS-PS planet yield. This is because our simplified
methodology for scaling-up $f$ is not intended to replicate the true $f$ around late M
dwarfs. We have adopted a simple scaling of $f$ as measurements
of $f$ around late M dwarfs are currently poorly constrained. Furthermore, there exist
some lines of empirical and theoretical evidence \citep{gillon17, pan17}
that the fraction of multi-planetary systems that
form low order mean-motion resonant chains is larger around late M dwarfs than is seen in the
Kepler sample \citep{lissauer11, fabrycky14}. In this case, the keplerian RV modelling used
in this study would not be applicable which may affect the resulting planet yield.
However, this exercise is intended to demonstrate that \emph{if} the cumulative planet occurrence
rate increases towards later M dwarfs, what is the nature of that effect
on our detection sensitivity and ultimately on the SPIRou planet yield.
Fortunately, the effect seems to be a positive
one for the detection of small planets around late M dwarfs with SPIRou.

\section{Measuring the RV Planet Frequency} \label{sect:measurements}
\subsection{Recovering planet frequency} \label{sect:occurrence}
With consistent RV monitoring of the target stars in the SLS-PS the resulting detections
from the survey will be able to provide independent constraints on the occurrence rate of
RV planets around M dwarfs.
Of course this calculation has previously been done using either
RV or transit survey data \citep[e.g.][]{bonfils13, dressing15a},
although those studies were limited to early M dwarfs whereas SPIRou is uniquely designed
to study mid-to-late M dwarfs whose planet occurrence rates are less certain \citep{demory16}.
Furthermore,
SPIRou is expected to uncover a larger set of planet detections than the HARPS M dwarf sample
presented in \cite{bonfils13} and thus provide stronger constraints on the full M dwarf planet
occurrence rate as a function of minimum planetary mass.

Here we wish to estimate the precision with which we expect to measure the
planet frequency based on the expected results of the SLS-PS. The planet frequency
$f$ differs somewhat from the planetary occurrence rate in that the planetary frequency does not
represent the number of a particular type of planet per host star but instead is the fraction
of stars which host a particular type of planet and is therefore only defined on the closed
interval $f \in [0,1]$.
To compute the planet frequency as a function of $P$ (or similarly $S$) and \msini{,}
we will adopt the formalism from \cite{carson06, lafreniere07b}. In our survey of $N$
stars denoted by $j=1,\dots,N$, we wish to compute the fraction of M dwarfs that host
a planet within a particular range of orbital periods and minimum masses; $f \to f(P,m_p\sin{i})$.
Firstly, we must estimate the probability that such a planet will be detected around
the $j^{\text{th}}$ star $p_j$, based on the star's known detection sensitivity (see
Sect.~\ref{sect:sensitivity}). The probability that a particular planet will be detected orbiting
the $j^{\text{th}}$ star is then $fp_j$ whereas the probability of a non-detection is $1-fp_j$.
Given the resulting planet yield from our simulated SLS-PS we can identify around which stars
a particular planet is detected. For a given range of $P$ and \msini{,} we denote
planet detections within that range around the $j^{\text{th}}$ star by $d_j$ which equals 1
if such a planet is detected and 0 otherwise. Now we can write down the likelihood of our
planet detections given $f$ as

\begin{equation}
  \mathcal{L}(d_j|f) = \prod_{j=1}^{N} (1-fp_j)^{1-d_j} (fp_j)^{d_j}.
  \label{eq:freqlike}
\end{equation}

In order to compute the value of $f(P,m_p\sin{i})$ in various $P$,\msini{} bins,
we will invoke Bayes theorem:

\begin{equation}
  \text{P}(f|d_j) = \frac{\mathcal{L}(d_j|f) p(f)}{\int_{0}^{1} \mathcal{L}(d_j|f) p(f) \text{d}f}
  \label{eq:freqpost}
\end{equation}

\noindent where $p(f)$ is the prior probability of measuring $f$ and 
P$(f|d_j)$ is the posterior PDF of measuring a frequency $f$ given the
observations $d_j$. To compute $f(P,m_p\sin{i})$ over the full grid of $P$ and
$m_p\sin{i}$, the above formalism is applied independently to each logarithmically spaced bin in
$P$ and \msini{.} The fraction of M dwarfs with a particular planet is defined as the
MAP value of the posterior PDF with its uncertainties characterized by the
$16^{\text{th}}$ and $84^{\text{th}}$ percentiles of the distribution.

Before computing $f(P,m_p\sin{i})$ from the results of our simulated SLS-PS, we must first assign planet
detections around each star in the target sample to integer values rather than the statistical
averages used to present the results of the full survey. To do so we round the number of detected planets
in each $P$, \msini{} bin to the nearest integer for each star individually and for the full distribution
of detected planets from the SLS-PS (see Fig.~\ref{fig:yield}). Rounding to integers
alters the total number of detected
planets so we use a small multiplicative correction factor on each star's detected planet population such
that we recover the correct total planet yield of 85 planets. By combining the results of all
realizations, the number of planets detected in each $P$, \msini{} bin over the full simulated survey
always exceeds the rounded average value shown in Fig.~\ref{fig:yield}. To account for this when
assigning the 85 planet detections, we assign planet detections to stars based on their sensitivity
to that particular type of planet. Specifically, in each $P$, \msini{} bin in which
we detect at least one planet, we identify the subset of stars which have at least one planet
detection in that bin and
sort those stars by their detection sensitivities within that bin. We then select the $n$ stars with the highest
detection sensitivity where $n$ is the rounded total number of planets detected within that bin.
Those stars have $d_j$ set to unity within that bin whereas $d_j=0$ for all remaining stars.
We note that by this routine stars
with the highest detection sensitivities are frequently chosen so we limit the number of planets that can be
detected around a single star to be $\leq 3$; the maximum number of planets detected by our automated
detection algorithm in the simulated SLS-PS. We also note that by selecting stars with the highest
detection sensitivity when assigning planet detections, we are maximizing the likelihood
(Eq.~\ref{eq:freqlike}) and consequentially computing the maximally constrained planet frequency values.

\begin{figure*}
  \centering
  \includegraphics[width=0.5\hsize]{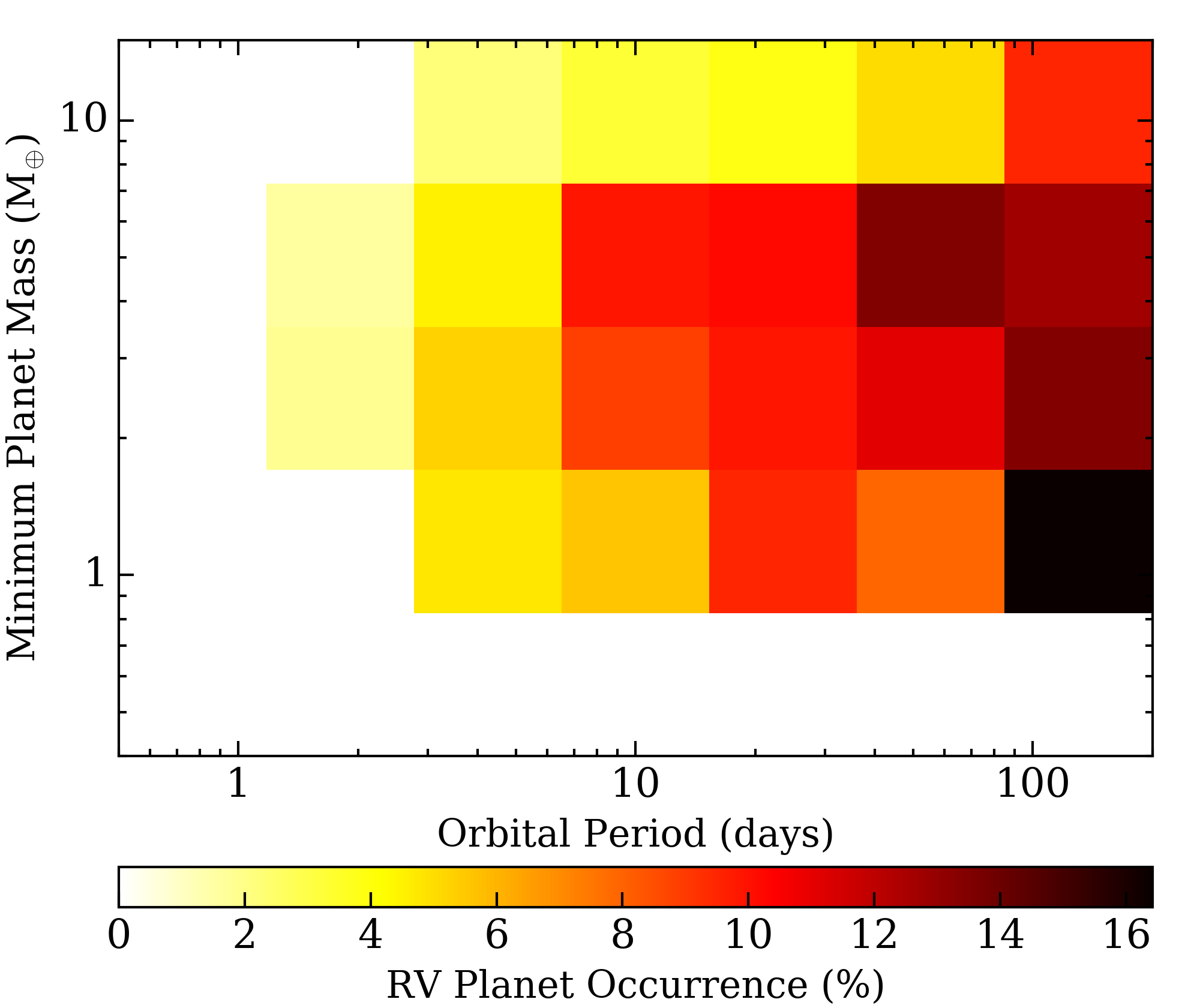}%
  \includegraphics[width=0.5\hsize]{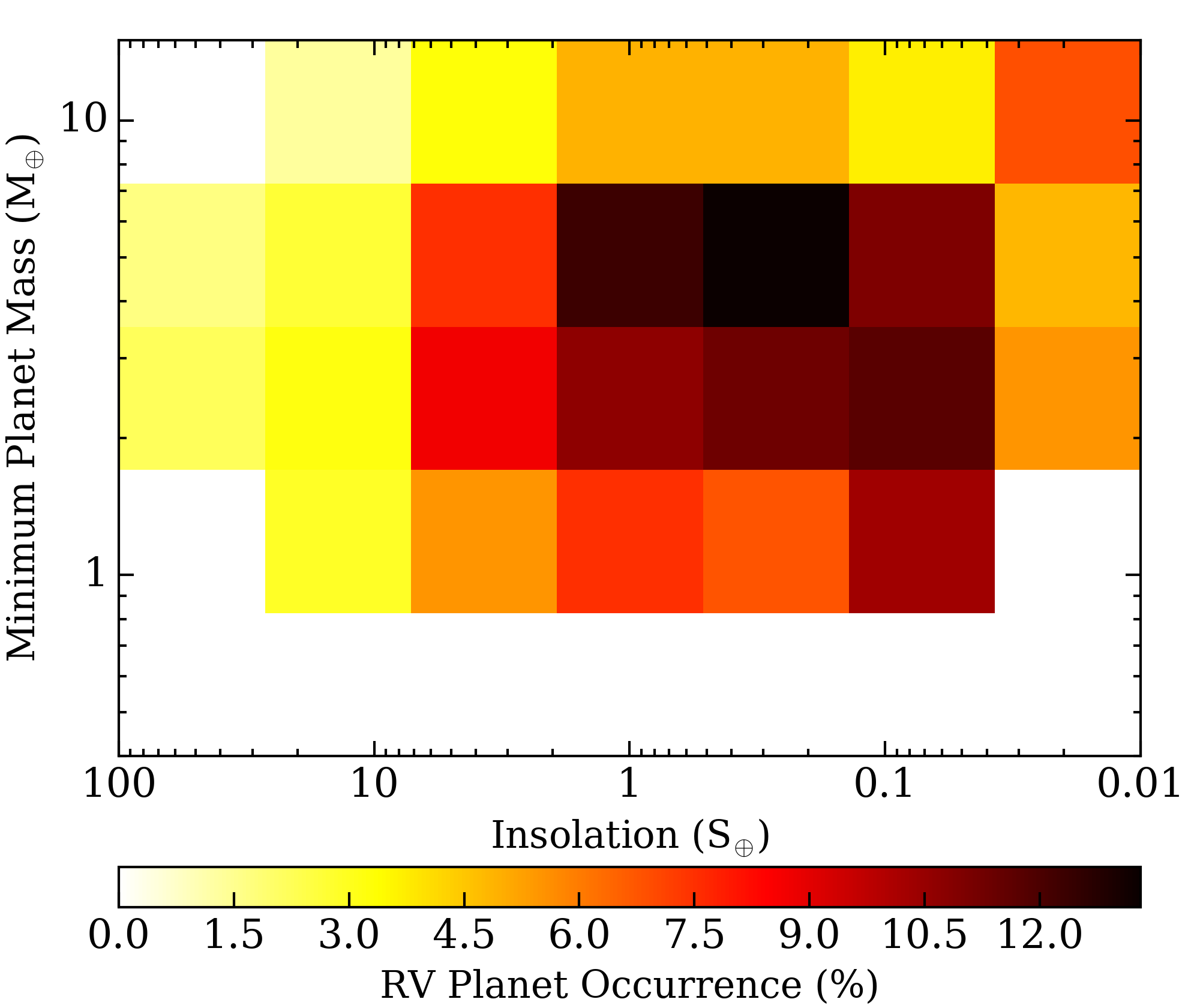}%
  \hspace{-\hsize}%
  \begin{ocg}{fig:postoff}{fig:postoff}{0}%
  \end{ocg}%
  \begin{ocg}{fig:poston}{fig:poston}{1}%
    \includegraphics[width=0.5\hsize]{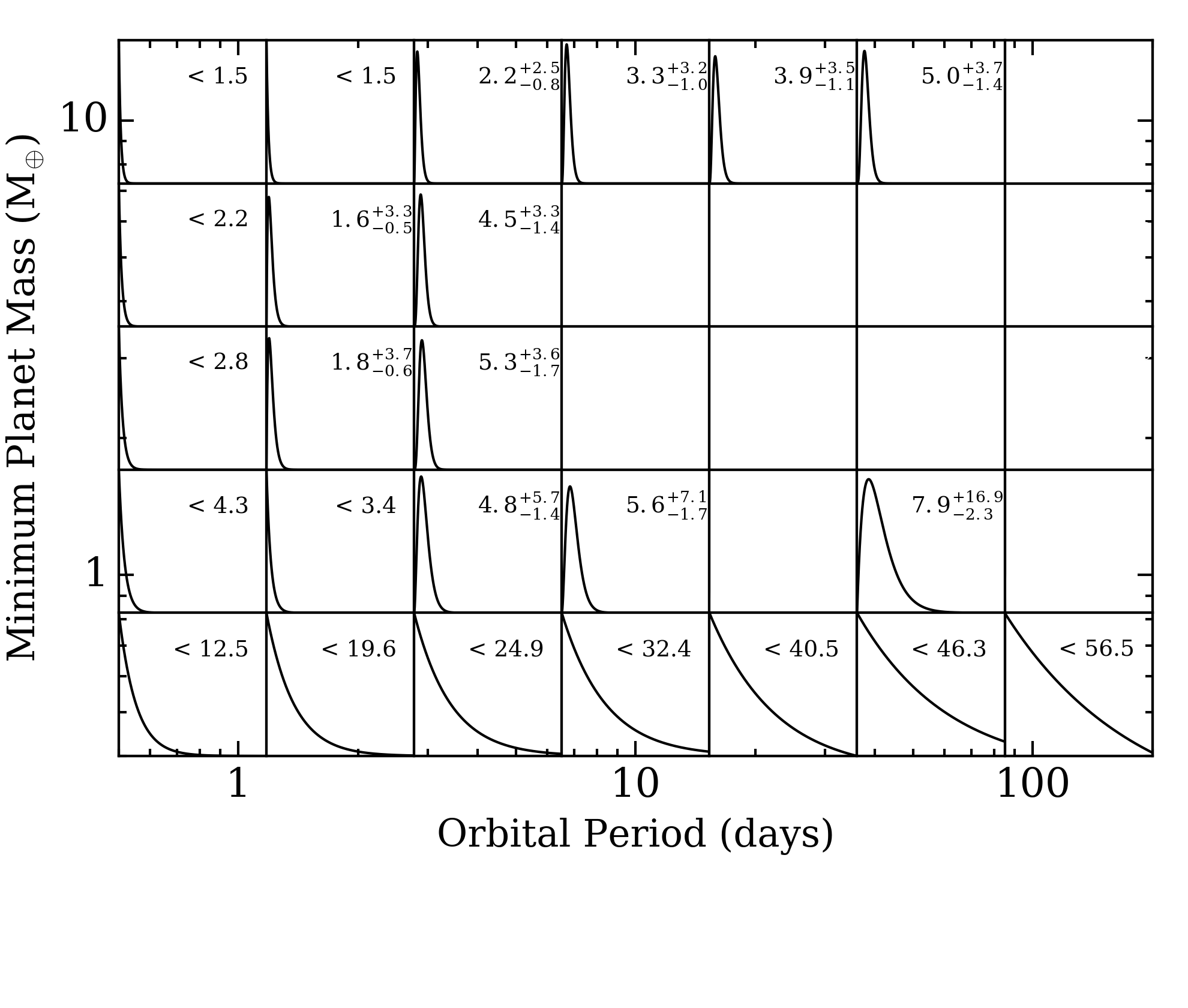}%
    \includegraphics[width=0.5\hsize]{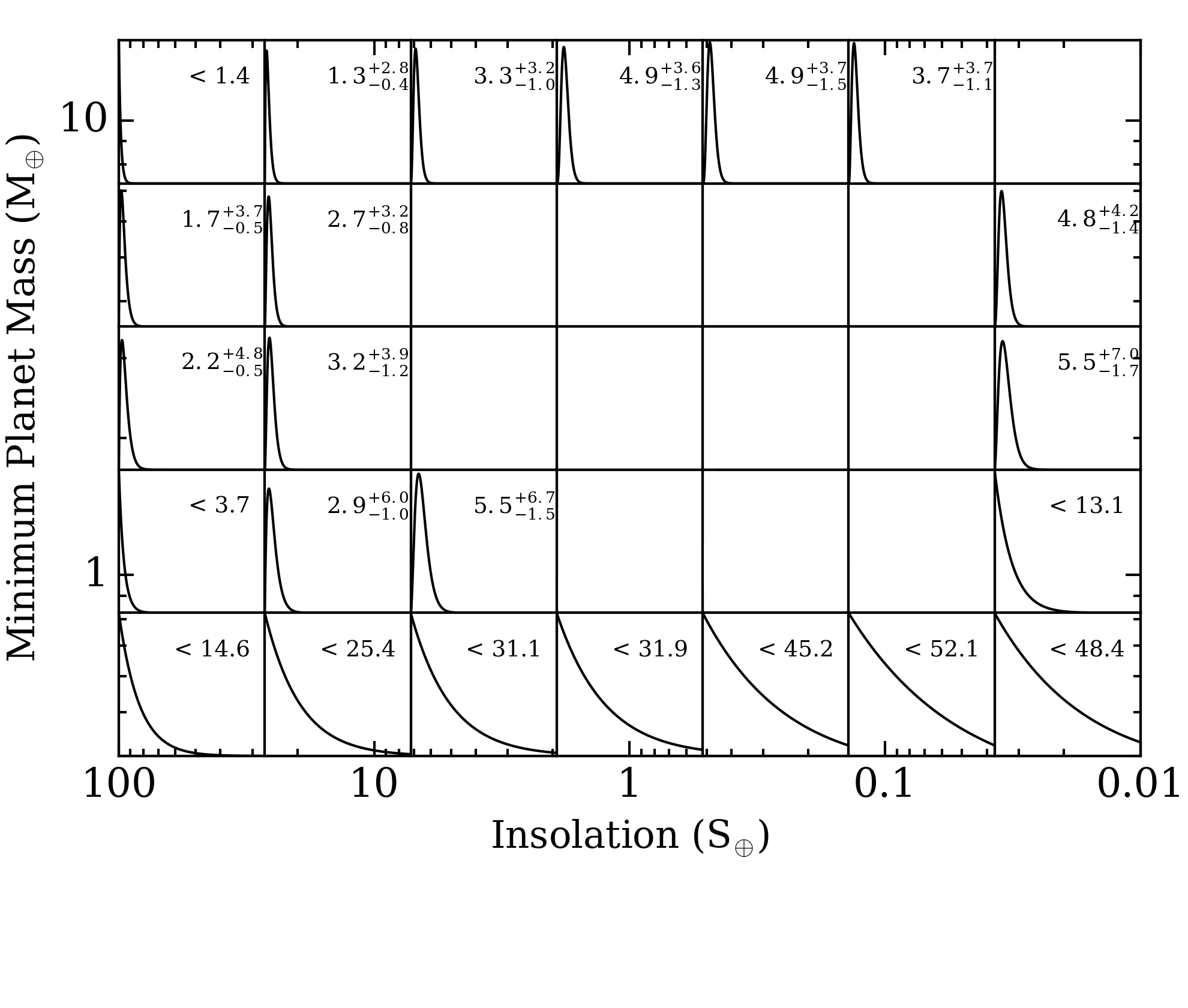}%
  \end{ocg}
  \hspace{-\hsize}%
  \begin{ocg}{fig:hzoff}{fig:hzoff}{0}%
  \end{ocg}%
  \begin{ocg}{fig:hzon}{fig:hzon}{1}%
    \includegraphics[width=0.5\hsize]{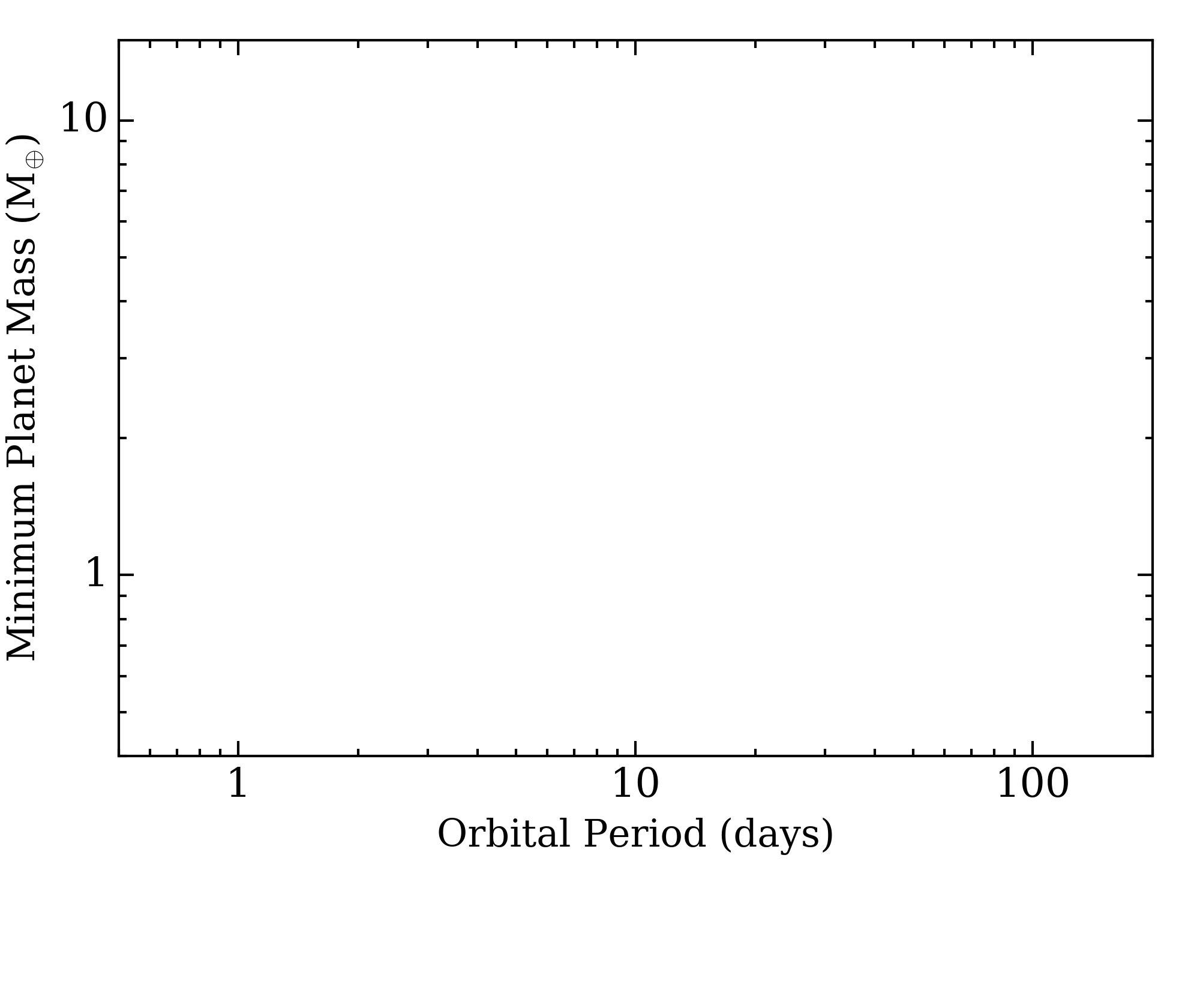}%
    \includegraphics[width=0.5\hsize]{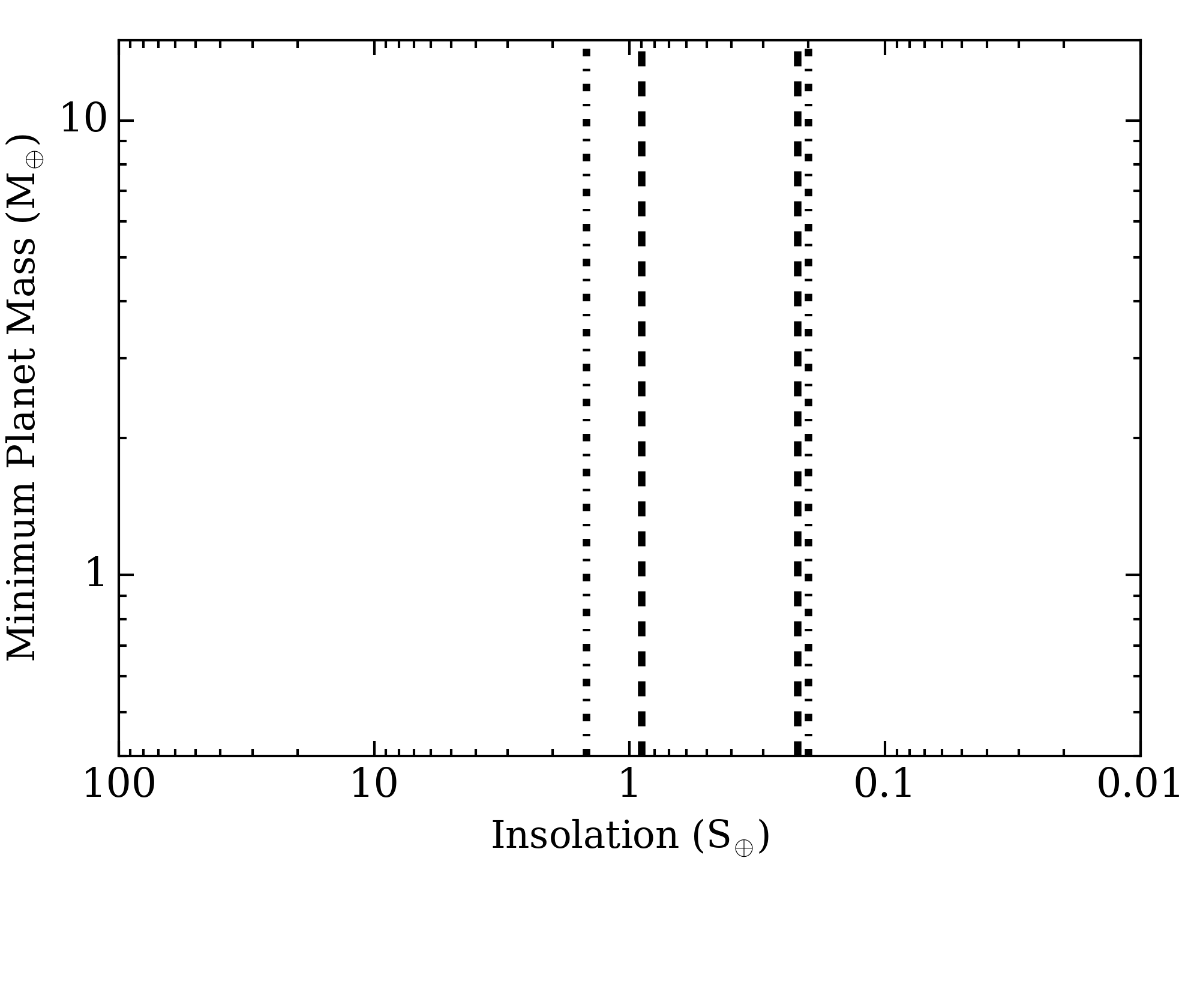}%
  \end{ocg}
  \hspace{-\hsize}%
  \caption{Coarsely binned maps of the RV planet frequency derived from the  
    SLS-PS as a function of minimum planet mass and orbital period 
    (\emph{left}) and insolation (\emph{right}). The \emph{dashed vertical lines} in the
    insolation panel indicate the approximate `water-loss' and `maximum-greenhouse' insolation limits
    of the \ToggleLayer{fig:hzon,fig:hzoff}{\protect\cdbox{HZ}} from \cite{kopparapu13}.
    The \emph{dashed-dotted vertical lines}
    indicate the less conservative `recent-Venus' and `early-Mars' HZ limits \citep{kopparapu13}. 
    Over-plotted in each bin are the   
    \ToggleLayer{fig:poston,fig:postoff}{\protect\cdbox{planet frequency posterior PDFs}} derived from
    the SLS-PS planet detections (Fig.~\ref{fig:yield}) and sensitivity (Fig.~\ref{fig:sensitivity}) using
    Eq.~\ref{eq:freqpost}. When the planet frequency PDF is consistent with 0 we
    report the $68^{\text{th}}$ percentile as an upper limit. When a non-zero planet frequency is detected
    we report the MAP value along with the $16^{\text{th}}$ and $84^{\text{th}}$ percentiles.}
  \label{fig:occurrencegrid}
\end{figure*}

The planet frequency derived using the predicted SPIRou planet detections spanning
$P \in [0.5,200]$ days, $S \in [0.01,100]$ S$_{\oplus}$, and $m_p\sin{i} \in [0.4,15]$ M$_{\oplus}$
are shown in Fig.~\ref{fig:occurrencegrid}. For bins in which the MAP value of the $f$ posterior PDF
is non-zero, we report the MAP value and its $16^{\text{th}}$ and $84^{\text{th}}$ percentiles in
Fig.~\ref{fig:occurrencegrid}. For bins
in which we detect only a small number of planets or have a low detection sensitivity, we find a MAP
$f=0$ and report the $68^{\text{th}}$ percentile as an upper limit.

By considering the SPIRou planet detections and our detection sensitivity within various ranges of planet
parameters, we can estimate the frequency of such planets and the precision of the measurement. We remind
the reader that it is the precision of the measurement that is meaningful here as the MAP values of $f$
are simply the result of combining the known Kepler occurrence rates, an empirical mass-radius relation,
and our detection sensitivity and therefore does not provide any new information regarding the planet
frequencies themselves. However, our adopted formalism (see Eqs.~\ref{eq:freqlike} and \ref{eq:freqpost})
computes the planet frequency over a specified range of planetary parameters.
Recall that the planet frequency is not equivalent to the planet occurrence rate when the occurrence
rate is greater than unity. Therefore, over the full range of $P$, $S$, and \msini{} presented in
Fig.~\ref{fig:occurrence}, we cannot use this formalism to compute the cumulative SPIRou planet occurrence
rate which is $>1$ at $\sim 2.5$ planets per star. Instead we must take the ratio of the 
SPIRou planet detection map shown in Fig.~\ref{fig:yield}---rounded to integer values---with the
sensitivity map shown in Fig.~\ref{fig:sensitivity}.
The resulting map depicts the SPIRou-derived planet frequency,
as a function of $P$, $S$, and \msini{,} which we can then integrate over to estimate the cumulative
planet occurrence rate of $1.8 \pm 0.2$ planets per M dwarf.

\subsection{Measuring $\eta_{\oplus}$}
\begin{figure}
  \centering
  \includegraphics[width=\hsize]{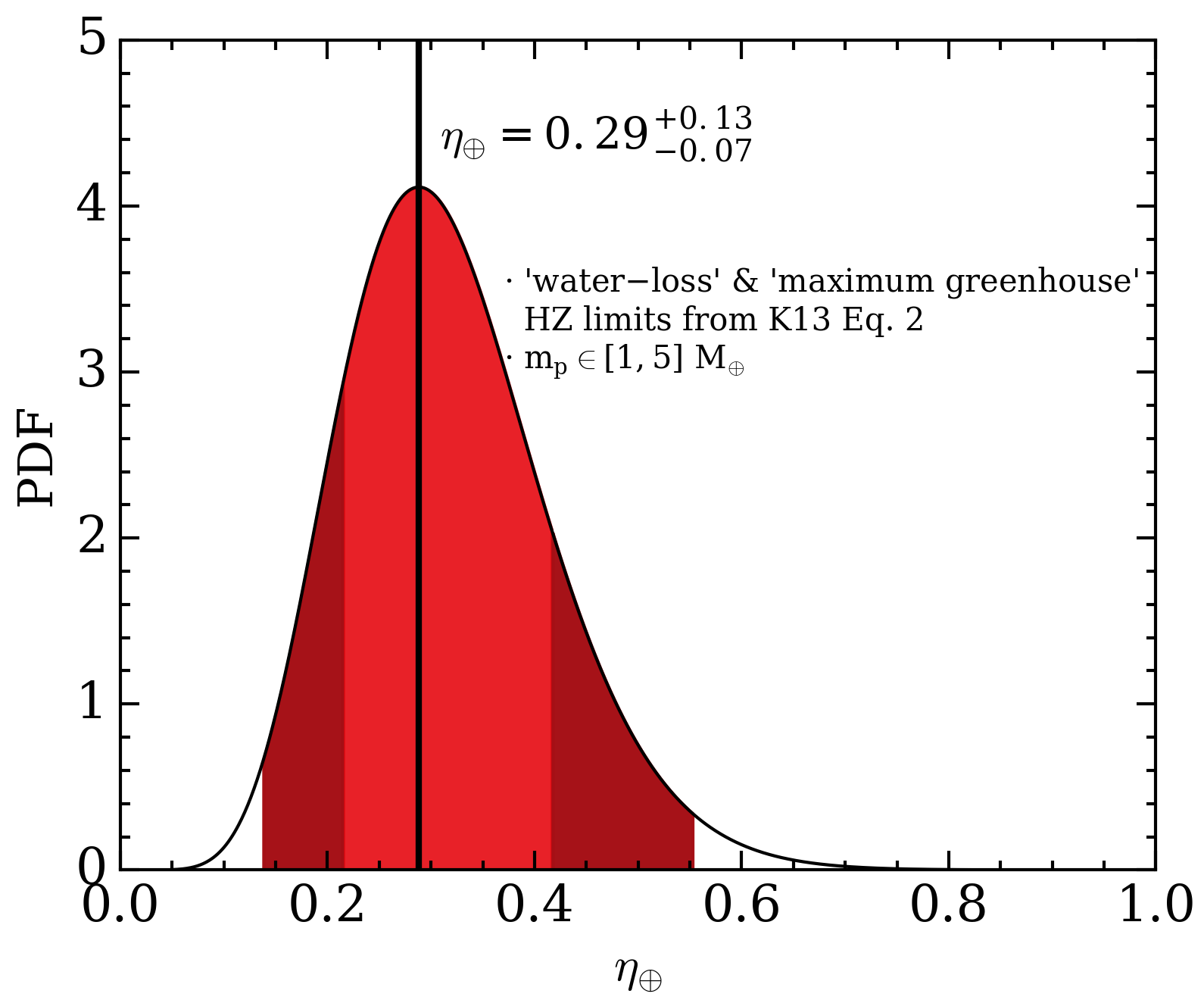}
  \caption{The probability density function of the RV value of $\eta_{\oplus}$ derived from
    the simulated SLS-PS. Here Earth-like HZ planets are defined according to the `water-loss'
    and 'maximum-greenhouse' HZ limits defined in \citealt{kopparapu13} (K13) and have absolute planet
    masses $\leq 5$ M$_{\oplus}$. The \emph{inner shaded} region corresponds to $16^{\text{th}}$ and
    $84^{\text{th}}$ percentiles whereas the outer regions mark the $2^{\text{nd}}$ and
    $98^{\text{th}}$ percentiles.}
  \label{fig:etaEarth}
\end{figure}

Using the formalism from Sect.~\ref{sect:occurrence} to compute the planet frequency,
we can estimate the frequency of any subset of planets. Of particular interest is the frequency 
of potentially habitable planets around M dwarfs; $\eta_{\oplus}$. We will define
potentially habitable---Earth-like---planets as those with $m_p \in [1,5]$
M$_{\oplus}$ and within the fiducial HZ period limits defined by the equations in
\cite{kopparapu13} for the `water-loss' and `maximum-greenhouse'. Here the absolute
planet mass is inferred from the detected population of minimum planet masses by
correcting for the geometrical effect of randomly orientated orbits following a
geometrical distribution.
The upper limit on $m_p \sim 5$ M$_{\oplus}$ approximately corresponds to the expected mass
of a 1.5 R$_{\oplus}$ planet which marks the approximate radius boundary
between rocky/Earth-like and gaseous planets \citep[e.g.][]{valencia13, lopez14, fulton17}.  
By our definition we expect to detect 9 potentially habitable planets
in the full SLS-PS of 100 stars. Based on our derived detection sensitivity to such planets,
we measure $\eta_{\oplus}=0.28^{+0.12}_{-0.07}$ derived from its posterior PDF shown in
Fig.~\ref{fig:etaEarth}.

Focusing solely on late M dwarfs (M5-M9) then we expect to detect
$3.4^{+3.2}_{-1.4}$ potentially habitable planets. As a result we measure a new 
$\eta_{\oplus}=0.29^{+0.24}_{-0.10}$ whose measurement uncertainty is effectively doubled 
relative to the uncertainty on the value of $\eta_{\oplus}$ derived from the full SPIRou input
catalog. This is because only 37/100 stars from the input catalog are
classified as a late M dwarf with spectral type later than M5. The result of a decreased sample size and the,
on-average, lower detection sensitivity around dim late M dwarfs, is a greater uncertainty
on $\eta_{\oplus}$ around late M dwarfs. However recall that existing estimates of $\eta_{\oplus}$
around M dwarfs have been limited to early-to-mid M dwarfs making the SPIRou estimate of
$\eta_{\oplus}$ around late M dwarfs potentially the first of its kind.

\section{Direct Imaging of Nearby Planetary Systems} \label{sect:imaging}
Assessing the habitability of exoplanets relies heavily on probing the planet's atmospheric
conditions. One potential avenue for studying the atmospheres of non-transiting exoplanets is
via high-contrast imaging
wherein photons from the spatially resolved planet are directly detected following the suppression
of quasi-static speckles associated with the bright host star. Various observational techniques such as
adaptive optics/coronagraphy and post-processing techniques (e.g. ADI; \citealt{marois06},
LOCI; \citealt{lafreniere07a}, and KLIP; \citealt{soummer12})
have enabled the direct detection of a number of exoplanets via high-contrast imaging.
However, the planet population for which this technique is currently amenable is limited to self-luminous
sub-stellar objects at large angular separations from their host star e.g. young gas giants on wide orbits with
planet-to-star contrasts of $\mathcal{O}(10^{-4})$.
However the large apertures on-board the up-coming generation of Extremely Large Telescopes (ELTs)
will offer sufficiently high spatial and spectral resolution to reach
the small planet-to-star contrasts in the nIR ($\sim \mathcal{O}(10^{-7})$) required to directly image
a small number of small HZ planets around the closest stars.
Time-resolved rotational color variations of small planets may even permit 
the detection of the number, reflectance spectra, sizes, and longitudinal positions of major surface
features such as continents and/or liquid oceans, and measure cloud properties
\citep{ford01,fujii10,fujii11,cowan09,cowan13}.

To obtain the set of direct images of HZ exoplanets, we must first find the closest habitable
worlds. Of particular focus are M dwarfs in the Solar neighbourhood because of their abundance and the
favorable contrasts of their HZ planets compared to HZ planets around Sun-like stars \citep{crossfield13}. 
The closest HZ exoplanet has already been discovered around Proxima Centauri
\citep[1.3 pc;][]{angladaescude16} but many more M dwarf HZ planets likely remain undetected
within $\lesssim 10$ pc. The SLS-PS will uncover many of these systems and potentially with a lesser total
observation time per planetary system than that which is required to detect M dwarf planetary systems
using optical velocimeters. To identify the subset of SPIRou
detections which are amenable to direct imaging, we adopt the nIR contrast performance expected for
a number of dedicated imagers and compare the detected SPIRou planet population to these contrast curves
in the planet's projected separation/contrast space.

With the up-coming ELT imagers optimized at nIR wavelengths, targeted planets are observed in
reflected light such that the expected planet-to-star contrast is

\begin{equation}
  C = 1.81 \times 10^{-7} A \left( \frac{r_p}{1\text{ R}_{\oplus}} \right)^2
  \left( \frac{a}{0.1\text{ AU}} \right)^{-2}
\end{equation}

\noindent where $r_p$ is the planet's radius, $a$ is the semi-major axis, and
$A$ is the geometric albedo which we assume is 0.3 for all
detected planets in the simulated SLS-PS. At such low planet-to-star contrasts, no existing
high-contrast imager is presently capable of imaging HZ planets. However there are
proposed techniques to achieve such small nIR contrasts which involve
coupling high-contrast imaging capabilities to spatially resolve the targeted planet followed
by the use of 
high-dispersion spectroscopy to filter out the stellar component as a result of the differentially
Doppler-shifted planet and stellar spectra \citep[e.g.][]{snellen15, lovis17}. This technique was
recently used with CRIRES on the VLT to measure carbon monoxide \citep{snellen10, brogi12, dekok13}
and water \citep{birkby13} in the atmospheres of hot Jupiters. This technique has also
become considerably more topical since the discovery of a HZ planet in the closest exoplanetary
system---Proxima Centauri---and the prospect of detecting the potential biosignature O$_2$ in this
system using the VLT \citep{lovis17}.

Similar promising
techniques have led to a suite of contrast curve predictions which indicate the types planets
which may be detected at a particular detection significance given the capabilities of the
imager and the expected S/N achievable based on the target properties and observational strategy. Of these,
we will consider the expected geometric mean of the E-ELT
EPICS\footnote{European-Extremely Large Telescope-Exoplanet Imaging Camera and Spectrograph. Since renamed the
Planetary Camera and Spectrograph; PCS.}
\citep{kasper10} and TMT PFI\footnote{Thirty Meter Telescope-Planet Formation Imager.} \citep{macintosh06}
$5\sigma$ $H$ band contrast curves.
We also consider the space-based hybrid Lyot coronagraph on-board WFIRST \citep{trauger15}.

\begin{figure*}
  \centering
  \includegraphics[width=\hsize]{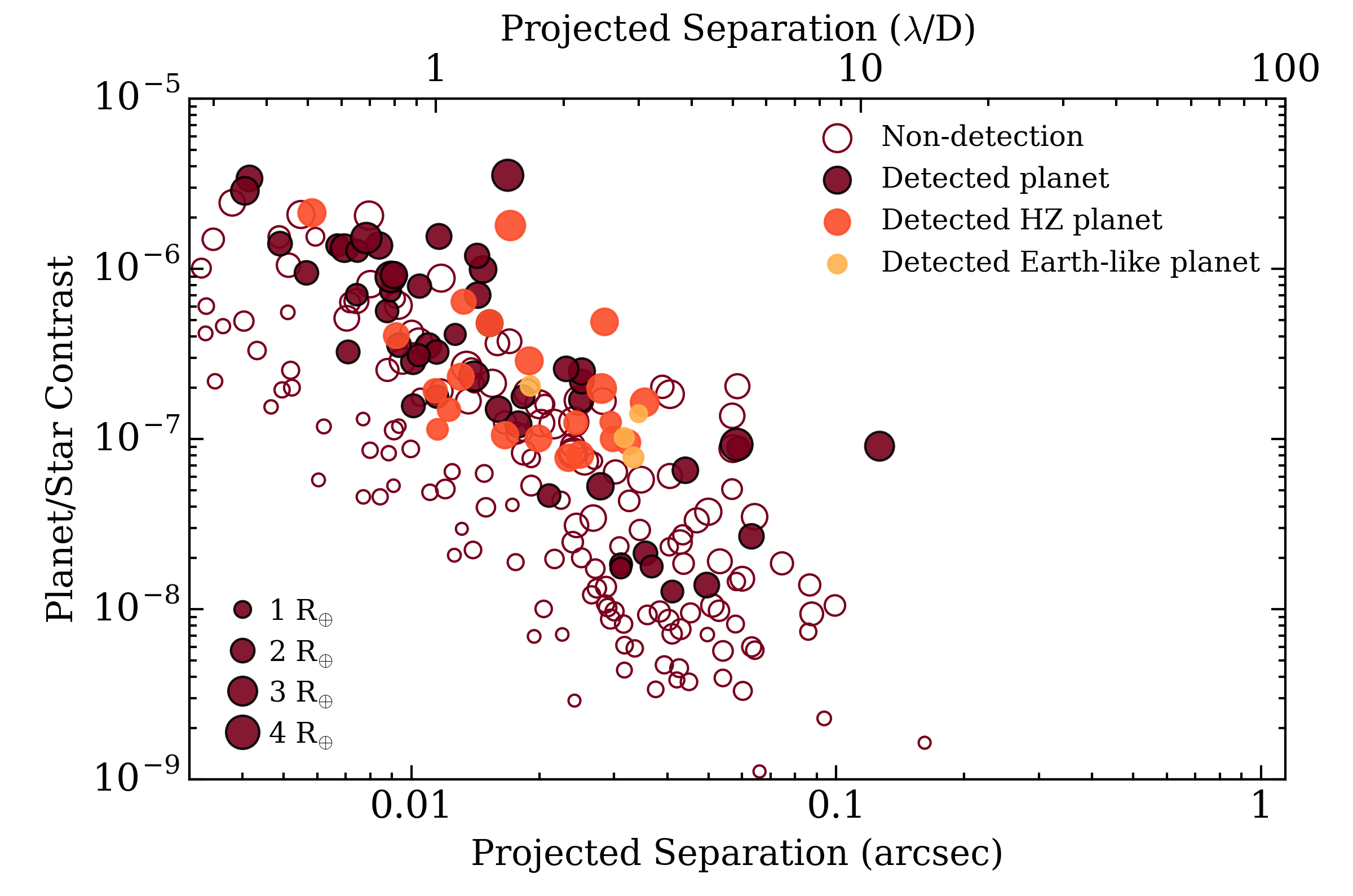}%
  \hspace{-\hsize}%
  \begin{ocg}{fig:Poff}{fig:Poff}{0}%
  \end{ocg}%
  \begin{ocg}{fig:Pon}{fig:Pon}{1}%
   \includegraphics[width=\hsize]{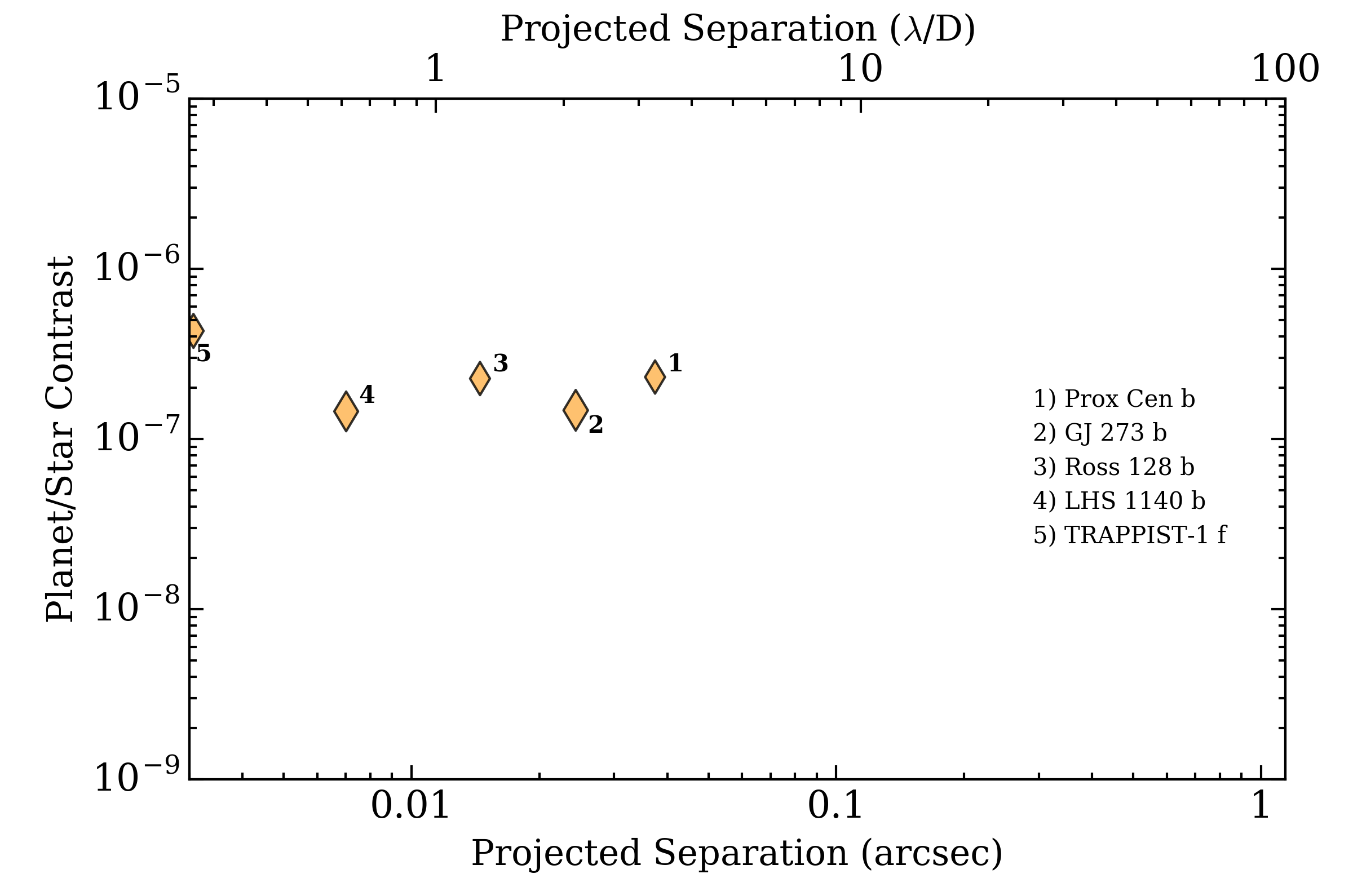}%
  \end{ocg}
  \hspace{-\hsize}%
  \begin{ocg}{fig:c1off}{fig:c1off}{0}%
  \end{ocg}%
  \begin{ocg}{fig:c1on}{fig:c21on}{1}%
   \includegraphics[width=\hsize]{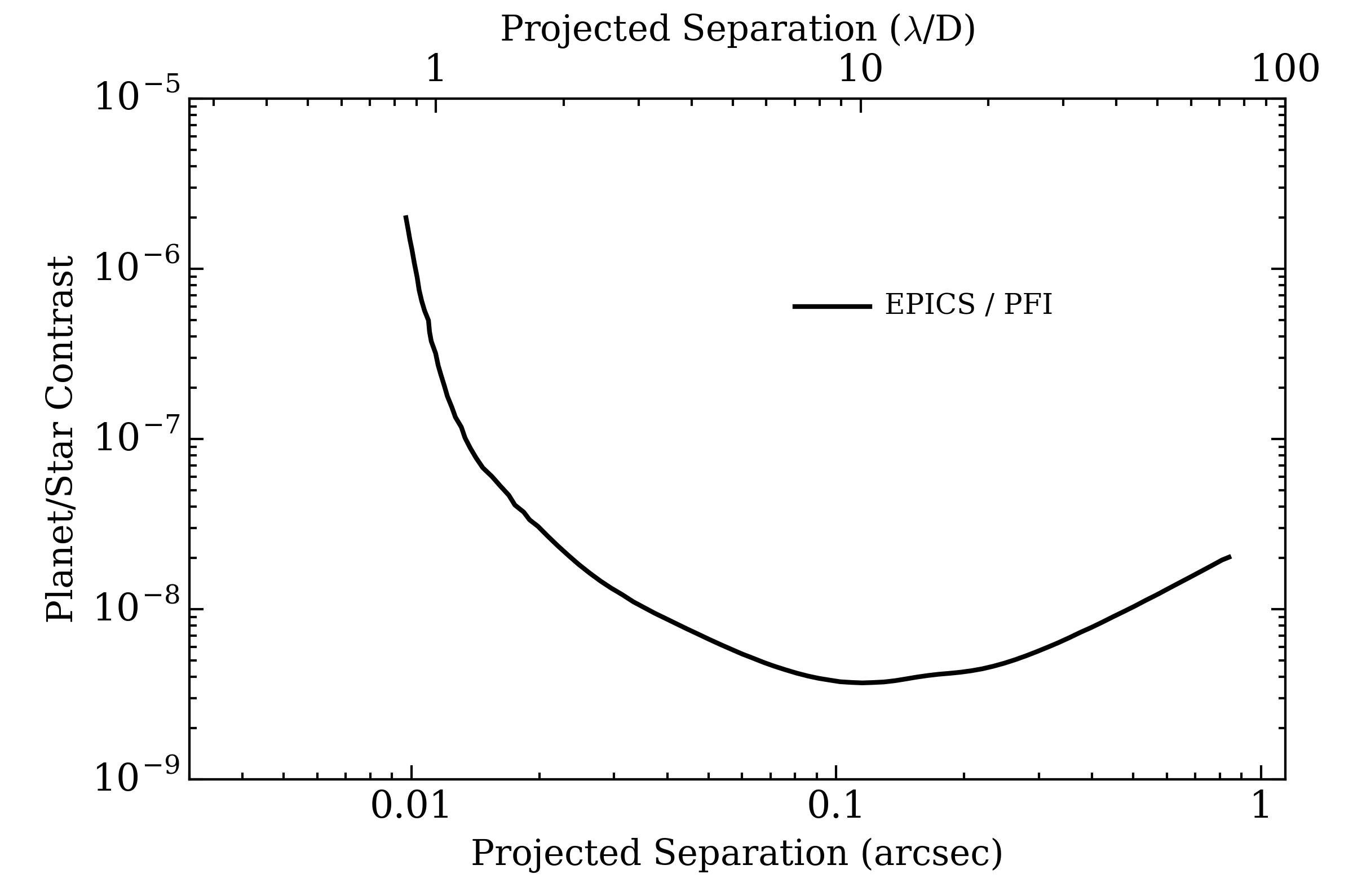}%
  \end{ocg}
  \hspace{-\hsize}%
  \begin{ocg}{fig:c2off}{fig:c2off}{0}%
  \end{ocg}%
  \begin{ocg}{fig:c2on}{fig:c2on}{1}%
   \includegraphics[width=\hsize]{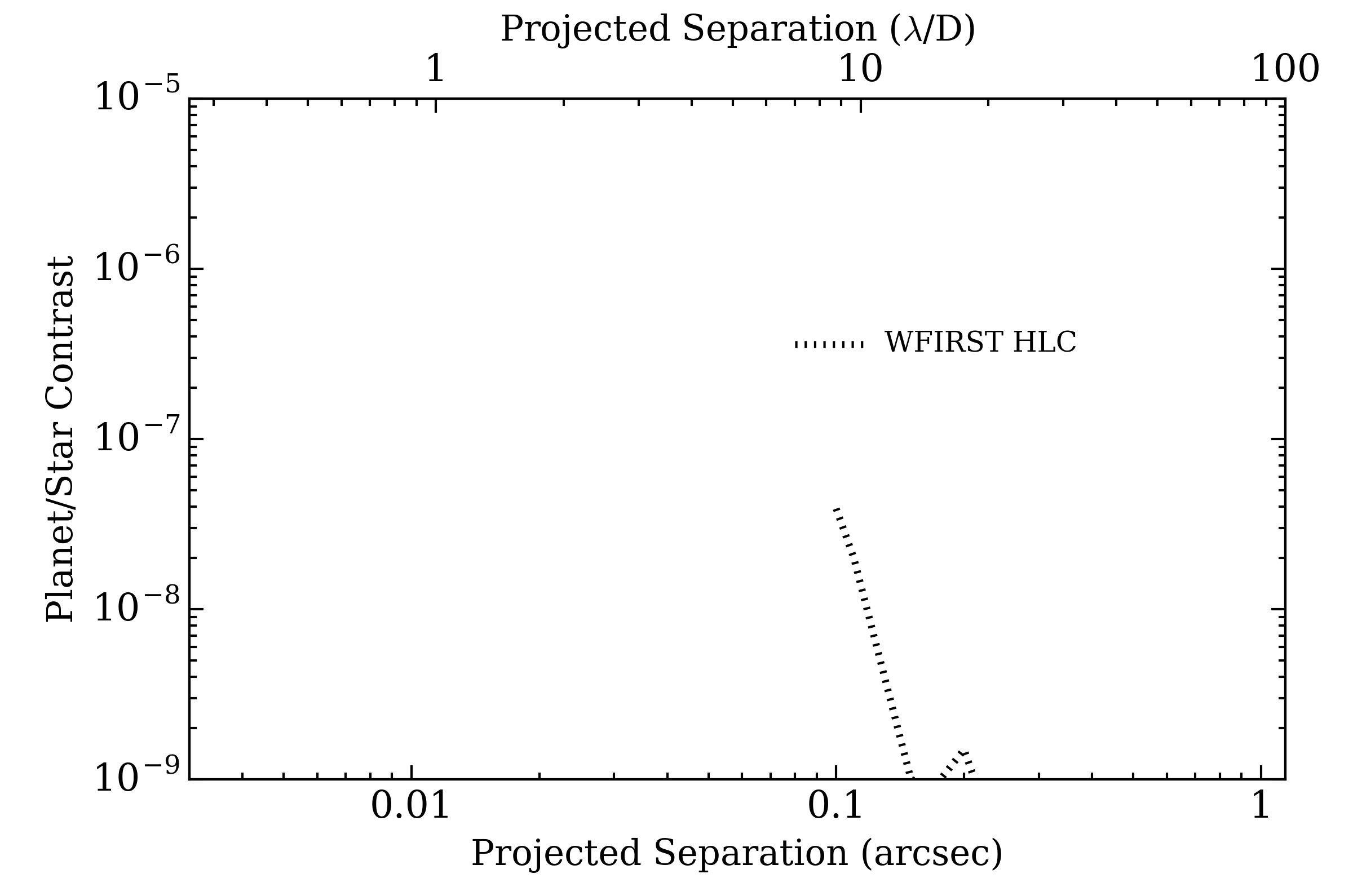}%
  \end{ocg}
  \hspace{-\hsize}%
  \caption{The same random subset of the simulated SPIRou planets from Fig.~\ref{fig:spiroudet}
    in the projected angular separation/reflected light contrast plane. The geometric albedo $A$ is set
    to 0.3 for all planets. 
    \emph{Yellow circles} highlight detected Earth-like planets ($m_p \in [1,5]$ M$_{\oplus}$), \emph{orange circles}
    highlight the remaining detected HZ planets, and \emph{red circles} highlight all non-HZ detected planets. 
    \emph{Open circles} represent planets that remain undetected by SPIRou.
    \ToggleLayer{fig:Poff,fig:Pon}{\protect\cdbox{\emph{Yellow Diamonds}}} labelled
    1-5 depict known, likely rocky planets at or near the HZ. 
    The size of each planet's marker is proportional to its radius.
    The planet population is compared to the geometric mean of the predicted $5\sigma$ $H$ band contrast curves for
    \ToggleLayer{fig:c1off,fig:c1on}{\protect\cdbox{EPICS and PFI}} (\emph{solid curve}) and
    the predicted performance for the 
    \ToggleLayer{fig:c2off,fig:c2on}{\protect\cdbox{WFIRST}} hybrid Lyot Coronagraph (\emph{dotted curve}). The
    projected separation is also depicted in units of $\lambda /D$ in the $H$ band ($\lambda = 1.66$ $\mu$m) for a
    $D=30$ m telescope.}
  \label{fig:spirouimaging}
\end{figure*}

A random subset of planets from the simulated SLS-PS are shown in Fig.~\ref{fig:spirouimaging}.
Comparing this population to the expected performance 
of various ELT imaging instruments, we expect $46.7^{+16.0}_{-6.0}$ SPIRou planets to be
imagable\footnote{Here the term `imagable' need not correspond exactly to an imaging observation as achieving
    the small planet/star contrasts exhibited by SPIRou planets will likely require high contrast imaging
    coupled with high-dispersion spectroscopy \citep[e.g.][]{snellen15}.}.
Here we have defined an imagable planet as one whose expected projected angular separation and contrast
lie above the geometric mean of the EPICS and PFI $5\sigma$ $H$ band contrast curves in
Fig.~\ref{fig:spirouimaging}. Note that this definition does not impose a minimum inner working
  angle although such a cut at say $3 \lambda /D$, would decrease the total number of imagable SPIRou planets by
  a factor of $\sim 2.7$.
The subset of imagable planets represents $\sim 55$\% of all SPIRou planets. Among the imagable planets
are $13.7^{+11.5}_{-4.9}$ HZ planets and $4.9^{+4.7}_{-2.0}$ HZ planets with $m_p \in [1,5]$ M$_{\oplus}$; the so-called
Earth-like planets. These SPIRou planets along 
with the known Proxima Centauri b, Ross 128b \citep{bonfils17a}, and GJ 273b \citep{astudillodefru17} 
will represent the best potential targets for imaging of small HZ exoplanets with ELTs.

\section{Comparison of Different Versions of the SLS-PS} \label{sect:surveys}
Table~\ref{table:summary} summarizes the main results of six simulated versions of the SLS-PS
including the fiducial version presented throughout this paper which we now refer to as the
\emph{optimized} version in Table~\ref{table:summary}. Brief descriptions and motivations
for each additional version of the SLS-PS are given below.

\begin{enumerate}
\item \emph{Optimized}: the SLS-PS version used throughout this study containing 100 stars
  in the input catalog. This version approximately represents the optimal compromise between maximizing
  detection sensitivity and producing a satisfactory number of planet detections, including
  a sizable fraction of planets that may be amenable to direct imaging with ELTs. 
\item \emph{Closest}: contains the 50 closest stars ($d \lesssim 6.8$ pc) from the \emph{optimized}
  input catalog. Here we target the closest M dwarfs to the Solar System in an effort to focus
  observational resources on a small number of target stars thus maximizing detection sensitivity and thus
  the number of detections of small planets that may be imagable with ELTs.
\item \emph{Large}: contains 360 stars in the input catalog where 360 is the number of targeted
  stars in the original
  \href{http://spirou.irap.omp.eu/content/download/232/1348/file/SPIRou-sciencecase-last.pdf}{SPIRou Science Case}
  proposal from 2013. Targeting a large sample of M dwarfs may result in the greatest planet yield
  which is desirable for putting tight constraints on the cumulative planet occurrence rates around M dwarfs
  and $\eta_{\oplus}$. 
\item \emph{Short}: has an identical input catalog to \emph{optimized} with 100 stars but includes
  only half of total available time on-sky; we obtain half as many RVs per star as in \emph{optimized}. This
  experimental setup is undesirable but the simulation is used to demonstrate by how much the SLS-PS detection
  sensitivity and yield suffer given fewer measurements.
\item \emph{Degraded}: has an identical input catalog and window functions to \emph{optimized} with 100
  stars but imposes a degraded noise floor on the RV measurement precision of 2 \mps{} compared to the 1 \mps{}
  noise floor assumed in \emph{optimized}. This survey version is used to access the impact of
  a degradation in the RV measurement precision on the detection sensitivity and planet yield.
\item \emph{Dark}: is identical to the \emph{optimized} survey with 100 stars but whose window functions are
  \emph{not} restricted to non-dark-time only. Here we include dark-time observations thus alleviating the
  strong aliasing of planets with $P=15$ or 30 days as in \emph{optimized}. 
\end{enumerate}

\subsection{Optimized: the optimal survey strategy}
The \emph{optimized} survey version represents an ideal compromise between i) achieving sufficient planet
detection sensitivity to put meaningful constraints on the planet occurrence rates---including
$\eta_{\oplus}$---and ii) to produce a large planet yield including a set of imagable planets and, in
particular, Earth-like imagable planets from the SLS-PS. For other RV planet search campaigns with similar
science goals to those aforementioned, we advocate for a similar survey strategy as \emph{optimized}. With that
said, at the time of writing of this manuscript, the SPIRou input catalog has not been absolutely defined and
will likely be altered between the time of these
simulations and the beginning of the actual SLS-PS. The results presented in this paper are therefore intended as
a guideline to inform how many stars should be included in the SLS-PS.

Here we detect $\sim 8$ Earth-like planets with a sensitivity of $\sim 33.5$\%. This will result in a
constraint on $\eta_{\oplus}$ at a level of precision of $\lesssim 45$. We also detect $\sim 5$ Earth-like
planets that may be amenable to direct imaging with ELTs. This is the largest number of imagable Earth-like
planets detected with SPIRou compared to any other simulated survey version other than \emph{dark}, whose
idealized experimental setup is not feasible with the suite of instruments on the CFHT.

\subsection{Closest: the closest M dwarfs}
The \emph{closest} survey version reduces the size of the SPIRou input catalog to 50 stars but maintains
the same volume of nights. Specifically, the 50 closest stars from the \emph{optimized} SPIRou
input catalog are retained.
By focusing on fewer stars with the same amount of total survey time as in the \emph{optimized}
survey, more observing time can be dedicated to each star. The result is
a higher detection sensitivity to all planets around each star compared to \emph{optimized}.
In this way, we are able to detect $\sim 5$ Earth-like planets, $\sim 4$ of which will be amenable to
direct imaging with ELTs due to their close proximity to the Solar System. 

Although the sensitivity to any given planet is maximized in the \emph{closest} survey version, the stellar
input catalog is too small to result in a larger total planet yield than in \emph{optimized}. The small number
of planet detections also has a detrimental effect on measuring the cumulative planet occurrence rate
around M dwarfs and particularly late M dwarfs.

\subsection{Large: lots of stars}
The \emph{large} survey version was originally considered as the tentative survey strategy for the SLS-PS.
It's aim is to discover the greatest number of exoplanets by surveying many more stars (360 stars) than
in any other considered survey version. Targeting so many stars comes at the expense of a reduced detection
sensitivity per star and a particularly low detection sensitivity to Earth-like planets which seem to require
$\gtrsim 150$ RV measurements to detect \citep[e.g.][]{astudillodefru17}. Although the
overall planet yield in the \emph{large} survey version is high, many of the most interesting
systems---Earth-like planets that may be imagable with ELTs---will remain largely undetected. For example,
we only detect $\lesssim 2$ imagable Earth-like planets in \emph{large} compared to the $\sim 5$ in
\emph{optimized}. Furthermore, the
increase in precision with which the cumulative planet occurrence rate can be measured in \emph{large} is only
marginal, and in our opinion, not worth the small yield of imagable Earth-like planets.

\subsection{Short: the effect of fewer observations}
The \emph{short} survey version features an identical input catalog to the \emph{optimized} version but
with half as many RV observations per star. This survey strategy is useful to characterize the loss in detection  
sensitivity and planet yield if fewer than the total number of possible measurements are
obtained throughout the campaign. The loss in detection sensitivity---and hence in planet yield---compared to
\emph{optimized} evolves
approximately as $\sqrt{n_{\text{obs}}}$ for all but the smallest, Earth-like planets. The loss in sensitivity
for Earth-like planets is worsened by the small number of measured RVs in \emph{small}. The rough scaling of
detection sensitivity with $\sqrt{n_{\text{obs}}}$ is the result of the use of our GP regression activity modelling
to model non-white noise in active time-series and that that modelling performs well on the majority of applicable
systems (see Fig.~\ref{fig:compareGPres}).

\subsection{Degraded: a degraded RV measurement precision}
In the \emph{optimized} version of the survey we had assumed that SPIRou will operate with a long-term RV precision
of $\sigma_{\text{RV}}=1$ \mps{.} This was imposed as an RV noise floor on all stars for which we are likely
to be able to achieve a photon-noise limited RV uncertainty of $<1$ \mps{.} However, given that the
long-term RV stability of SPIRou has yet to be tested on-sky, it is conceivable that a degraded level of RV precision
$>1$ \mps{} may instead be realized. Here we repeat the simulation of the \emph{optimized} survey but increase
the RV noise floor to 2 \mps{.}

Similarly to in \emph{short}, the degradation in detection sensitivity approximately scales as
$1/\sqrt{\sigma_{\text{RV}}}$. The results for \emph{degraded} are therefore closely related to the results for
\emph{short}.

\subsection{Dark: no window function restrictions}
The \emph{dark} survey version relaxes the assumption used in \emph{optimized} that SPIRou observations may
only be obtained during non-dark-time. Although such a scenario is unlikely to be realized for SPIRou on
CFHT---given the scheduling of other instruments on the telescope---window functions that \emph{do} include
dark-time observations may be obtained with other nIR velocimeters like
NIRPS on the ESO 3.6m telescope at La Silla \citep{bouchy17}. This experimental setup represents a best-case
scenario for SPIRou and illustrates how many more planets can be detected with dark-time observations
including a number of HZ planets around M2-4 dwarfs.

We can detect $\sim 3$ more planets in total compared to \emph{optimized}. Most of these planets lie within
the HZ. The number of imagable planets detected is only increased by $\sim 2$ new planets with less than one
additional imagable Earth-like planet in \emph{dark} compared to \emph{optimized}. These modest increases in
the planet yield result in only slight improvements to the measured cumulative planet occurrence rate and
$\eta_{\oplus}$.

\acknowledgements
RC thanks the Canadian Institute for Theoretical Astrophysics for use of the Sunnyvale
computing cluster throughout this work.
RC is partially supported in this work by the National Science and Engineering Research
Council of Canada.
JFD thanks the IDEX initiative of Universit\'e F\'ed\'erale Toulouse Midi-Pyr\'en\'ees (UFTMiP) for generous
fund allocation without which SPIRou could not have been built. JFD also thanks the ERC for funding the
NewWorlds project focused on SPIRou-related science.
X. Dumusque acknowledges the Society in Science - The Branco Weiss Fellowship for its financial support.

\appendix
\section{Computing $\sigma_K$ from the Fisher Information Matrix} \label{app:fisher}
As discussed in Sect.~\ref{sect:vett}, we vet putative planet detections by insisting that
all bona fide planet detections have an RV semi-amplitude detection significance of at least
$3\sigma$; $K/\sigma_{\text{K}} \geq 3$. 
In order to estimate the precision with which a planet's semi-amplitude $K$ can be measured 
in a given RV time-series $y_k=y(t_k)$ we must estimate the semi-amplitude measurement
uncertainty $\sigma_{\text{K}}$ from the Fisher information matrix. The Fisher information matrix $B$
encodes the amount of information about $K$ contained within the dataset $y_k$ and is computed
analytically
from the lnlikelihood given in Eq.~\ref{eq:like} under some simplifying assumptions. Namely,
we treat each detected planet in a given system individually thus restricting the keplerian
model $\mu_k=\mu(t_k)$ to a fixed number of parameters describing a single keplerian orbital
solution. Secondly, we adopt the simplifying assumption that the planet is on a circular
orbit such that only three model parameters (i.e. $\boldsymbol{\theta}=\{P,T_0,K\}$) need be
constrained by the time-series. When considering planets individually we
must absorb all noise sources from unmodelled RV activity and additional planets
into an effective RV uncertainty $\sigma_{\text{eff}}$ equal to the rms of the RVs after the
removal of the planet's keplerian model. Lastly we assume that the noise properties of the
time-series are Gaussian distributed such that we can write the lnlikelihood as

\begin{equation}
  \ln{\mathcal{L}} = -\frac{1}{2\sigma_{\text{eff}}^2} \sum_{k=1}^{n_{\text{obs}}} (y_k - \mu_k)^2 + c, \label{eq:ll2}
\end{equation}

\noindent where $c$ is a constant that is independent of the model parameters in $\boldsymbol{\theta}$
and \nobs{} is the number of observations in the time-series.
Recall that for a circularized planet the keplerian orbital solution reduces to a simple sinusoid which
we write as

\begin{equation}
  \mu_k = -K \sin{\phi_k} \label{eq:kep}
\end{equation}

\noindent where $\phi_k = 2\pi (t_k-T_0) / P$ and $T_0$ represents the epoch of inferior
conjunction for a circularized planet. The elements of the matrix $B$ are then computed via

\begin{equation}
  B_{ij} = -\frac{\partial^2 \ln{\mathcal{L}}}{\partial \theta_i \partial \theta_j}.
  \label{eq:fish}
\end{equation}

\noindent Taking $\boldsymbol{\theta}=\{P,T_0,K\}$, the symmetric Fisher matrix takes the
form

\begin{equation}
  B =
 \begin{bmatrix}
   B_{P,P} & B_{P,T_0} & B_{P,K} \\
   B_{T_0,P} & B_{T_0,T_0} & B_{T_0,K} \\
   B_{K,P} & B_{K,T_0} & B_{K,K} 
 \end{bmatrix} \label{eq:B}
\end{equation}

\noindent and contains six independent terms.
Here we explicitly compute the analytical forms of each independent element of $B$
calculated using Eqs.~\ref{eq:ll2},~\ref{eq:kep}, and~\ref{eq:fish}. The first partials
of the lnlikelihood with respect to each of the model parameters in $\boldsymbol{\theta}$ are

\begin{align}
  \frac{\partial \ln{\mathcal{L}}}{\partial P} &=
  -\left( \frac{\pi K}{P^2 \sigma_{\text{eff}}^2} \right) \sum_{k=1}^{N} (K \sin{2\phi_k} - 2y_k \cos{\phi_k}) (t_k-T_0), \label{eq:partP} \\
  \frac{\partial \ln{\mathcal{L}}}{\partial T_0} &=
  -\left( \frac{\pi K}{P \sigma_{\text{eff}}^2} \right) \sum_{k=1}^{N} (K \sin{2\phi_k} - 2y_k \cos{\phi_k}), \label{eq:partT0} \\
  \frac{\partial \ln{\mathcal{L}}}{\partial K} &=
  \left( \frac{1}{\sigma_{\text{eff}}^2} \right) \sum_{k=1}^{N} \left( K\sin^2{\phi_k} - y_k\sin{\phi_k}  \right). \label{eq:partK}
\end{align}

\noindent We are now in a position to compute the six independent elements of the Fisher matrix using Eq.~\ref{eq:fish}
and the first partials given in Eqs.~\ref{eq:partP},~\ref{eq:partT0}, and~\ref{eq:partK}.

\begin{align}
  \begin{split}
    B_{P,P} &= -\frac{\partial}{\partial P} \left( \frac{\partial \ln{\mathcal{L}}}{\partial P} \right) \\
    &= \frac{2\pi K}{P^3 \sigma_{\text{eff}}^2} \left[ \sum_{k=1}^{N} \left( 2y_k \cos{\phi_k}
      - \frac{2\pi K (t_k-T_0)}{P}\cos{2\phi_k} -\frac{2\pi y_k (t_k-T_0)}{P} \sin{\phi_k} -K \sin{2\phi_k} \right) (t_k-T_0) \right]
  \end{split} \\
  \begin{split}
    B_{T_0,P} &= -\frac{\partial}{\partial T_0} \left( \frac{\partial \ln{\mathcal{L}}}{\partial P} \right) \\
    &= \frac{\pi K}{P^2 \sigma_{\text{eff}}^2} \left[ \sum_{k=1}^{N} \left( 2y_k \cos{\phi_k}
      -\frac{4\pi K (t_k-T_0)}{P} \cos{2\phi_k} -\frac{4\pi y_k (t_k-T_0)}{P} \sin{\phi_k} - K \sin{2\phi_k}  \right) \right]
  \end{split} \\
  B_{K,P} &= -\frac{\partial}{\partial K} \left( \frac{\partial \ln{\mathcal{L}}}{\partial P} \right)
  = \frac{2\pi}{P^2 \sigma_{\text{eff}}^2} \left[ \sum_{k=1}^{N} \left( K\sin{2\phi_k} -y_k\cos{\phi_k} \right) (t_k-T_0) \right] \\
  B_{T_0,T_0} &= -\frac{\partial}{\partial T_0} \left( \frac{\partial \ln{\mathcal{L}}}{\partial T_0} \right)
  = \frac{4\pi^2 K}{P^2 \sigma_{\text{eff}}^2} \left[ \sum_{k=1}^{N} \left( -K\cos{2\phi_k} - y_k \sin{\phi_k} \right) \right] \\
  B_{K,T_0} &= -\frac{\partial}{\partial K} \left( \frac{\partial \l{\mathcal{L}}}{\partial T_0} \right)
  = \frac{2\pi}{P \sigma_{\text{eff}}^2} \left[ \sum_{k=1}^{N} \left( K \sin{2\phi_k} -y_k\cos{\phi_k}  \right) \right] \\
  B_{K,K} &= -\frac{\partial}{\partial K} \left( \frac{\partial \ln{\mathcal{L}}}{\partial K} \right)
  = -\frac{1}{\sigma_{\text{eff}}^2} \left[ \sum_{k=1}^{N} \sin^2{\phi_k} \right] 
\end{align}

Using the above expressions to compute the elements of $B$ we can then compute the covariance matrix $C$ of the model
parameters in $\boldsymbol{\theta}$ via $C=|B^{-1}|$. The diagonal elements of the $3 \times 3$ matrix $C$ are the
estimated measurement variances of the 3 model parameters $\boldsymbol{\theta}$. Therefore the measurement uncertainty of
the planet's semi-amplitude is $\sigma_{K} = \sqrt{C_{K,K}}$.

\bibliographystyle{apj}
\bibliography{refs}

\clearpage
\begin{turnpage}
\begin{deluxetable*}{lcccccc}
  \tabletypesize{\scriptsize}
  \tablecaption{Overview of SLS-PS Versions\label{table:summary}}
  \tablewidth{0pt}
  \tablehead{ & \emph{Optimized} & \emph{Closest} & \emph{Large} & \emph{Short} & \emph{Degraded} & \emph{Dark}}
  \startdata
  Fraction of total available time on-sky\tablenotemark{a} & 1 & 1 & 1 & 0.5 & 1 & 1 \\
  Number of target stars & 100 & 50 & 360 & 100 & 100 & 100 \\
  Average number of RVs per star & 198.1 & 396.0 & 62.0 & 99.5 & 198.1 & 198.1 \\
  Median $\sigma_{\text{RV}}$ [\mps{]} & 1.33 & 1.21 & 1.88 & 1.33 & 2.52 & 1.33 \\
  Median expected $K$ measurement uncertainty [\mps{]} & 0.19 & 0.14 & 0.40 & 0.25 & 0.26 & 0.20 \\
  \hline
  Average detection sensitivity\tablenotemark{b} [\%] & $44.9 \pm 0.5$ & $53.6 \pm 1.1$ (1.2)\tablenotemark{c} & $19.5 \pm 0.2$ (0.4) & $34.4 \pm 0.4$ (0.8) & $34.6 \pm 0.6$ (0.8) & $47.2 \pm 0.7$ (1.1) \\
  Average detection sensitivity to HZ planets\tablenotemark{d} [\%] & $43.1 \pm 1.0$ & $51.2 \pm 2.1$ (1.2) & $19.6 \pm 0.3$ (0.5) & $34.8 \pm 0.9$ (0.8) & $34.0 \pm 1.2$ (0.8) & $48.1 \pm 1.3$ (1.1) \\
  Average detection sensitivity to Earth-like planets\tablenotemark{e} [\%] & $33.5 \pm 1.2$ & $45.4 \pm 2.9$ (1.4) & $8.1 \pm 0.3$ (0.2) & $21.4 \pm 1.0$ (0.6) & $20.7 \pm 1.3$ (0.6) & $35.2 \pm 1.5$ (1.1) \\
  Average detection sensitivity to imagable planets [\%]  & $47.7 \pm 0.7$ & $52.8 \pm 1.3$ (1.1) & $24.5 \pm 0.4$ (0.5) & $37.1 \pm 0.6$ (0.8) & $36.0 \pm 0.9$ (0.8) & $49.7 \pm 1.0$ (1.0) \\
  Average detection sensitivity to imagable HZ planets [\%] & $46.1 \pm 1.3$ & $50.3 \pm 2.3$ (1.1) & $27.9 \pm 1.0$ (0.6) & $37.8 \pm 1.2$ (0.8) & $35.2 \pm 1.6$ (0.8) & $51.5 \pm 1.8$ (1.1)  \\
  Average detection sensitivity to imagable Earth-like planets [\%] & $33.7 \pm 1.6$ & $42.5 \pm 3.1$ (1.3) & $11.4 \pm 1.0$ (0.3) & $20.4 \pm 1.2$ (0.6) & $18.1 \pm 1.7$ (0.5) & $35.7 \pm 2.1$ (1.1) \\
  \hline
  Total planet yield & $85.3^{+29.3}_{-12.4}$ & $50.6^{+15.7}_{-6.5}$ (0.6) & $142.7^{+77.7}_{-29.5}$ (1.7) & $65.7^{+27.8}_{-11.1}$ (0.8) & $65.2^{+27.1}_{-11.0}$ (0.8) & $88.7^{+30.6}_{-13.0}$ (1.0) \\
  Total yield of HZ planets & $20.0^{+16.8}_{-7.2}$ & $12.1^{+10.1}_{-4.3}$ (0.6) & $35.3^{+29.5}_{-12.6}$ (1.8) & $16.4^{+13.7}_{-5.9}$ (0.8) & $15.9^{+13.3}_{-5.7}$ (0.8) & $22.9^{+19.1}_{-8.2}$ (1.1) \\
  Total yield of Earth-like planets & $8.1^{+7.6}_{-3.2}$ & $5.4^{+5.0}_{-2.1}$ (0.7) & $7.6^{+7.2}_{-3.0}$ (0.9) & $5.2^{+4.9}_{-2.1}$ (0.6) & $5.0^{+4.8}_{-2.0}$ (0.6) & $8.6^{+8.1}_{-3.4}$ (1.1) \\
  Total yield of imagable planets & $46.7^{+16.0}_{-6.0}$ & $33.5^{+10.4}_{-3.8}$ (0.7) & $38.3^{+20.9}_{-7.0}$ (0.8) & $36.5^{+15.4}_{-5.4}$ (0.8) & $34.4^{+14.3}_{-5.1}$ (0.7) & $48.3^{+16.7}_{-6.2}$ (1.0) \\
  Total yield of imagable HZ planets & $13.7^{+11.5}_{-4.9}$ & $9.7^{+8.2}_{-3.5}$ (0.7) & $8.2^{+6.9}_{-2.9}$ (0.6) & $11.3^{+9.5}_{-4.0}$ (0.8) & $10.1^{+8.5}_{-3.6}$ (0.7) & $15.5^{+12.9}_{-5.5}$ (1.1) \\
  Total yield of imagable Earth-like planets & $4.9^{+4.7}_{-2.0}$ & $4.0^{+3.7}_{-1.6}$ (0.8) & $1.5^{+1.4}_{-0.6}$ (0.3) & $3.0^{+2.8}_{-1.2}$ (0.6) & $2.5^{+2.3}_{-1.0}$ (0.5) & $5.3^{+5.0}_{-2.1}$ (1.1) \\
  \hline
  Cumulative planet occurrence rate\tablenotemark{b}[planets per star] & $1.8 \pm 0.2$ & $1.8 \pm 0.3$ & $1.8 \pm 0.2$ & $1.8 \pm 0.2$ & $1.8 \pm 0.2$ & $1.8 \pm 0.2$ \\
  Frequency of Earth-like planets, $\eta_{\oplus}$ & $0.29^{+0.13}_{-0.07}$ & $0.26^{+0.16}_{-0.08}$ & $0.32^{+0.18}_{-0.09}$ & $0.30^{+0.18}_{-0.09}$ & $0.26^{+0.18}_{-0.09}$ & $0.27^{+0.11}_{-0.07}$
  \enddata
  \tablenotetext{a}{For a notional survey duration of $\sim 300$ nights over $\sim 3$ years.}
  \tablenotetext{b}{Over range of $P \in [0.5,200]$ days and $m_p\sin{i} \in [0.4,20]$ M$_{\oplus}$.}
  \tablenotetext{c}{Numbers in parentheses indicate the fractional value of the corresponding quantity relative to the \emph{optimized} survey version.}
  \tablenotetext{d}{Based on the `water-loss' and `maximum greenhouse' limits of the HZ from \cite{kopparapu13}.}
  \tablenotetext{e}{Earth-like planets are defined by the `water-loss' and `maximum greenhouse' HZ limits
    \citep{kopparapu13} and have $m_p \in [1,5]$ M$_{\oplus}$.}
\end{deluxetable*}
\clearpage
\end{turnpage}

\end{document}